\documentclass{ws-mod-ehs}
\usepackage{color}
\usepackage{graphics}
\usepackage{epsfig}

\begin{document}

\newcommand{\lae}{\stackrel{<}{\sim}}
\newcommand{\gae}{\stackrel{>}{\sim}}

\def\EPJ{{\em E. Phys. J.} C}
\def\RMP{{\em Rev. Mod. Phys.}}
\def\IJMA{{\em Int. J. Mod. Phys.} A}
\def\PTP{{\em Prog. Theor. Phys.}}
\def\ARNPS{{\em Ann. Rev. Nucl. Part. Sci.}}
\def\PTP{{\em Prog. Theor. Phys.}}
\def\PRr{{\em Phys. Rev.}}
\def\MPLa{{\em Mod. Phys. Lett.}A}

\title{TOP PHYSICS\footnote{\bf Lectures presented at
TASI 2000, {\it Flavor Physics for the Millennium}, June 4-30, 2000,
University of Colorado, Boulder, CO. Preprint number BUHEP-00-23.}}

\author{Elizabeth H. Simmons}

\address{Department of Physics, Boston University \\ 590 Commonwealth
Avenue, Boston, MA   02215,
USA\\and \\
Radcliffe Institute for Advanced Study, Harvard University \\
34 Concord Avenue, Cambridge, MA   02138\\ e-mail: simmons@bu.edu}

 
\maketitle

\abstracts{The Run I experiments at the Fermilab Tevatron Collider
discovered the top quark and provided first measurements of many of its
properties.  Run II (and eventually the LHC and NLC experiments) promise to
extend our knowledge of the top quark significantly.  Understanding the top
quark's large mass, and indeed the origin of all mass, appears to require
physics beyond the Standard Model.  Thus, the top quark may have unusual
properties accessible to upcoming experiments.}


\section{Within the Standard Model}

The three-generation Standard Model (SM) of particle physics came into
existence with the discoveries of the tau lepton\cite{taudisc} and b
quark.\cite{bdisc}  Completing the model required a weak partner for b.
Several important properties of this hypothetical ``top'' quark could be
deduced from measurements of bottom quark characteristics.  The electric
charge of the b quark was related to the ratio
\begin{equation}R = \frac{\sigma(e^+e^- \to hadrons)}
{ \sigma(e^+ e^- \to \mu^+\mu^-)} = \Sigma_q\, (3 Q_q^2).
\end{equation}
The increment in the measured\cite{qb} value $\delta R^{expt} = 0.36
\pm 0.09 \pm 0.03$ at the b threshold agreed with
the predicted $\delta R^{SM} = \frac{1}{3}$, confirming $Q^b = -
\frac{1}{3}$.  Likewise, data\cite{afb} on the front-back asymmetry
for electroweak b-quark production
\begin{equation}
A_{FB} = \frac{\sigma(b\,,\theta>90^\circ)\ -\ \sigma(b\,,\theta<90^\circ)}
{\sigma(b\,,\theta>90^\circ)\ +\ \sigma(b\,,\theta<90^\circ)}
\end{equation}
where $\theta$ is the angle between the incoming electron and outgoing b
quark, showed $A^{expt}_{FB} = -(22.8\pm6.0\pm2.5)\%$ while $A^{SM}_{FB} =
-.25$ was predicted.  Since the $Zb\bar{b}$ coupling depends on the weak
isospin of the b quark, the measurement confirmed that $T_3^b =
-\frac{1}{2}$.  Therefore, the b quark's weak partner in the SM 
was required to be a color-triplet, spin-$\frac{1}{2}$ fermion with
electric charge $Q = \frac{2}{3}$ and weak charge $T_3 = \frac{1}{2}$.

\begin{figure}[tb]
\centerline{\scalebox{.2}{\includegraphics{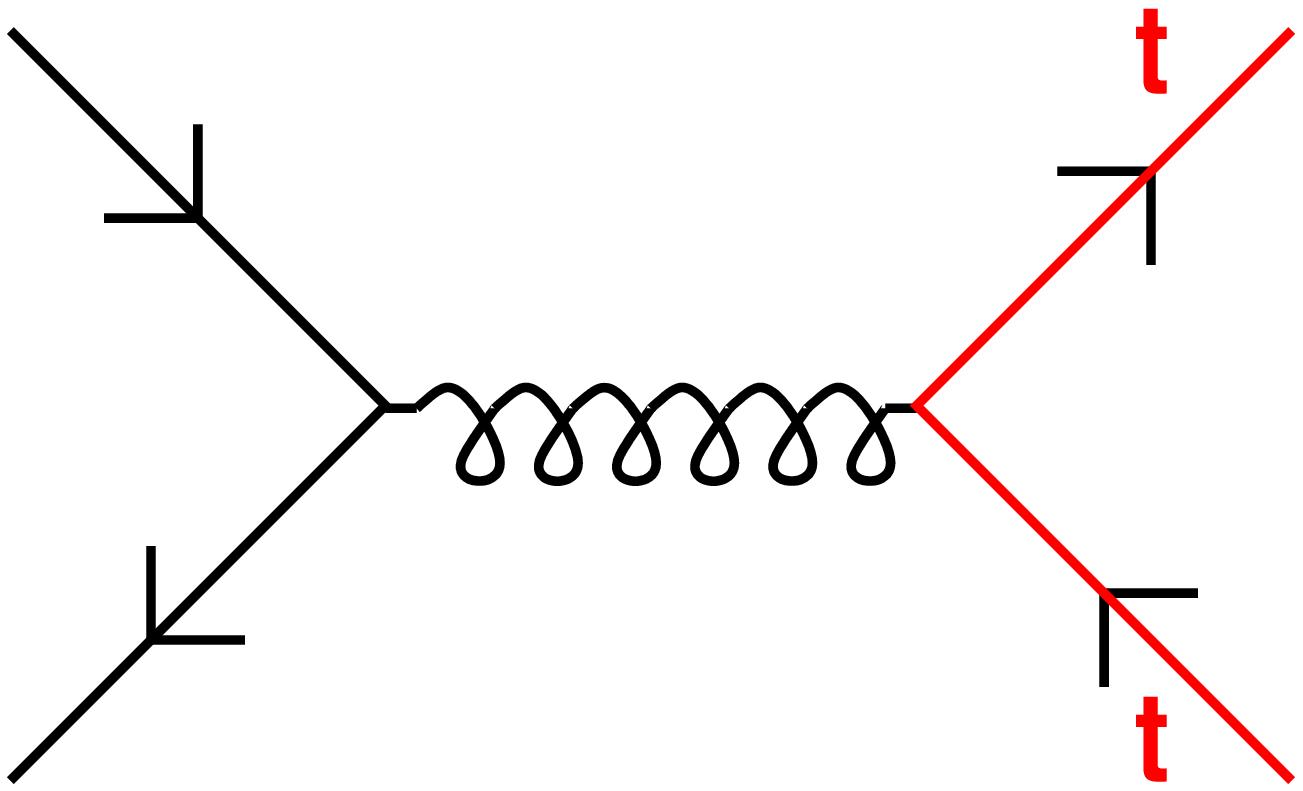}}
\hspace{.75cm}\scalebox{.2}{\includegraphics{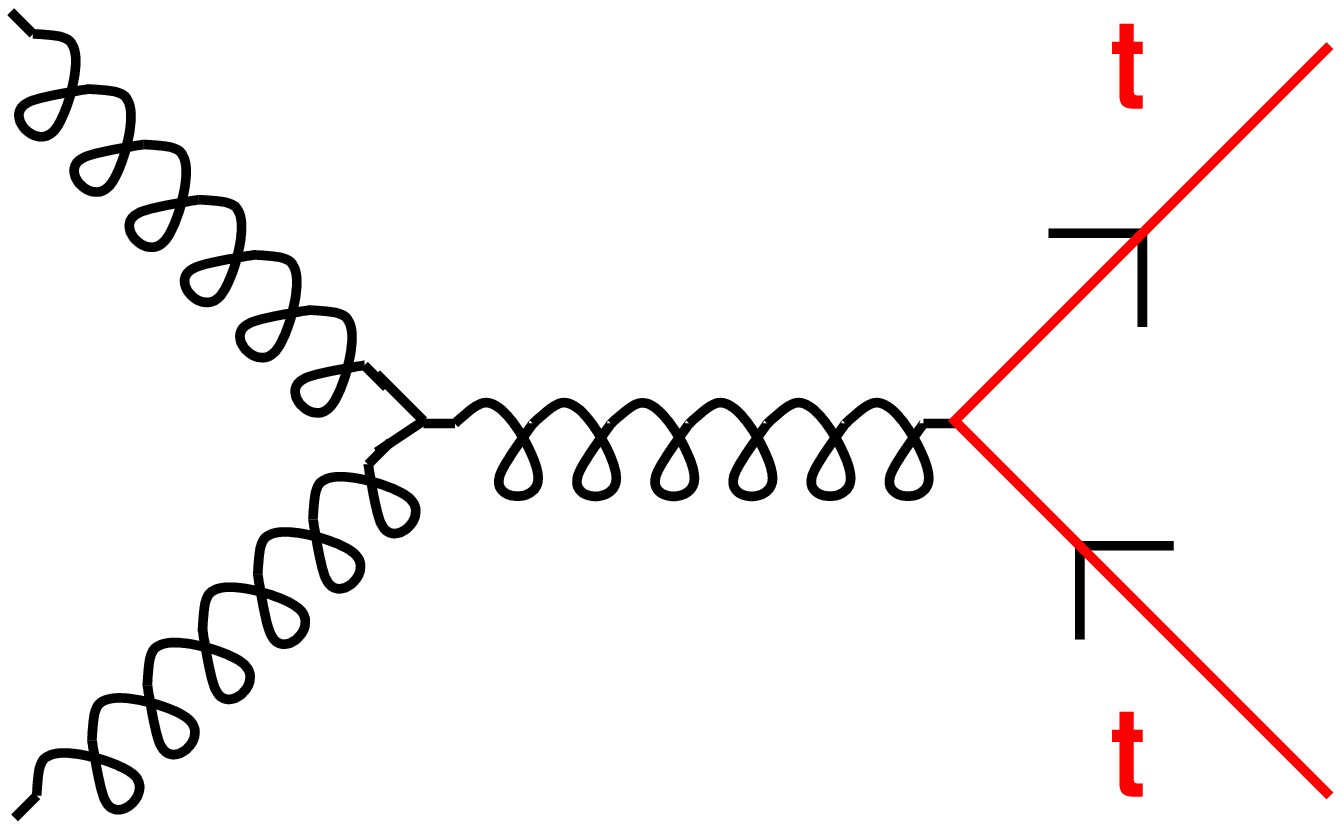}}
\hspace{.75cm}\scalebox{.25}{\includegraphics{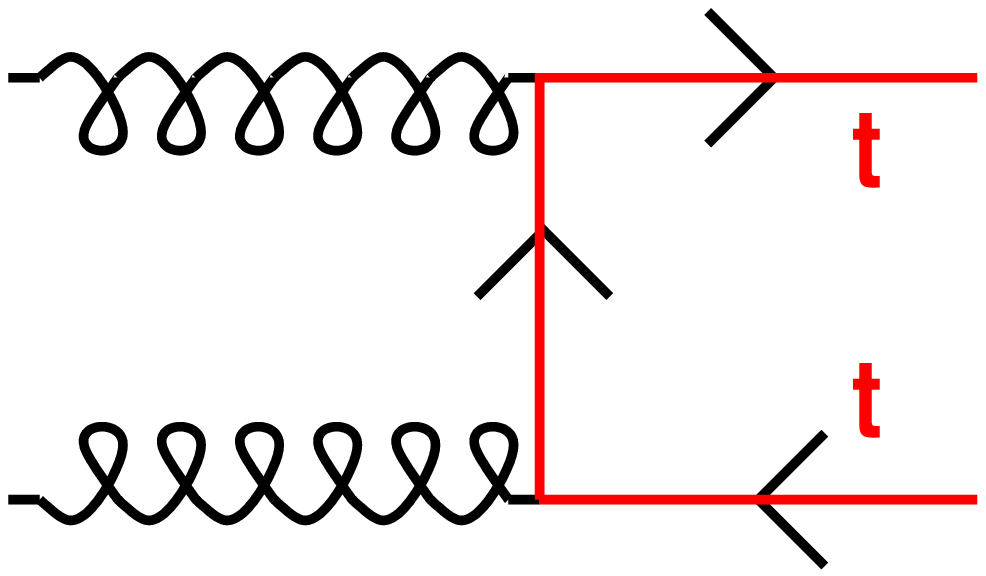}}
\hspace{.75cm}\scalebox{.25}{\includegraphics{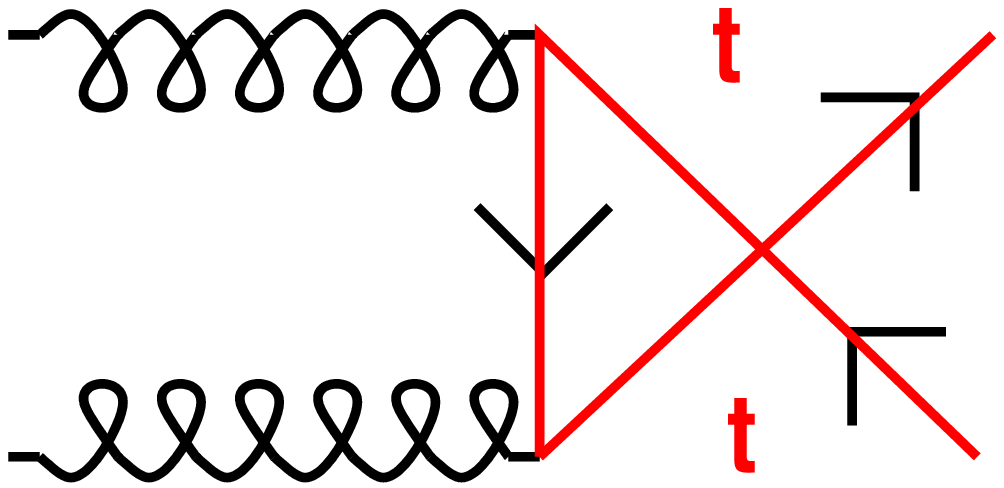}}}
\caption[aa]{Feynman diagrams for QCD pair-production of top quarks.}
\label{fig:qcdprod}
\end{figure}

Such a particle is readily pair-produced by QCD processes involving
quark/anti-quark annihilation or gluon fusion, as illustrated in Figure
\ref{fig:qcdprod}.  At the Tevatron's collision energy $\sqrt{s} = 1.8$ TeV,
a 175 GeV top quark is produced 90\% through $q\bar{q} \to t\bar{t}$ and
10\% through $gg \to t\bar{t}$; at the LHC with $\sqrt{s} = 14$ TeV, the
opposite will be true.  This is because the incoming partons must carry a
momentum fraction of order $m_t / E_{beam}$, a large fraction at the
Tevatron and a small one at LHC, and because the parton distribution
function of gluons is softer than that of valence quarks.  Note that had
the size of $m_t$ been different, weak (single top)
production would have rivaled QCD (pair) production: for $m_t \sim 60$ GeV,
the process $q\bar{q} \to W \to t \bar{b}$ is competitive while for $m_t
\sim 200$ GeV, $W g \to t \bar{b}$ dominates.\cite{franklintop}

In the three-generation SM, the top quark decays primarily to
$W+b$ because $\vert V_{tb} \vert \approx 1$.  As the $W$ can decay into
leptons or hadrons, there are three main classes of final states from top
pair production.  In the ``dilepton'' events (5\% of all $t\bar{t}$
events), both $W$'s decay to $\ell \nu_\ell$ (where $\ell \equiv e,\,\mu$)
and the event includes two b-jets, two leptons and missing energy from two
neutrinos.  In the ``lepton+jets'' events (30\%), there are two b-jets,
two other jets from $W$ decay, one energetic lepton, and missing energy.
The ``all jets'' events (44\%) have multiple jets (including 2 b-jets) and
no hard leptons.  The remaining 21\% of events would include tau leptons
which are harder to identify in high-energy hadron collider experiments.

In 1995, the CDF\cite{cdfdisc} and D\O\cite{d0disc} experiments at
Fermilab discovered a new particle answering the above description and
having a pair-production cross-section consistent with that predicted
for a SM top quark with $m_t$ = 175 GeV.  During Tevatron Run I, each
experiment gathered $\approx$ 125 pb$^{-1}$ of integrated luminosity,
measured some top quark properties in detail and took a first look at
others.  In this section of the talk, we will review the measured
characteristics of the top quark, considered primarily as a Standard
Model particle\footnote{Another useful reference on this topic is
ref.\cite{rmpwill}}.  We will discuss the Run I results on the top
quark mass, width, pair and single production cross-sections, spin
correlations, and decays.  We will also describe the increases in
measurement precision anticipated at Run II and future accelerators
and discuss what we hope to learn.

\begin{figure}[b]
\centerline{\lower15pt\hbox{\scalebox{.25}{\includegraphics{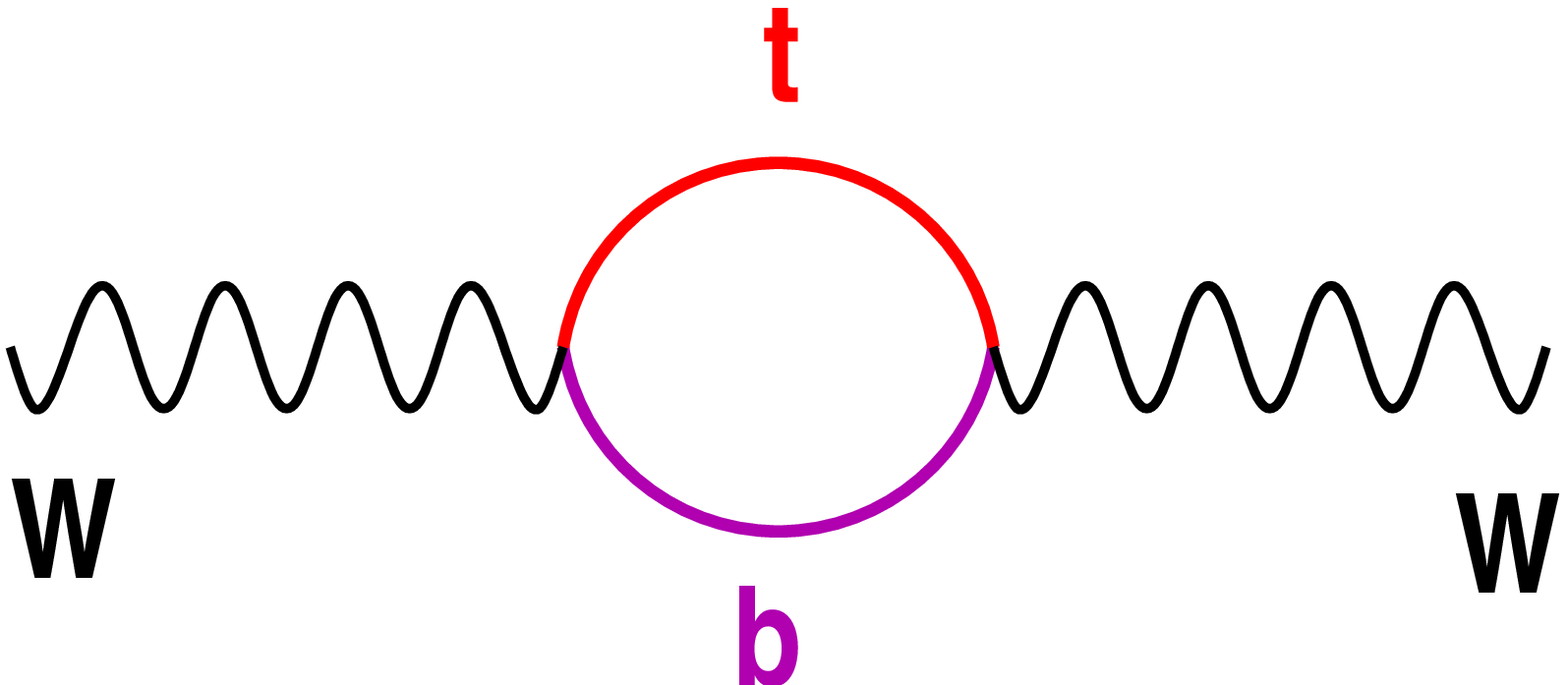}}} \hspace{2.5cm}  \lower25pt\hbox{\scalebox{.21}{\includegraphics{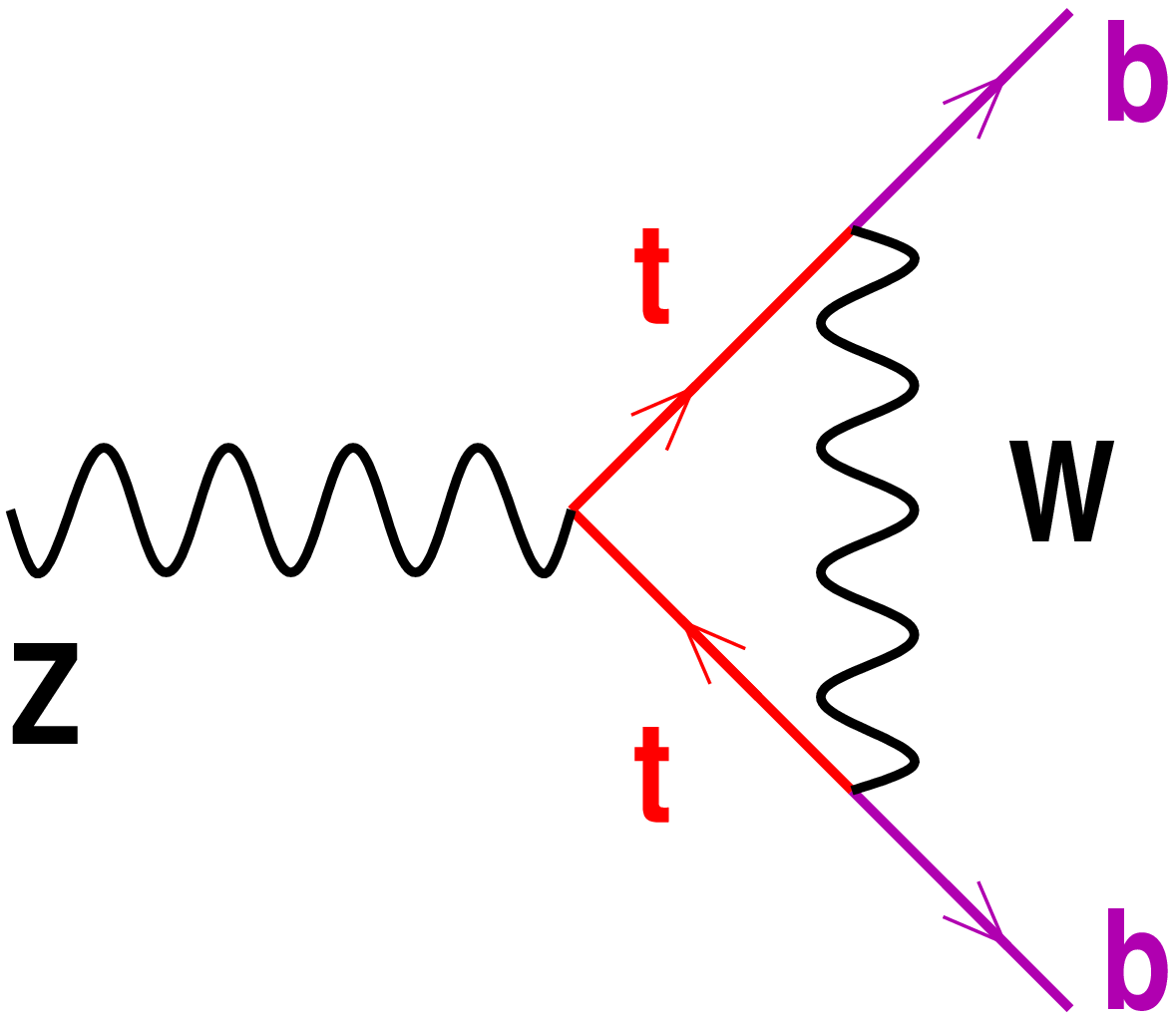}}} + ... }
\caption[ab]{Examples of SM radiative corrections sensitive to $m_t$:
(left) $\Delta\rho$ (right) $Zb\bar{b}$.}
\label{fig:tvirrad}
\end{figure}

\subsection{Mass}\label{subsec:mass}

The top quark mass has been measured\cite{tmasscdf,tmassd0} by
reconstructing the decay products of top pairs produced at the Tevatron.
The most precise measurements use lepton+jets decay channel which affords
both a large top branching fraction and full event reconstruction.  The
combined measurement from CDF and D\O\ is $m_t = 174.3 \pm 5.1$ GeV, as
shown in Table \ref{tab:masscross}.  This implies that the top Yukawa
coupling $\lambda_t = 2^{3/4} G_F^{1/2} m_t$ is approximately 1, so that
the top is the only quark to have a Yukawa coupling of ``natural'' size.

\begin{table}[bt]
\caption[ll]{Measured\protect\cite{wattshf8} $m_t$ and $\sigma_{tt}$ from CDF
and D\O .}
\begin{center}
\begin{tabular}{|l|l|c|c|}
\hline
experiment & channel & $m_t$ (GeV)& $\sigma_{tt}$ (pb)\\
\hline
CDF & dilepton      & 167.4 $\pm$ 11.4 & 8.4$^{+4.5}_{-3.5}$\\
    & lepton + jets & 175.9 $\pm$ 7.1  & 5.1 $\pm$ 1.5\ \ \ \ \ \ \ 
9.2 $\pm$ 4.3\\
    &               &        &  (SVX b-tag) \ \ \ (soft lepton tag) \\
    & all jets      & 186.0 $\pm$ 11.5 & 7.6$^{+3.5}_{-2.7}$  \\
    & combined      & 176.0 $\pm$ 6.5  & 6.5$^{+1.7}_{-1.4}$ ($m_t = 175$)\\
\hline
D\O & dilepton      & 168.4 $\pm$ 12.8 & \\
    & lepton + jets & 173.3 $\pm$ 7.8  & 4.1 $\pm$ 2.1\ \ \ \ \ \ \ 
8.3 $\pm$ 3.6\\
    &               &        & (topological)\ \ \ (soft lepton tag)\\
    & all jets      &        & 7.1 $\pm$ 3.2\\
    & combined      & 172.1 $\pm$ 7.1  & 5.9 $\pm$ 1.7 ($m_t = 172$)\\
\hline
Tevatron & combined & 174.3 $\pm$ 5.1  &    \\
\hline
\end{tabular}
\end{center}
\label{tab:masscross}
\end{table}

\begin{figure}[tb]
\begin{center}
$\ $ \vspace{-1cm}
$\ $\hspace{-1cm}\scalebox{.33}{\includegraphics{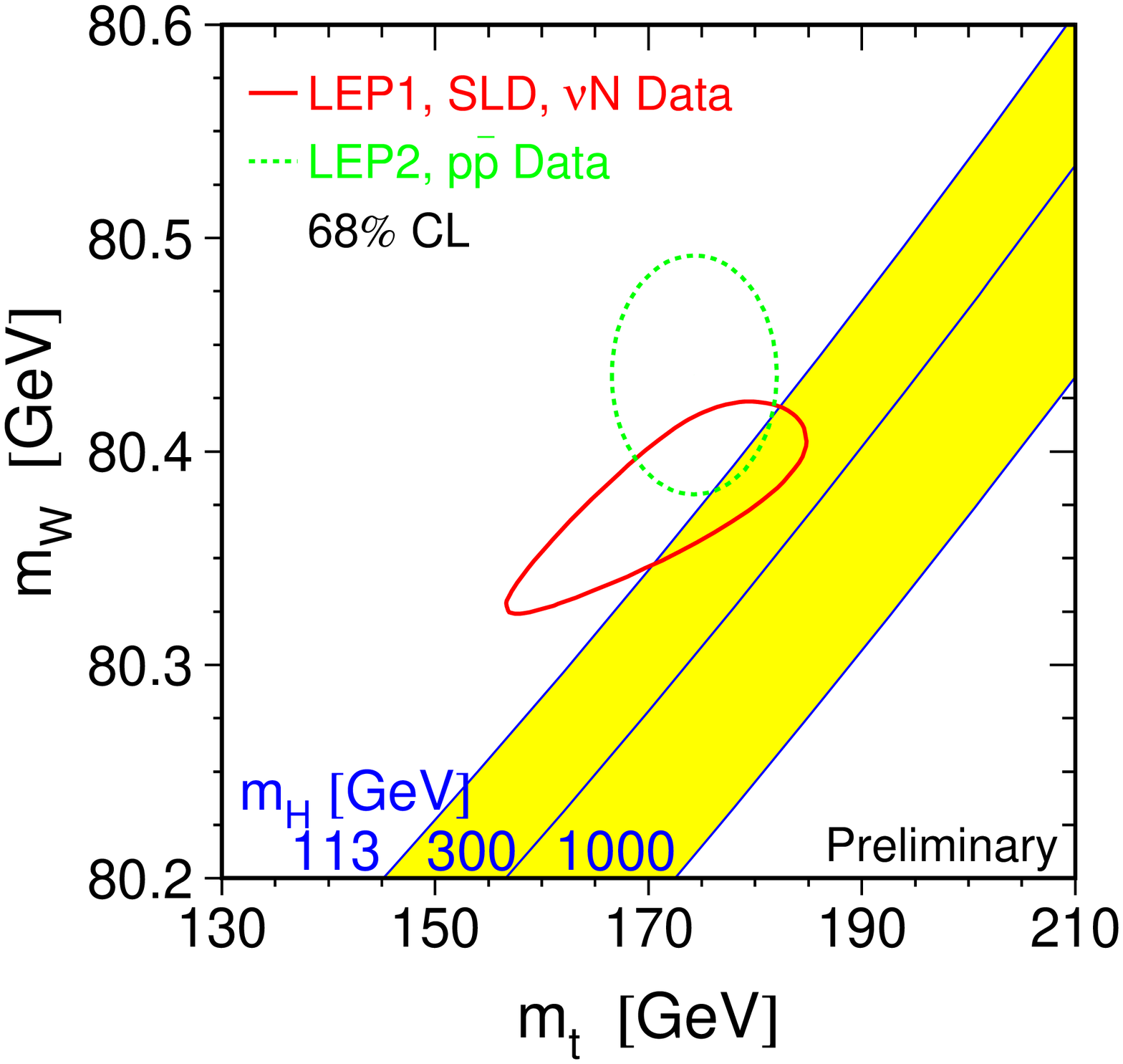}}
\end{center}
\vspace{-.3cm}
\caption[abh]{Predicted\cite{degrassi} $M_W(m_h, m_t)$ in the SM
compared\cite{lepewwg} to data on $M_W$, $m_t$.}
\label{fig:mwmtmh}
\end{figure}

The top quark's mass is already known to $\pm$3\%, comparable to the
precision with which $m_b$ is measured and better than that for the
light quarks.\cite{pdg} This is quite impressive given that the top
quark was discovered nearly 20 years after the bottom!  This precision
is also quite useful in interpreting other measurements because many
electroweak observables are subject to radiative corrections sensitive
to $m_t$.  As illustrated in Figure \ref{fig:tvirrad}, for example,
the $W$ mass (which enters $\Delta \rho$) and the $Zb\bar{b}$ coupling
(which enters $R_b$) are affected by virtual top quarks.  Comparing
the experimental constraints on $M_W$ and $m_t$ with the SM
prediction\cite{degrassi} for $M_W (m_t, m_{Higgs})$ provides an
opportunity to test the consistency of the SM and to constrain
$m_{Higgs}$.  As Figure \ref{fig:mwmtmh} shows, the current data are
suggestive, but not precise enough to provide a tightly-bounded value
for $m_{Higgs}$.  Run II measurements of the $W$ and top masses are
expected\cite{wattshf8} to yield $\delta M_W \approx$ 40 MeV (per
experiment) and $\delta m_t \approx$ 3 GeV (1 GeV in Run IIb or LHC).
With this precision, it should be possible to obtain a much tighter
bound\cite{wattshf8} on the SM Higgs mass: $\delta M_H / M_H \leq
40$\%.

A far more precise measurement, with $\delta m_t \approx 150$ MeV,
could in principle be extracted from near-threshold NLC\cite{nlcmt} data on 
$\sigma(e^+e^- \to t\bar{t})$.  The calculated line shape shows
a distinct rise at the remnant of what would have been the
toponium 1S resonance if the top did not decay so quickly.  The location of
the rise depends on $m_t$; the shape and size, on the decay width
$\Gamma_t$.  This measurement has the potential for good precision
because it is based on counting color-singlet $t\bar{t}$ events, making it
relatively insensitive to QCD uncertainties.  

\begin{figure}[t]
\begin{center}
 \scalebox{.2}{\includegraphics{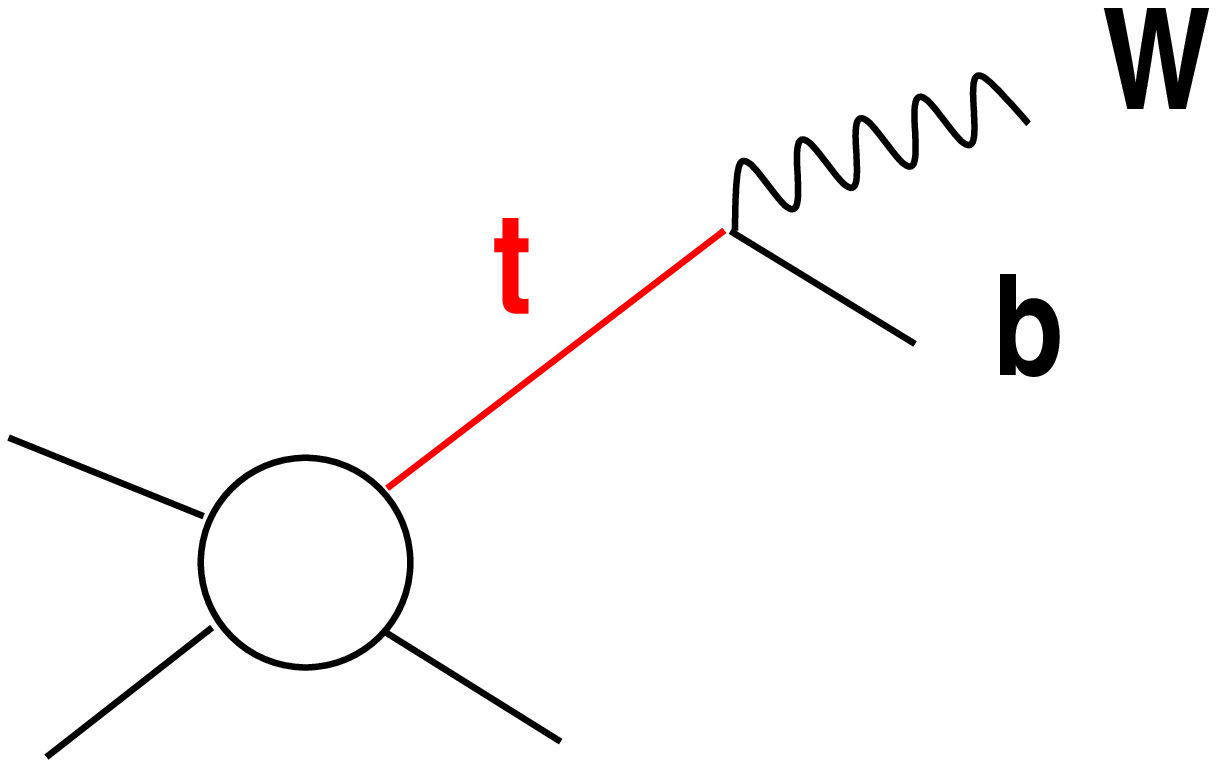}}\hspace{1cm}\scalebox{.2}{\includegraphics{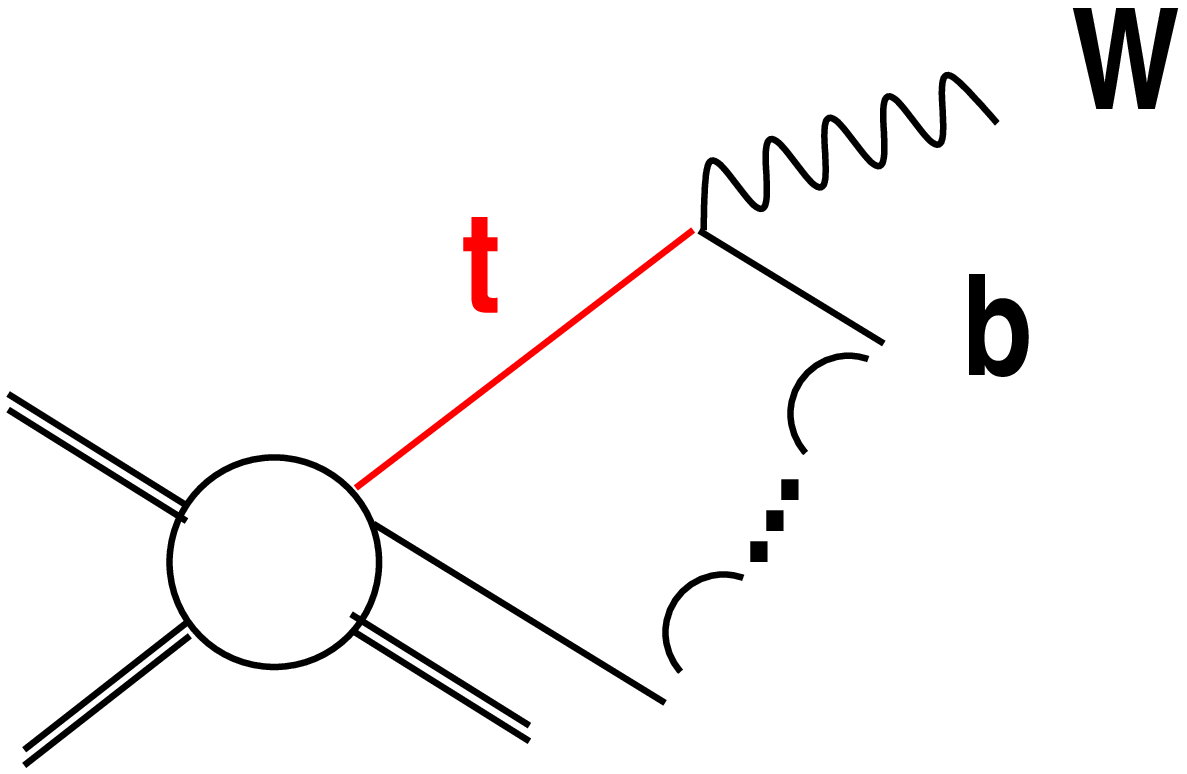}}\hspace{1cm}$\Sigma^\infty_{n=0}$ \scalebox{.15}{\includegraphics{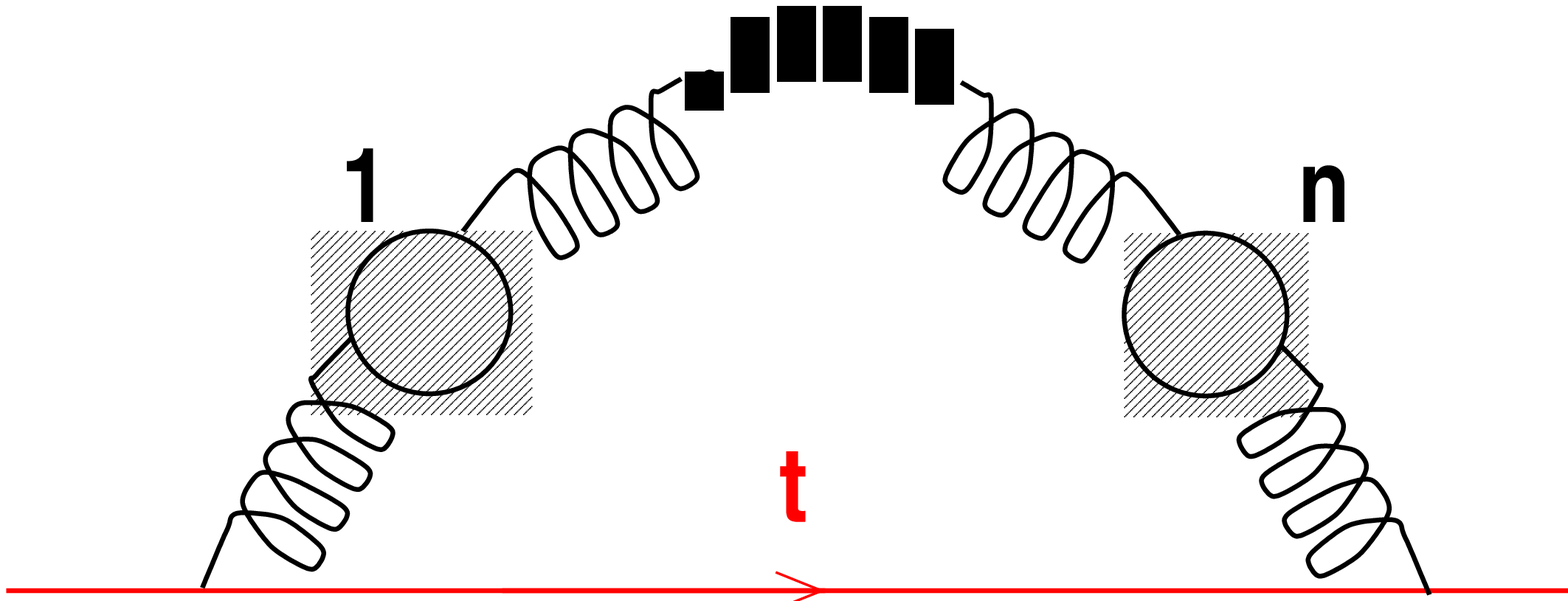}}
\end{center}
\caption[ab]{(left) Top production and decay. (center) Same, with b-quark
hadronization indicated. (right) Soft gluon resummation in the top
propagator. After ref. \cite{willen}.}
\label{fig:massdiag}
\end{figure}

Taking advantage of this requires a careful choice of the definition of
$m_t$ used to extract information from the data.  Consider, for example,
the mass appearing in the propagator $D(p\!\!\!/) = i / (p\!\!\!/ - m_R -
\Sigma(p\!\!\!/))$ .  In principle, one can reconstruct this mass from the
four-vectors of the top decay products, as is done in the current Tevatron
measurements.  But this pole mass is inherently uncertain to ${\cal
O}(\Lambda_{QCD})$.  For example, the clean top production and decay
process sketched in Figure \ref{fig:massdiag}(left) is, in reality,
complicated by QCD hadronization effects which connect the b-quark from top
decay to other quarks involved in the original scattering,\cite{willen} as
in Figure \ref{fig:massdiag}(center).  Attempting to sum the soft-gluon
contributions to the top propagator sketched in Figure \ref{fig:massdiag}
(right) yields the same conclusion.  Taking the Borel transform of the
self-energy allows one to effect the summation,\cite{borel} but real-axis
singularities (infrared renormalons) in the Borel-transformed self-energy
impede efforts to invert the transform.\cite{willen}  The ambiguity
introduced in distorting the integration contour of the inverse Borel
transform around the singularities is of order\cite{ambig} $\Lambda_{QCD}$.

\begin{figure}[tb]
\begin{center}
{\hspace{0cm}\scalebox{.8}{\includegraphics*[40mm,205mm][100mm,250mm]{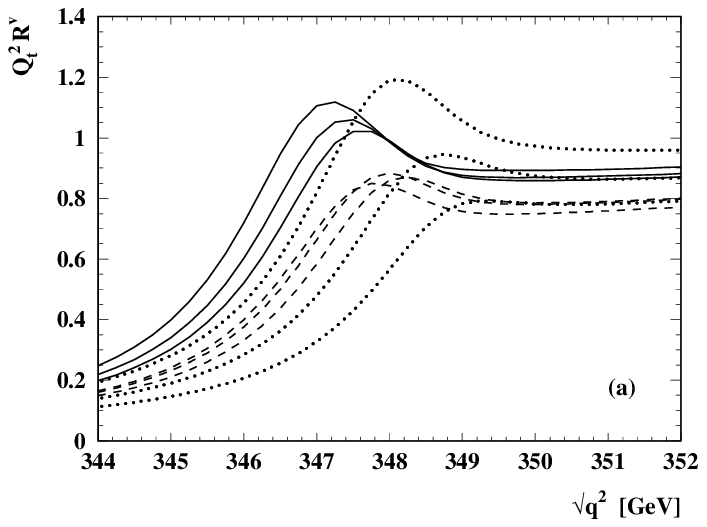}}} \hspace{2cm}
\hbox{\scalebox{.8}{\includegraphics*[40mm,205mm][100mm,250mm]{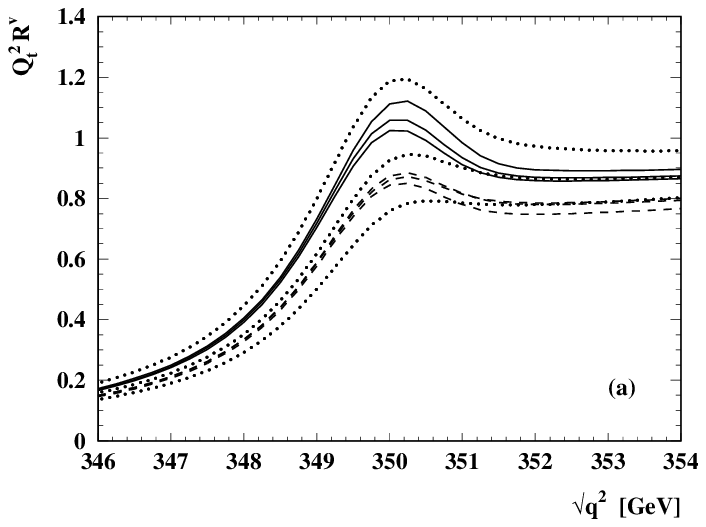}}}
\end{center}
\caption[ab]{Near-threshold cross-section for photon-induced top production
at an NLC\cite{hoang} calculated (left) in the pole mass scheme and (right)
in the 1S mass scheme.  Leading-order (dotted), NLO (dashed) and NNLO
(solid) curves are shown with renormalization scales $\mu$ = 15 (topmost),
30, and 60 GeV}
\label{fig:phoindtt}
\end{figure}

Using a short-distance mass definition avoids these
difficulties.\cite{hoangsum}  For example, one could adopt the $\overline{MS}$
mass definition
\begin{equation}
\bar{m}(\bar{m}) = m_{pole}\left( 1 + \frac{4 \bar{\alpha}_s}{3\pi} + 8.3 \left(\frac{\bar{\alpha}_s}{\pi}\right)^2 + ...\right)^{-1}
\end{equation}
although the numerical value lies about 10 GeV below $m_{pole}$,
which is inconvenient for data analysis.  Another is the 1S mass\cite{hoang}
\begin{equation}
m_{1S} = m_{pole} - \frac{2}{9}\alpha_s^2 m_{pole} + ... 
\end{equation}
where $2 m_{1S}$ is naturally near the peak of $\sigma(e^+e^- \to
t\bar{t})$ .  Others include the potential-subtracted\cite{psms} or
kinetic\cite{kinetic} masses.  Figure \ref{fig:phoindtt} compares the
photon-induced $t\bar{t}$ cross-section near threshold as calculated
in the pole mass and 1S mass schemes (for $m_t = 175$ GeV and
$\Gamma_t = 1.43$ GeV).  In the pole mass scheme, the location and
height of the peak vary with renormalization scale and order in
perturbation theory; this choice introduces QCD uncertainties into
what should be a color-singlet process.  Using the short-distance mass
renders the peak location stable and large higher-order corrections
are avoided.

\subsection{Top Decay Width}\label{subsec:width}

In the 3-generation SM, data on the lighter quarks combined with CKM matrix
unitarity implies\cite{pdg} ${ 0.9991 < \vert V_{tb} \vert < 0.9994}$.  Thus the top
decays almost exclusively through $t \to W b$.  At tree level, in the
approximation where $M_W = m_b = 0$ and setting $\vert V_{tb} \vert = 1$, the
decay width is 
\begin{equation}
\Gamma_o(t\to Wb) \equiv \frac{G_F\, m_t^3}{8 \pi \sqrt{2}} = 1.76\, {\rm GeV}\,.
\end{equation}
More precise calculations yield similar results.  Including
$M_W \neq 0$ gives 
\begin{equation}
\Gamma_t/\vert V_{tb}\vert^2 = \Gamma_o (1 - 3\frac{M_W^4}{m_t^4} + 2
\frac{M_W^6}{m_t^6}) = 1.56\, {\rm GeV}\,.\,
\end{equation}
while including the b-quark mass and radiative corrections refines this
to\cite{lhctop} $\Gamma_t/\vert V_{tb}\vert^2 = 1.42\, {\rm GeV}$ .

As a result, the top decays in $\tau_t \approx 0.4 \times 10^{-24}$ s.
Since this is appreciably shorter than the characteristic QCD time scale
$\tau_{QCD} \approx 3 \times 10^{-24}$ s, the top quark decays before it
can hadronize.  Therefore, unlike the $b$ and $c$ quarks which offer rich
spectra of bound states for experimental study, the top quark is not
expected to provide any interesting spectroscopy.

A precise measurement of the top quark width could, in principle, be made
at an NLC running at $\sqrt{s} \sim 350$ GeV by exploiting the fact that
$\Gamma_t$ controls the threshold peak height in $\sigma(e^+e^- \to
t\bar{t})$.  Until recently, the NNLO calculations were plagued by a 20\%
normalization ambiguity which made the realization of this aim
uncertain;\cite{hoangsum} preliminary new results\cite{hoang2} suggest the
issue has been favorably resolved.

\subsection{Pair Production}\label{subsec:pairprod}

The top pair production cross-section has been measured in all available
channels by CDF\cite{tcrosscdf} and D\O .\cite{tcrossd0}  As with $m_t$, the
lepton+jets channel, with its combination of statistics and full
reconstruction, gives the single most precise measurement (see Table
\ref{tab:masscross}). The combined average of $\sigma_{tt}(m_t=172\, {\rm
GeV}) = 5.9 \pm 1.7$ pb is consistent with SM predictions including
radiative corrections.\cite{tcrosstheory}

\begin{figure}[tb]
\begin{center}
\scalebox{1}{\includegraphics{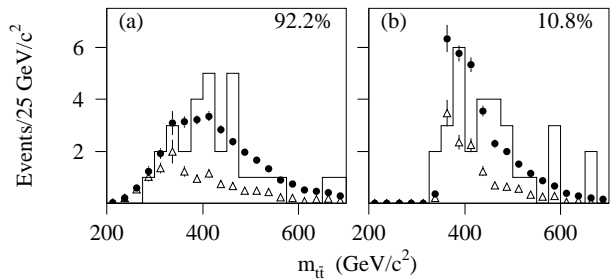}}
\end{center}
\vspace{-.5cm}
\caption[ab]{Invariant mass distribution\protect\cite{tmassd0} for top
pairs: D\O\ data (histogram), simulated background (triangles), simulated
S+B (dots).  In (a) $m_t$ unconstrained; in (b) $m_t = 173$ GeV.}
\label{fig:d0ttdist}
\end{figure}

\begin{figure}[tb]
\begin{center}
\scalebox{.3}{\includegraphics{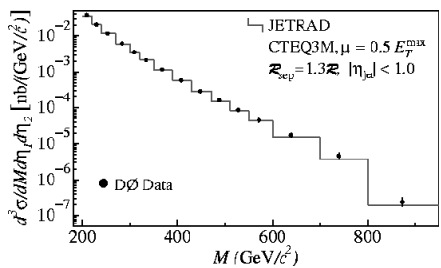}}
\end{center}
\vspace{-.05cm}
\caption[abc]{Light dijet invariant mass distribution\protect\cite{dijets}:
prediction (solid) and D\O\ data (dots).}
\label{fig:qcdjj}
\end{figure}

\begin{figure}[tb]
\begin{center}
\scalebox{1}{\includegraphics*[0mm,0mm][100mm,50mm]{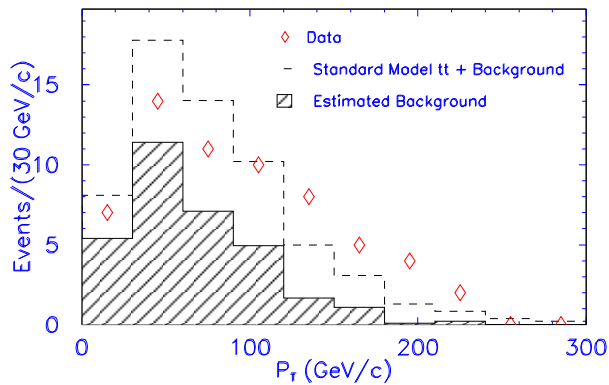}}
\end{center}
\caption[abb]{$P_T$ distribution for hadronically-decaying tops in
lepton+jets events from CDF.\protect\cite{topptcdf}}
\label{fig:hiptcdf}
\end{figure}

Initial measurements of the invariant mass ($M_{tt}$) and transverse
momentum ($p_T$) distributions of the produced top quarks have been made,
as shown in Figure \ref{fig:d0ttdist}.  While a comparison with the
measured $M_{jj}$ distribution for QCD dijets (Figure \ref{fig:qcdjj})
illustrates how statistics-limited the Run I top sample is, some
preliminary limits on new physics are being extracted.  It has been noted,
e.g. that a narrow 500 GeV Z' boson is inconsistent with the observed shape
of the high-mass end of CDF's $M_{tt}$ distribution.\cite{leptophobe}  The
$p_T$ distribution for the hadronically-decaying top in fully-reconstructed
lepton + jets events (Figure \ref{fig:hiptcdf}) constrains non-SM physics
which increases increase the number of high-$p_T$ events.  The fraction
$R_4 = 0.000^{+0.031}_{-0.000} (stat)^{+0.024}_{-0.000}(sys)$ of events in
the highest $p_T$ bin ($225 \leq p_T \leq 300$ GeV) implies\cite{topptcdf}
a 95\% c.l. upper bound $R_4 \leq 0.16$ as compared with the SM prediction
$R_4$ = 0.025.

\begin{figure}[tb]
\begin{center}
\scalebox{.35}{\includegraphics*[0mm,50mm][200mm,230mm]{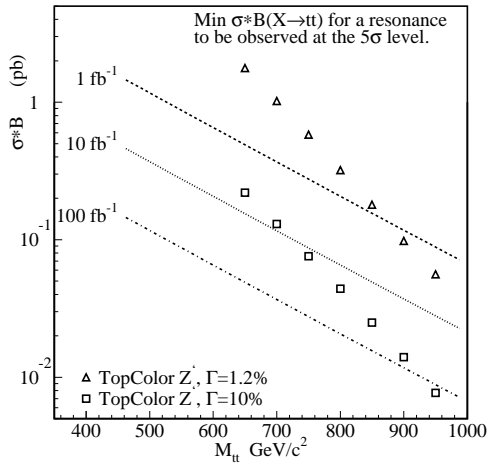}}
\end{center}
\vspace{.2cm}
\caption[ab]{Anticipated\protect\cite{tev2000} Run II limits on
$\sigma\cdot B(X \to t\bar{t})$.}
\label{fig:sigbxtt}
\end{figure}

In Run II, the $\sigma_{tt}$ measurement will be dominated by systematic
uncertainties; the collaborations will use the large data sample to reduce
reliance on simulations.\cite{tev2000}  Acceptance issues such as initial
state radiation, the jet energy scale, and the b-tagging efficiency will be
studied directly in the data.  The background uncertainty for the
lepton+jets mode will be addressed by measuring the heavy-flavor content of
W+jets events.  It is anticipated\cite{tev2000} that an integrated
luminosity of 1 (10, 100) fb$^{-1}$ will enable $\sigma_{tt}$ to be
measured to $\pm$ 11 (6, 5) \%.  The $M_{tt}$ distribution will then
constrain $\sigma\cdot B$ for new resonances decaying to $t\bar{t}$ as
illustrated in Figure \ref{fig:sigbxtt}.

\subsection{Spin Correlations}\label{subsec:spincorr}

When a $t\bar{t}$ pair is produced, the spins of the two fermions are
correlated.\cite{bargero}  This can be measured at lepton or hadron
colliders, and provides another means of testing the predictions of the SM
or looking for new physics.

One starts from the fact that the top quark decays before its spin can
flip.\cite{bigi}  The spin correlations between $t$ and $\bar{t}$
therefore yield angular correlations among their decay products.  If $\chi$
is angle between the top spin and the momentum of a given decay product,
the differential top decay rate (in the top rest frame) is
\begin{equation}
\frac{1}{\Gamma} \frac{d \Gamma}{d \cos\chi} = 
\frac{1}{2} (1 + \alpha \cos\chi)
\end{equation}
The factor $\alpha$ is computed\cite{alphaval} to be 1.0 (0.41, -0.31, -0.41)
if the decay product is $\ell$ or $d$ ($W$, $\nu$ or $u$, $b$).  A
final-state lepton is readily identifiable and has 
largest value of $\alpha$; thus, dilepton events are best for
studying $t\bar{t}$ spin correlations.

\begin{figure}[tb]
\begin{center}
\null\hspace{-.5cm}\scalebox{.2}{\includegraphics{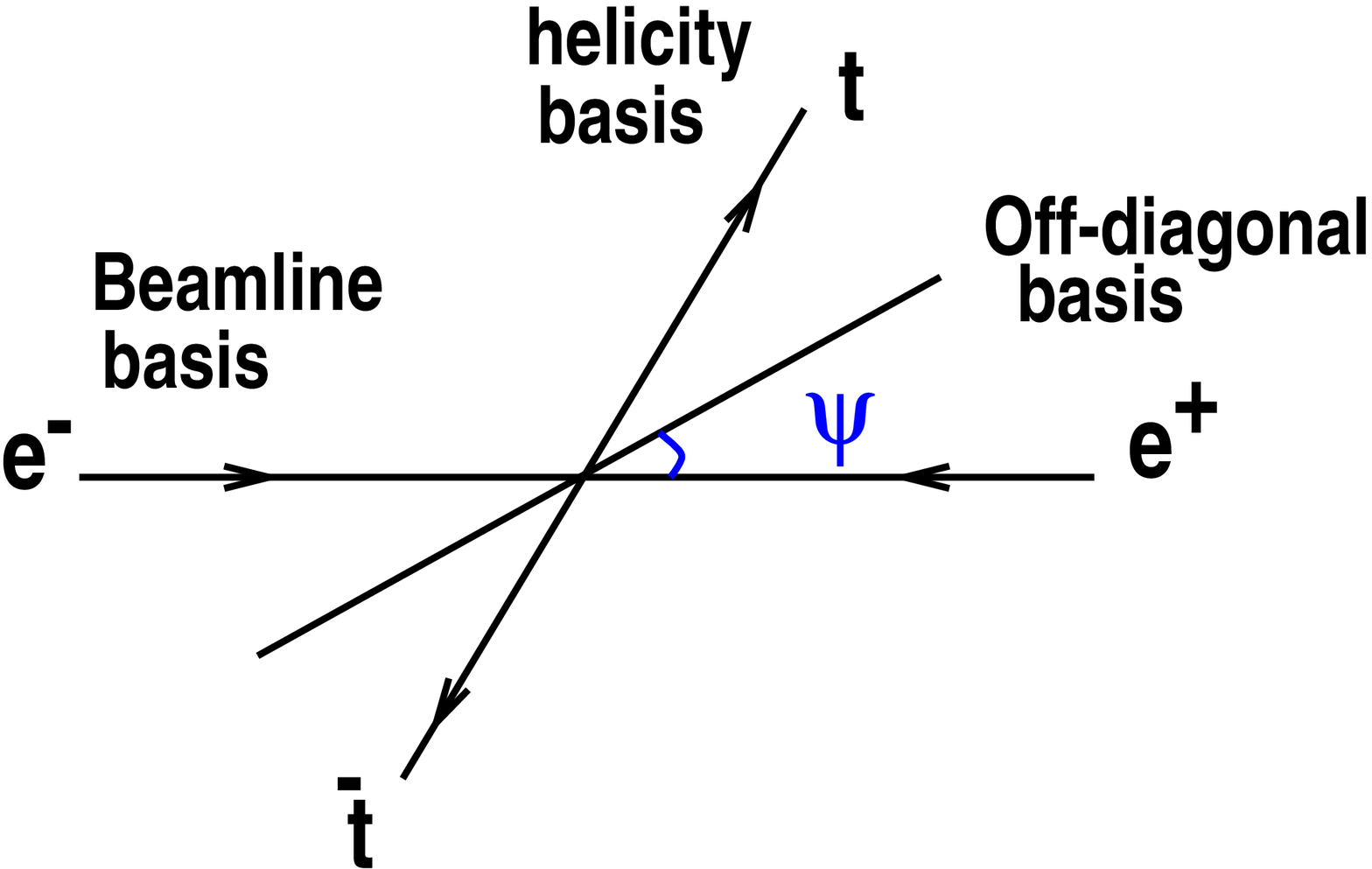}}\hspace{1.5cm}\scalebox{.22}{\includegraphics{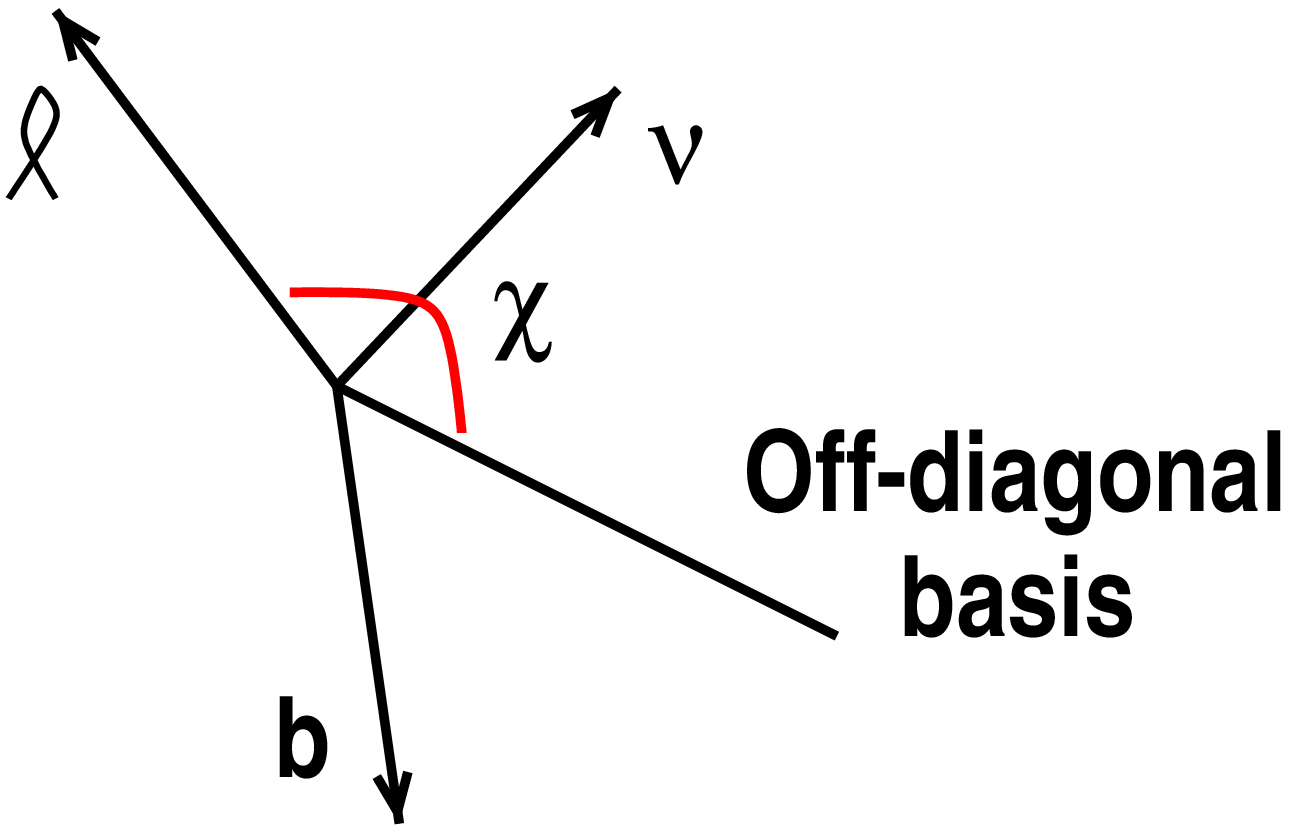}}
\end{center}
\caption[abn]{Definitions of the off-diagonal basis and decay lepton angles
for studying top spin correlations.}
\label{fig:spindiag}
\end{figure}

Choosing a good basis along which to project the spin variables is key to
extracting information from the data.  For example, consider $e^+e^- \to
t\bar{t}$ at the NLC.  If the beams are polarized, using a helicity basis
seems logical, but near the $t\bar{t}$ threshold helicity is not very
useful.  Fortunately, there is an optimal ``off-diagonal''
basis\cite{parkeshadmi} which gives a clean prediction for spin
correlations: in leading order the spins are purely anti-correlated
($t_\uparrow \bar{t}_\downarrow + t_\downarrow \bar{t}_\uparrow$).  One
projects the top spin along an axis identified by angle $\psi$
\begin{equation}
\tan\psi = \frac{\beta^2 \sin\theta^* \cos\theta^*}{1 - \beta^2
\sin^2\theta^*}
\end{equation}
where $\beta$ is the top quark's speed in the center-of-momentum
scattering frame and $\theta*$ is the top scattering angle in that frame.
The basis angle $\psi$ and decay lepton angle $\chi$ are illustrated in
Figure \ref{fig:spindiag}.

\begin{figure}[tb]
\begin{center}
\null\hspace{-1cm}\scalebox{.45}{\includegraphics{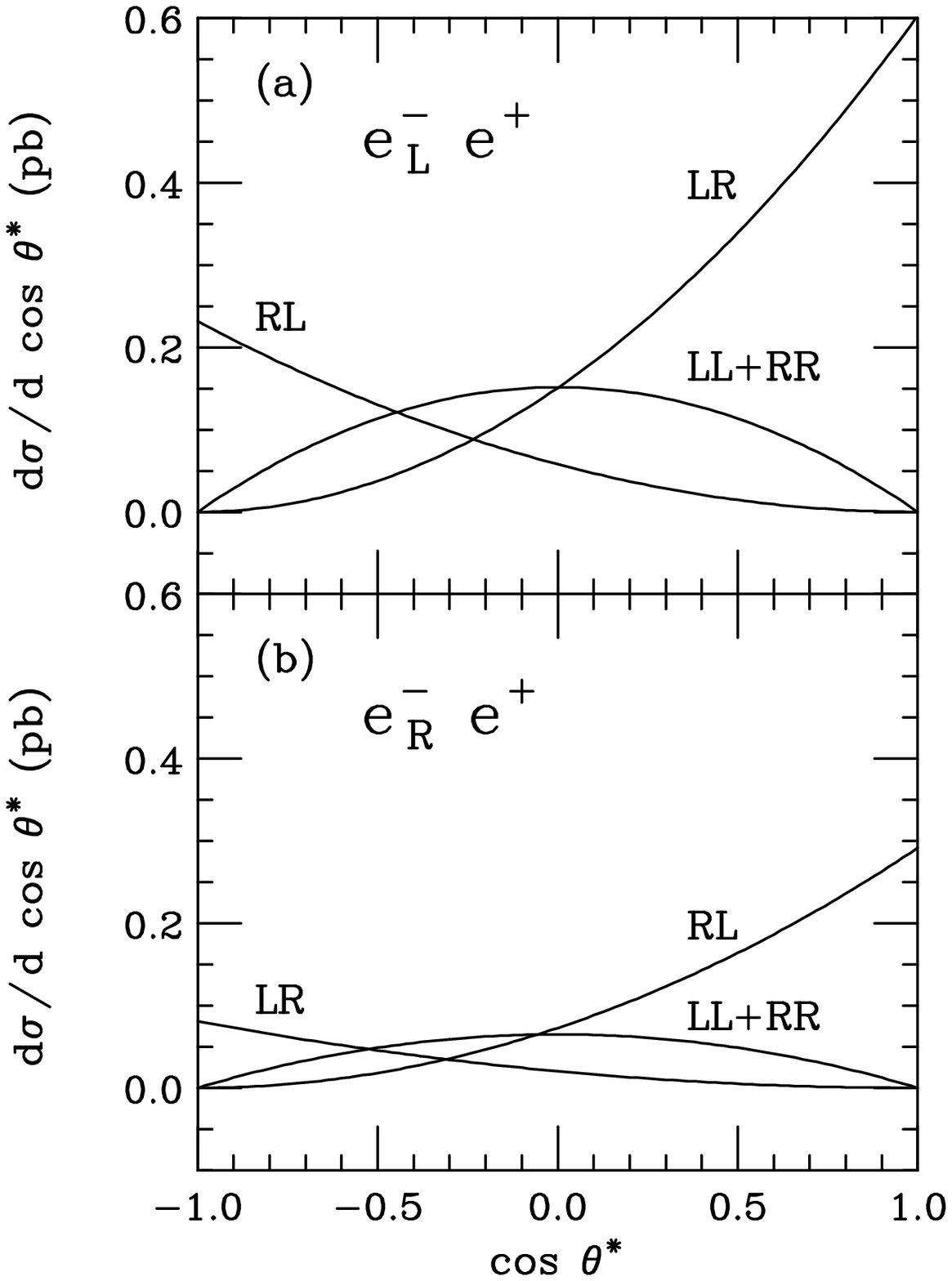}} \hspace{1cm} 
\scalebox{.45}{\includegraphics{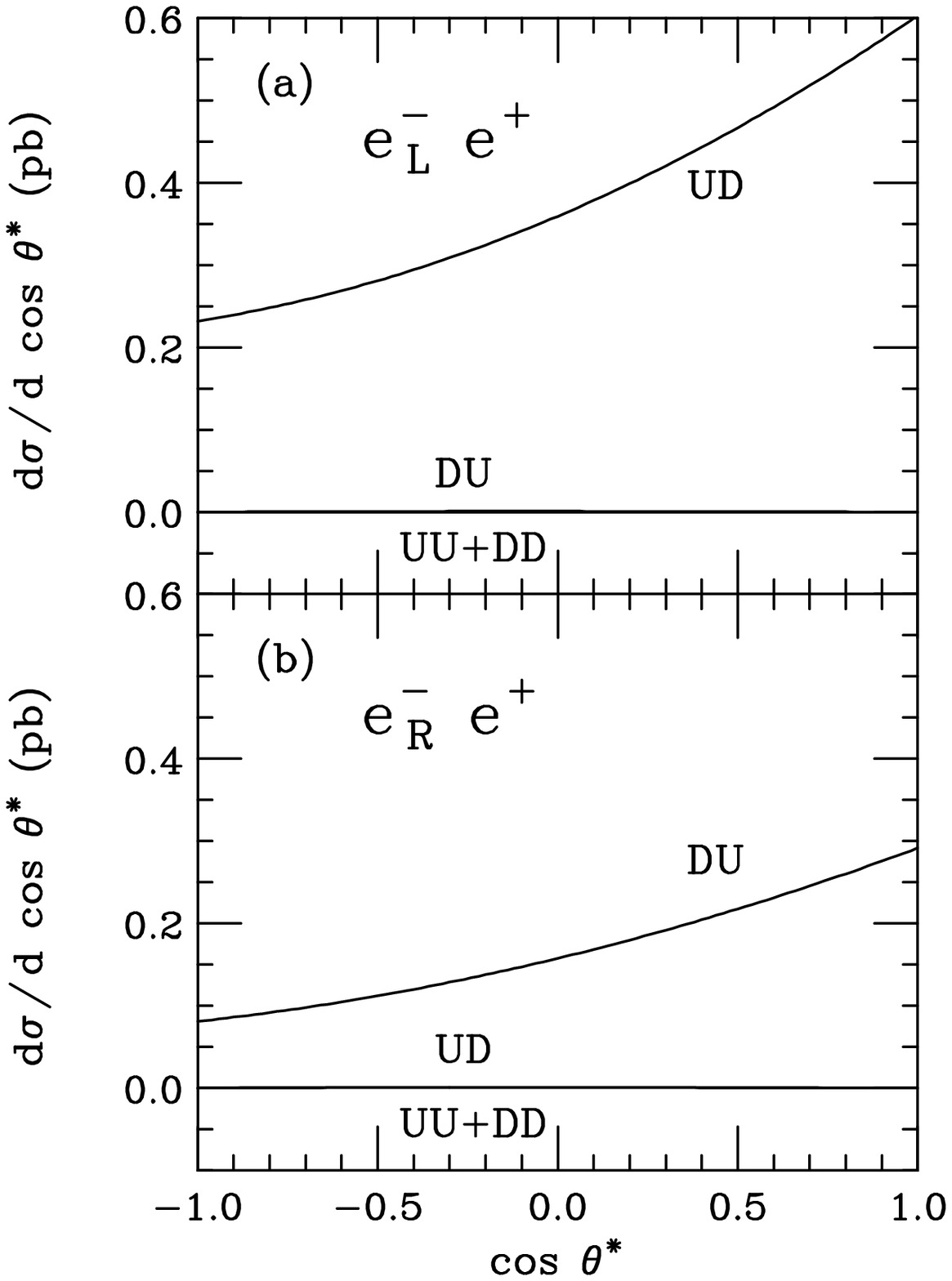}}
\end{center}
\caption[abnn]{Differential cross-section for top production at 400 GeV NLC
with\protect\cite{parkeshadmi} (a) LH and (b) RH electron beams.  At left spins are projected onto
the helicity basis; at right, onto the off-diagonal basis.}
\label{fig:spinhel}
\end{figure}

\begin{figure}[tb]
\begin{center}
\scalebox{.45}{\includegraphics{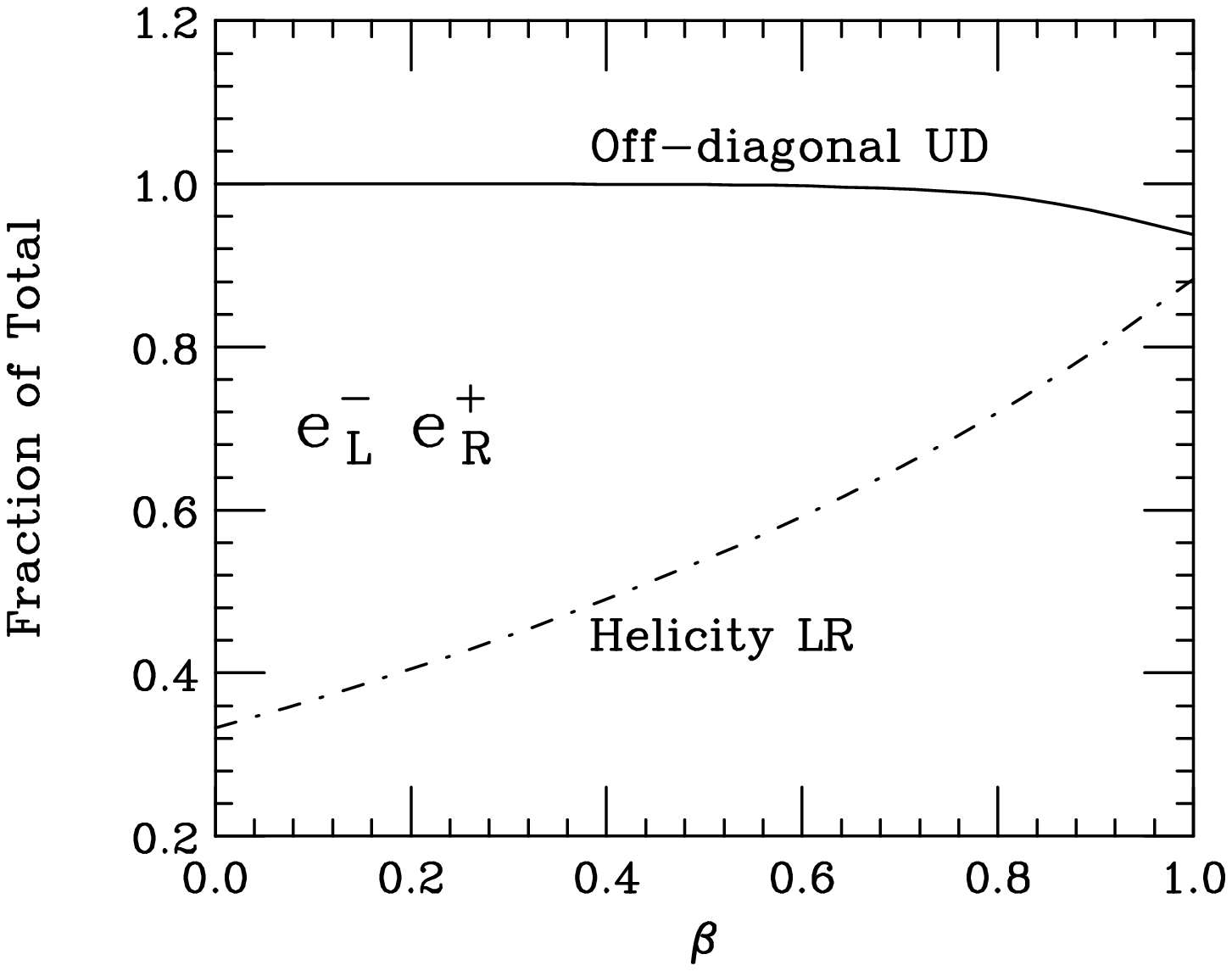}}
\end{center}
\caption[abnm]{Fraction of top quarks in the dominant spin
configuration\protect\cite{parkeshadmi} for $e^-_L e^+_R \to t\bar{t}$.}
\label{fig:spinbas}
\end{figure}

The advantages of an appropriate basis are clear from Figure
\ref{fig:spinhel}: for a given data sample, discerning the clean
predication of the off-diagonal basis should be far easier than untangling
the several possible spin configurations in the helicity basis.  Moreover,
while the fraction of top quarks in the dominant spin configuration in
$e^-_L e^+_R \to t\bar{t}$ approaches unity in the helicity basis at large
$\beta$, it is always nearly one in the off-diagonal basis (Figure
\ref{fig:spinbas}).

This idea carries over to the Tevatron.  In the helicity basis, 70\% of
$t\bar{t}$ pairs have opposite helicities\cite{stelwillcorr}: threshold
production via $q\bar{q}$ annihilation puts the tops in a $^3S_1$
state\cite{hara} where their spins tend to be aligned.  But the
off-diagonal basis still does better\cite{mahlonparke}: 92\% of the top
pairs have anti-aligned spins.  The larger spin correlation translates into
larger and more measurable correlations among the decay leptons.  Writing
the differential cross-section in terms of the angular positions $\chi^\pm$
of the decay leptons $\ell^\pm$
\begin{equation}
\frac{1}{\sigma} \frac{d^2 \sigma}{d(\cos\chi_+) d(\cos\chi_-)} =
\frac{1}{4} (1 + \kappa \cos\chi_+ \cos\chi_-)
\label{eq:difsig}
\end{equation}
one finds $\kappa \approx 0.9$ in the SM for $\sqrt{s} = 1.8$ TeV.  As D\O\
recorded only six dilepton events in Run I, they set\cite{spincorrd0}
merely the 68\% c.l. limit $\kappa \geq -0.25$.  Nonetheless, the
possibility of making a top spin correlation measurement in a hadronic
environment has been established and Run IIa promises $\sim$150 dilepton
events.\cite{tev2000}

At the LHC, the top dilepton sample will be of order $4\times 10^5$
events per year\cite{lhctop} -- but no spin basis with nearly 100\%
correlation at all $\beta$ has been identified.  Pair production
proceeds mainly through $gg \to t\bar{t}$, putting the tops in a
$^1S_0$ state\cite{hara} at threshold.  Near threshold, angular
momentum conservation favors like helicities; far above threshold,
helicity conservation favors opposite helicities.\cite{lhctop} In the
helicity basis, one conventionally studies a differential
cross-section of the same form as Eq. (\ref{eq:difsig}), in which the
coefficient $\kappa$ is renamed $-C$ and the angle $\chi_\pm$ refers
to the angle between the $t$ ($\bar{t}$) momentum in the
center-of-momentum frame and the $\ell^\pm$ direction in the $t$
($\bar{t}$) rest frame.  The SM predicts $C \approx 0.33$ in leading
order at the LHC.  Physically, $C$ corresponds\footnote{This
expression also holds for $-\kappa$ at the Tevatron if $L$ and $R$ are
taken to refer to the off-diagonal rather than the helicity basis.} to
the ratio\cite{stelwillcorr}
\begin{equation}
 C= \frac{N(t_L \bar{t}_L +t_R \bar{t}_R) -
                              N(t_L \bar{t}_R +t_R \bar{t}_L)} 
                             {N(t_L \bar{t}_L +t_R \bar{t}_R) + 
                              N(t_L \bar{t}_R +t_R \bar{t}_L)}\ .
\end{equation}
The effects of radiative corrections and the likely measurement precision
achievable remain to be evaluated.\cite{lhctop}

\subsection{Single Production}\label{subsec:singleprod}

\begin{figure}[tb]
\centerline{\scalebox{.2}{\includegraphics{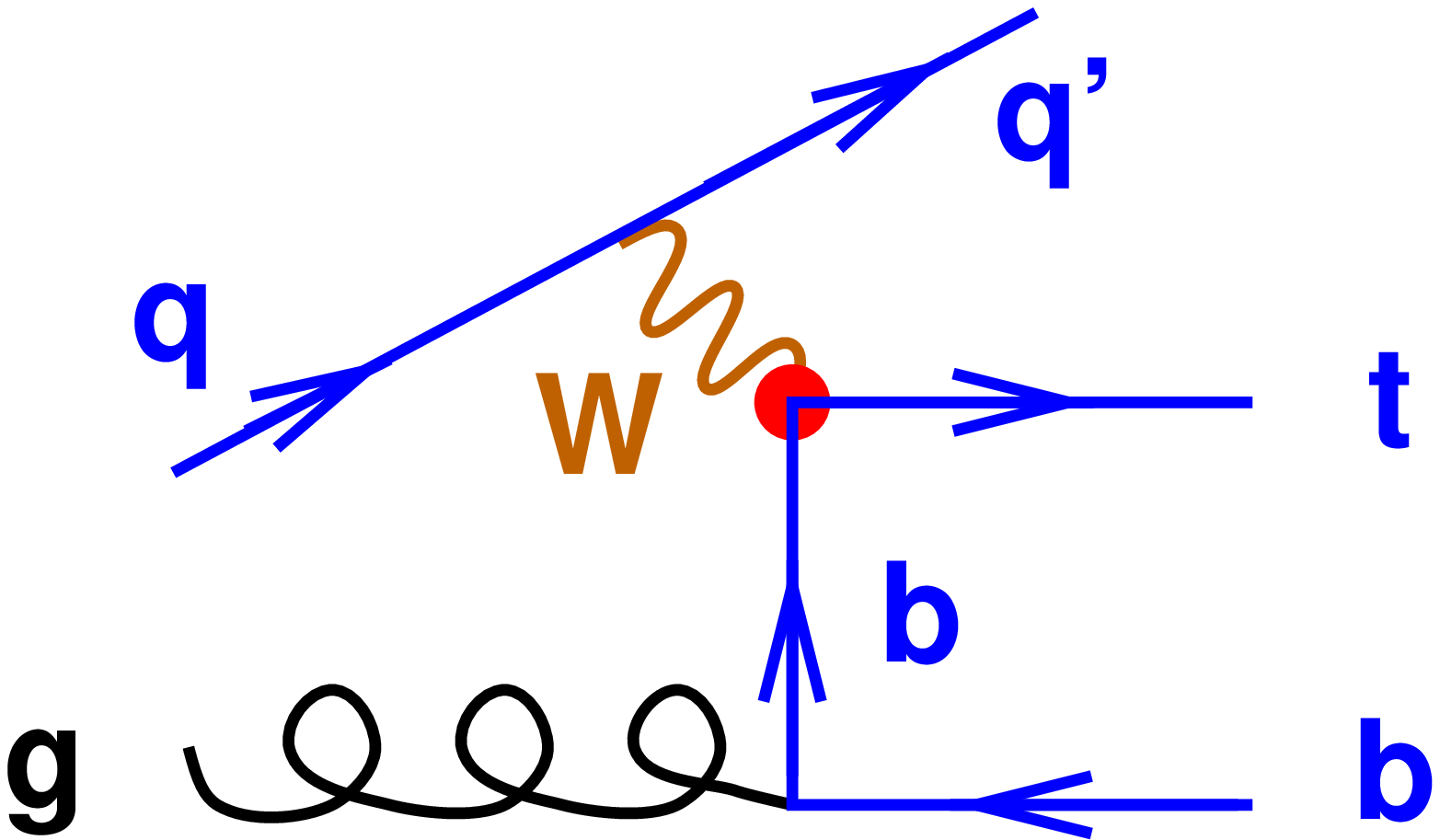}}
\hspace{.8cm}\scalebox{.22}{\includegraphics{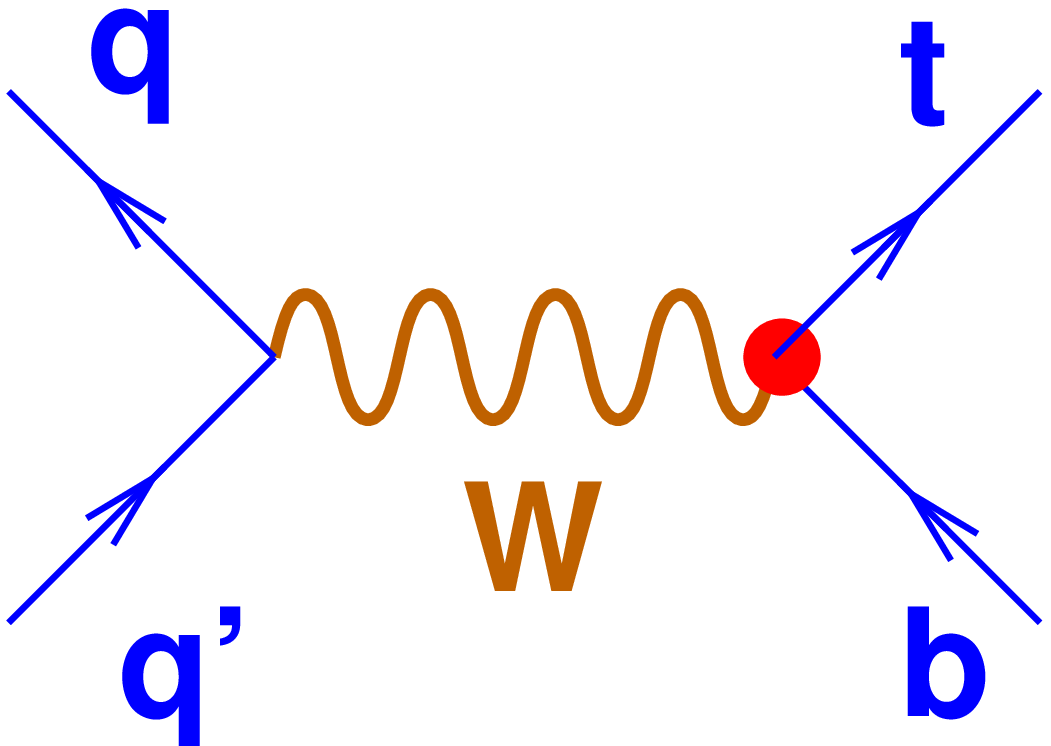}}
\hspace{.8cm}\scalebox{.2}{\includegraphics{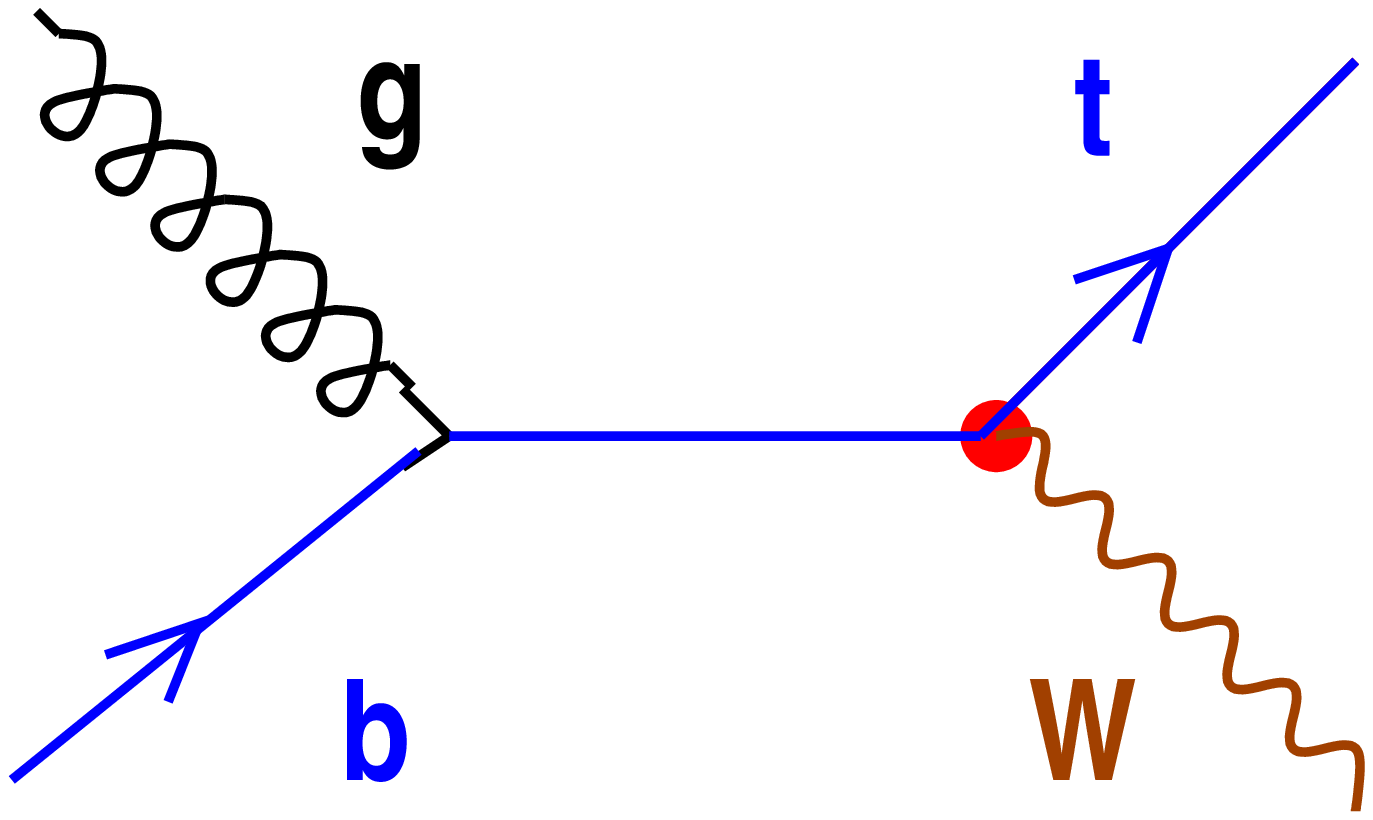}}}
\caption[aan]{Feynman diagrams for single top quark production.}
\label{fig:singtdiag}
\end{figure}

The three SM channels for single top production are $Wg$ fusion, $q\bar{q}$
annihilation through an off-shell W, and $gb \to t W$; the Feynman diagrams
are shown in Figure \ref{fig:singtdiag}.  The $Wg$ fusion events are
characterized by one hard and one soft b-jet, an additional jet and a $W$;
the SM Run I cross-section is calculated\cite{singtopcalc} to be
$1.70\pm0.9$ pb.  The $W^*$ events, in contrast, include two hard b quarks
and a $W$ from top decay; the calculated\cite{singtopcalc} SM Run I
cross-section is $\sigma = 0.73\pm0.04$ pb.  The $gb \to t W$ process is
highly suppressed at the Tevatron.

\begin{figure}[tb]
\begin{center}
\vspace{-.5cm}
\scalebox{.45}{\includegraphics{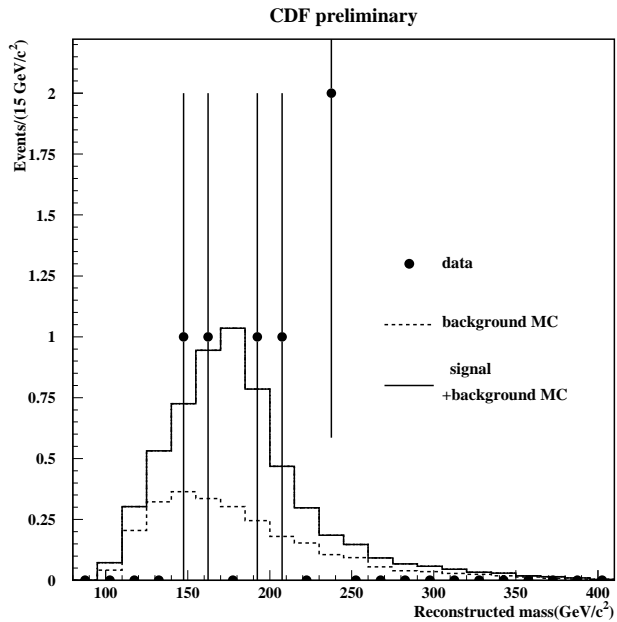}}
\end{center}
\vspace{-.3cm}
\caption[ab]{Reconstructed $m_t$ for the CDF\cite{cdfsinglet} $Wbb$
sample. ``Signal'' is single top production.}
\label{fig:wbbsamp}
\end{figure}

\begin{figure}[tb]
\begin{center}
\scalebox{.45}{\includegraphics{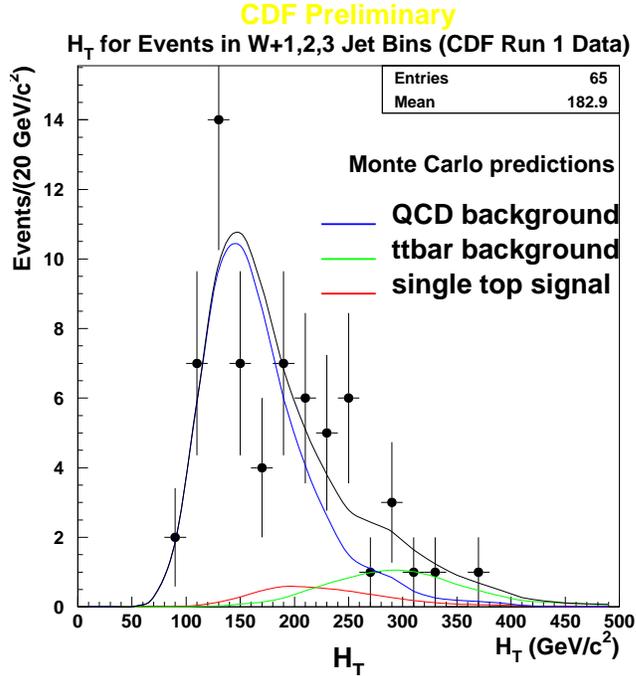}}
\end{center}
\vspace{-.3cm}
\caption[ab]{$H_T$ distribution in single top production\cite{cdfsinglet}:
CDF data and simulated backgrounds.}
\label{fig:htsum}
\end{figure}

Searches for single top production generally focus on
leptonically-decaying $W$ bosons.  The principle backgrounds come from
top pair production, $W$+jets events, QCD multijet events in which a
jet fakes an electron, and $WW$ events.  While the D\O\ analysis of
single top production is still in progress\footnote{The D\O\
limit\cite{dosing} became available after these lectures were given.
It is not stronger than the CDF limits.}, CDF has set two
limits.\cite{cdfsinglet} The first is based on reconstructing a top
quark mass for the six events with $Wbb$ identified in the final
state, as illustrated in Figure \ref{fig:wbbsamp}.  Using Run I data,
CDF finds $\sigma_{tb} < 18.6$ pb (the SM prediction is 2.43 pb). The
higher luminosity in Run II should provide $S/\sqrt{B} \geq 4$ in this
channel.  The second limit exploits the differences among the $H_T$
distributions in signal and background $W$+jet events; $H_T$ is the
scalar sum of the jet, lepton, and missing transverse-energies.  Each
event is required to include 1-3 jets (one of which is b-tagged), a
lepton from $W$ decay, and a reconstructed top mass in the range 140 -
210 GeV.  The cross-section limit set with Run I data shown in Figure
\ref{fig:htsum} is $\sigma_{tb} < 13.5$ pb.

\subsection{Decays}\label{subsec:decays}

\smallskip
\noindent{\tt W helicity in top decay}
\smallskip

The SM predicts the fraction (${\cal F}_0$) of top quark decays to
longitudinal (zero-helicity) $W$ bosons will be quite large, due to the top
quark's big Yukawa coupling:
\begin{equation}
{\cal F}_0 = \frac{m_t^2/2 M_W^2}{1 + m_t^2/2 M_W^2} = (70.1 \pm 1.6)\%\ .
\end{equation}
One can measure ${\cal F}_0$ in dilepton or lepton+jet events by exploiting
the correlation of the $W$ helicity with the momentum of the decay leptons.
For $W^+ \to \ell^+\nu$, the spins of the decay leptons align with that of
the $W$; for massless leptons, the $\ell^+$ ($\nu$)momentum points along
(opposite) its spin.  Then a positive-helicity $W$ (boosted along its spin)
yields harder charged leptons than a negative-helicity $W$.  The
longitudinal $W$ gives intermediate results.

\begin{table}[bh]
\caption[]{Predicted\cite{tev2000} precision of Run II $W$ helicity
measurement for several $\int {\cal L} dt$.}
\begin{center}
\begin{tabular}{|l|c|c|c|}
\hline
& 1 fb$^{-1}$ & 10 fb$^{-1}$ & 100 fb$^{-1}$ \\
\hline
$\delta {\cal F}_0$ & 6.5\% & 2.1\% & 0.7\%\\
$\delta {\cal F}_+$ & 2.6\% & 0.8\% & 0.3\%\\
\hline
\end{tabular}
\end{center}
\label{tab:helic}
\end{table}

\begin{figure}[tb]
\scalebox{.4}{\includegraphics{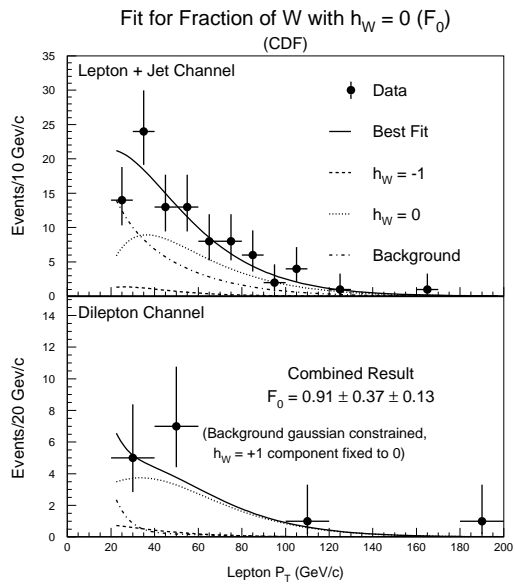}}
\vspace{-.7cm}
\caption[bb]{Measured lepton $p_T$ spectra and fits to $W$ helicity by CDF.\protect\cite{whelicitycdf} }
\label{fig:helic}
\end{figure}

CDF has measured\cite{whelicitycdf} the lepton $p_T$ spectra for dilepton
and lepton + jet events and performed fits as shown in Figure
\ref{fig:helic}.  There is insufficient data to permit forming conclusions
about all three helicity states simultaneously.  By assuming no
positive-helicity $W$'s are present, CDF obtains the limit ${\cal F}_0 =
0.91\pm0.37\pm0.13$.  By setting ${\cal F}_0$ to its SM value of 0.70, they
obtain the 95\% c.l. upper limit ${\cal F}_+ < 0.28$.  Note, however, that
the first limit essentially states only that no more than 100\% of the
decay $W$'s are longitudinal and the second, that no more than $1 - {\cal
F}_0$ have positive helicity.  More informative constraints are expected
from Run II (see Table \ref{tab:helic}).

\smallskip
\noindent{\tt $b$ quark decay fraction}
\smallskip

The top quark's decay fraction to $b$ quarks is measured by
CDF\cite{bbmeascdf} to be $B_b \equiv {\Gamma(t \to b W)} / {\Gamma(t \to q
W)} = 0.99 \pm 0.29$.  In the three-generation SM, $B_b$ is related to CKM
matrix elements as
\begin{equation}
B_b \equiv {\vert V_{tb}\vert^2 \over {\vert V_{tb}\vert^2 + \vert V_{ts}\vert^2
 + \vert V_{td}\vert^2}} \, .
\label{bbref}
\end{equation}
Three-generation unitarity dictates that the denominator of (\ref{bbref}) is
1.0, so that the measurement of $B_b$ implies\cite{bbmeascdf}
$\vert V_{tb} \vert > 0.76$ at 95\% c.l.
However, within the 3-generation SM, data on the light quarks
combined with CKM unitarity has already provided\cite{pdg} 
the much tighter constraints $ 0.9991 < \vert V_{tb} \vert < 0.9994$.

If we add a fourth generation of quarks, the analysis differs.  A search by
D\O\ has constrained\cite{pdg} any 4-th generation $b^\prime$ quark to have
a mass greater than $m_t - m_W$, so that the top quark could not readily
decay to $b^\prime$.  This means that the original expression (\ref{bbref})
for $B_b$ is still valid.  However, once there are four generations, the
denominator of the RHS of (\ref{bbref}) need not equal 1.0.  Then the CDF
measurement of $B_b$ implies $ \vert V_{tb} \vert \gg \vert V_{td} \vert\ ,
\ \vert V_{ts} \vert$.  In contrast, light-quark data combined with
4-generation CKM unitarity allows $\vert V_{tb}\vert$ to lie in the
range\cite{pdg} $0.05 < \vert V_{tb} \vert < 0.9994$.  While the
measurement of $B_b$ gives only qualitative information about $\vert
V_{tb}\vert$, that information is new and useful in the context of a
4-generation model.

Direct measurement of $\vert V_{tb} \vert$ in single top-quark production
(via $q\bar{q} \to W* \to t\bar{b}$ and $gW \to t\bar{b}$) at the Tevatron
should reach an accuracy\cite{tev2000} of 10\% in Run IIa (5\% in Run IIb).

\smallskip
\noindent{\tt FCNC decays}
\smallskip

CDF has set limits\cite{fcnccdf} on the flavor-changing decays $t \to Zq,
\gamma q$ which are GIM-suppressed in the SM.  In seeking $t \to Zq$ they
looked at $p\bar{p}\to t\bar{t} \to qZbW, qZbZ \to \ell \ell + 4$ jets with
high jet $E_T$.  The SM background from $WW$, $ZZ$ and $WZ$ events is
predicted to be $0.6\pm0.2$ events.  The data contains a single candidate
(in which the $Z$ decayed to muons).  On this basis, the 95\% c.l. upper
limit $B(t\to Zq) < 0.33$ was set.  To study $t \to \gamma q$, CDF examined
$p\bar{p}\to t\bar{t} \to W b \gamma q$ events.  If the $W$ decayed
leptonically, the signature was $\gamma + \ell + E_T\!\!\!/ + (\geq 2)$
jets; if hadronically, the signature was $\gamma + (\geq 4)$ jets with one
jet b-tagged.  The expected SM background is a single event.  Finding a
single candidate event (with a leptonic $W$ decay), CDF set the 95\%
c.l. upper bound $B(t \to \gamma q) < 0.032$. Run II will provide much
greater sensitivity to these decays,\cite{tev2000} as indicated in Table
\ref{tab:fcnc}.

\begin{table}[tb]
\caption{Run II sensitivity\protect\cite{tev2000} to FCNC top decays as a
function of $\int {\cal L}dt$. }
\begin{center}
\begin{tabular}{|l|c|c|c|}
\hline
 & 1 fb$^{-1}$ & 10 fb$^{-1}$ & 100 fb$^{-1}$ \\
\hline 
$BR(t\to Z q)$  & 0.015 & $3.8\times 10^{-3}$ & $6.3\times 10^{-4}$ \\
$BR(t\to \gamma q)$ & $3.0\times 10^{-3}$ & $4.0\times 10^{-4}$ & $8.4\times 10^{-5}$\\
\hline
\end{tabular}
\end{center}
\label{tab:fcnc}
\end{table}

\subsection{Summary}\label{subsec:summary-sm}

The Run I experiments at the Tevatron discovered the top quark and provided
the first measurements of a variety of properties including $m_t,\
\Gamma_t,\ \sigma_{tt},\ \frac{d\sigma}{dM_{tt}}, \frac{d\sigma}{d p_T},
\kappa, \sigma_{tb}, {\cal F}_0, {\cal F}_{+}, B_b, \Gamma(t\to Zq),$
and $\Gamma(t\to \gamma q)$.  As we have seen, most of the measurements
were limited in precision by the small top sample size.  This will be
ameliorated at Run II and future colliders.

As a starting point for further discussion, we note that each property
measured has been seen to have multiple implications for theory.  Moreover,
the interpretation of the measurement can depend critically on the
theoretical context.  In some cases, measurements may even shed more light
on the merits of proposed non-standard physics than on the Standard Model
itself.  This is the line of thought we shall take up in the second section
of the talk.

\section{Beyond the Standard Model}

\begin{figure}[bt]
\begin{center}
\scalebox{.36}{\includegraphics{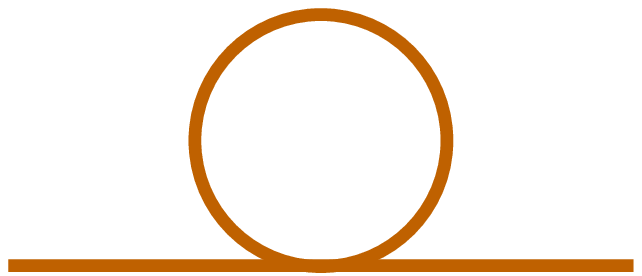}}
\hspace{3cm} \scalebox{.4}{\includegraphics{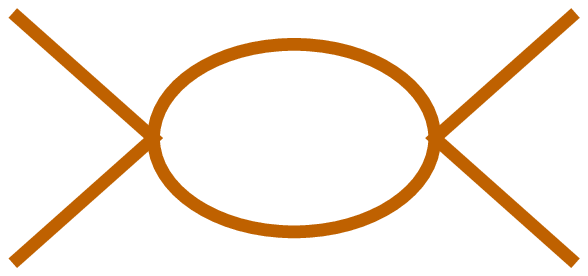}}
\end{center}
\caption{(left) Naturalness problem: $ M_H^2\ \propto\ \Lambda^2$. (right)
Triviality: $\beta(\lambda)\ = \ {{3\lambda^2}\over{2\pi^2}}\ > \
  0$.}
\label{fig2}
\end{figure}

Two central concerns of particle theory are finding the cause of
electroweak symmetry breaking and identifying the origin of flavor
symmetry breaking by which the quarks and leptons obtain their diverse
masses.  The Standard Higgs Model of particle physics, based on the
gauge group $SU(3)_c \times SU(2)_W \times U(1)_Y$ accommodates both
symmetry breakings by including a fundamental weak doublet of scalar
(``Higgs'') bosons ${\phi = {\phi^+ \choose \phi^0}}$ with potential
function $V(\phi) = \lambda \left({\phi^\dagger \phi - \frac12
v^2}\right)^2$.  However the SM does not explain the dynamics
responsible for the generation of mass.  Furthermore, the scalar
sector suffers from two serious problems.  The scalar mass is
unnaturally sensitive to the presence of physics at any higher scale
$\Lambda$ (e.g. the Planck scale), as shown in Figure \ref{fig2}.
This is known as the gauge hierarchy problem.  In addition, if the
scalar must provide a good description of physics up to arbitrarily
high scale (i.e., be fundamental), the scalar's self-coupling
($\lambda$) is driven to zero at finite energy scales as indicated in
Figure \ref{fig2}.  That is, the scalar field theory is free (or
``trivial''). Then the scalar cannot fill its intended role: if
$\lambda = 0$, the electroweak symmetry is not spontaneously broken.
The scalars involved in electroweak symmetry breaking must therefore
be a party to new physics at some finite energy scale -- e.g., they
may be composite or may be part of a larger theory with a UV fixed
point.  The SM is merely a low-energy effective field theory, and the
dynamics responsible for generating mass must lie in physics outside
the SM.

\noindent

\begin{figure}[tb] 
\begin{center}
\scalebox{.4}{\includegraphics{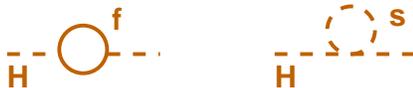}}
\end{center}
\caption{$\delta
      M_H^2 \sim {g_f^2\over{4\pi^2}} (m_f^2 - m_s^2) + m_s^2 log
      \Lambda^2$}\label{fig4}
\end{figure}

One interesting possibility is to introduce supersymmetry.\cite{susyrev}
The gauge structure of the minimal supersymmetric SM (MSSM) is identical to
that of the SM, but each ordinary fermion (boson) is paired with a new
boson (fermion) called its ``superpartner'' and two Higgs doublets are
needed to provide mass to all the ordinary fermions.  As sketched in Figure
\ref{fig4}, each loop of ordinary particles contributing to the Higgs
boson's mass is countered by a loop of superpartners.  If the masses of the
ordinary particles and superpartners are close enough, the gauge hierarchy
can be stabilized.\cite{susystab}  Supersymmetry relates the scalar
self-coupling to gauge couplings, so that triviality is not a concern.

Another intriguing idea, dynamical electroweak symmetry
breaking,\cite{dynamrev} is that the scalar states involved in
electroweak symmetry breaking could be manifestly composite at scales
not much above the electroweak scale $v \sim 250$ GeV.  In these
theories, a new strong gauge interaction with $\beta < 0$ (e.g.,
technicolor) breaks the chiral symmetries of massless fermions $f$ at
a scale $\Lambda \sim 1$ TeV.  If the fermions carry appropriate
electroweak quantum numbers (e.g. LH weak doublets and RH weak
singlets), the resulting condensate $\langle \bar f_L f_R \rangle \neq
0$ breaks the electroweak symmetry as desired.  The Goldstone Bosons
(technipions) of the chiral symmetry breaking simply become the
longitudinal modes of the $W$ and $Z$. The logarithmic running of the
strong gauge coupling renders the low value of the electroweak scale
(i.e.  the gauge hierarchy) natural.  The absence of fundamental
scalars obviates concerns about triviality.

Once we are willing to consider physics outside the SM, seeking
experimental evidence is imperative.  One logical place to look is in the
properties of the most recently discovered state, the top quark. The fact
that $m_t \sim v_{weak}$ suggests that the top quark may afford us insight
about non-standard models of electroweak physics and could even play a
special role in electroweak and flavor symmetry breaking.  Since the sample
of top quarks accumulated in Tevatron Run I was small, many of the top
quark's properties are still only loosely constrained.  The top quark may
yet prove to have properties that set it apart from the other quarks, such
as light related states, low-scale compositeness, or unusual gauge
couplings.  

The Run II experiments will help us evaluate these ideas.  One approach
would be to classify measurable departures from SM predictions and identify
the theories which could produce them.\cite{symptom}  For example, an
unexpectedly large rate for $t\bar{t}$ production could signal the presence
of a coloron resonance, a techni-eta decaying to $t\bar{t}$ or a gluino
decaying to $t\tilde{t}$.  The approach we adopt here, is to consider
general classes of theoretical models and identify signals characteristic
of each.  We will discuss two-higgs and SUSY models, dynamical symmetry
breaking, new gauge interactions for the top quark and the phenomenology of
strong top dynamics.

\subsection{Multiple-Scalar-Doublet Models}\label{subsec:twohiggs}

Many quite different kinds of models include relatively light charged
scalar bosons, into which top may decay: $t \to H^+ b$. The general class
of models that includes multiple Higgs bosons\cite{higgshunt} features
charged scalars that can be light.  Dynamical symmetry breaking models with
more than the minimal two flavors of new fermions (e.g.  technicolor with
more than one weak doublet of technifermions) typically possess
pseudoGoldstone boson states, some of which can couple to third generation
fermions.  SUSY models must include at least two Higgs doublets in order to
provide mass to both the up and down quarks, and therefore have a charged
scalar in the low-energy spectrum.

Experimental limits on charged scalars are often phrased in the
language of a two-higgs-doublet model.  In addition to the usual input
parameters $\alpha_{em}$, $G_F$ and $M_Z$ required to specify the
electroweak sector of the SM, two additional quantities are relevant
for the process $t \to H^+ b$: $\tan\beta$ (the ratio of the vev's of
the two scalar doublets) and $M_{H^\pm}$.

\begin{figure}[bt]
\begin{center}
\scalebox{.30}{\includegraphics*[20mm,50mm][230mm,212mm]{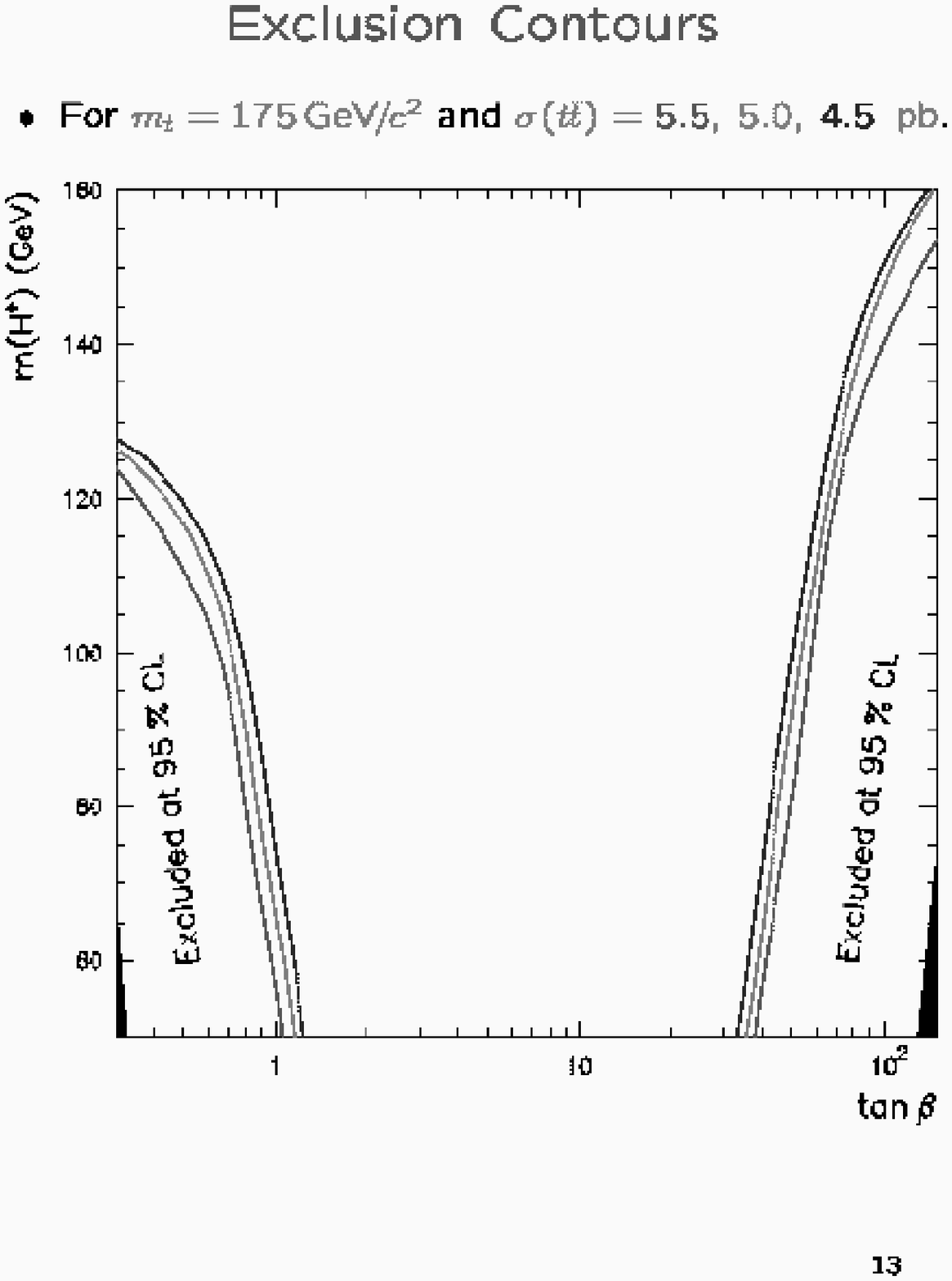}}   
\hspace{-1cm} \raisebox{10pt}{\scalebox{.35}{\includegraphics*[30mm,40mm][210mm,166mm]{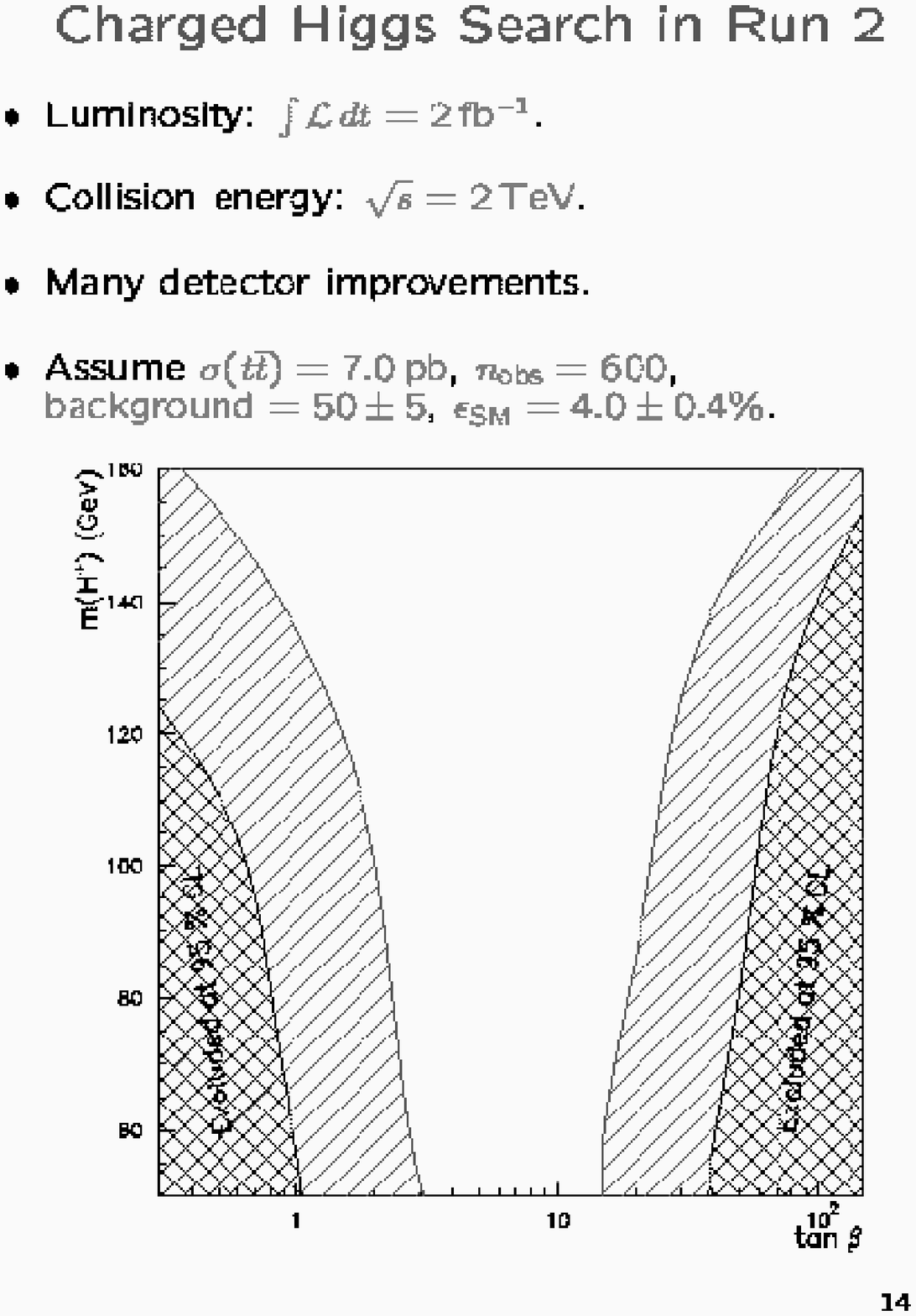}}}
\end{center}
\vspace{1.cm}
\caption{D\O\ charged scalar searches\protect\cite{chgdhiggsd0} in $t\to
H^\pm b$.  (left) Run I limits for $m_t = 175$ GeV.  The region below the
top (middle, bottom) contours is excluded if $\sigma(t\bar{t})$ = 4.5, 5.0,
5.5 pb.  (right) Projected Run II reach\protect\cite{tampere} assuming
$\sqrt{s} = 2$ TeV, $\int{\cal L} dt = 2 {\rm fb}^{-1}$, and
$\sigma(t\bar{t})$ = 7pb.}
\label{fig:chgddh}
\end{figure}

If the mass of the charged scalar is less than $m_t - m_b$, then the decay
$t \to H^+ b$ can compete with the standard top decay mode $t \to W b$.
Since the $t b H^\pm$ coupling depends on $\tan\beta$ as\cite{higgshunt}
\begin{equation}
g_{tbH^+} \propto m_t\cot\beta(1+\gamma_5) + m_b\tan\beta(1-\gamma_5)\ ,
\end{equation}  
the additional decay mode is significant for either large or small
values of $\tan\beta$.  The charged scalar, in turn, decays as $H^\pm
\to c s$ or $H^\pm \to t^*b \to Wbb$ if $\tan\beta$ is small and as
$H^\pm \to \tau \nu_\tau$ if $\tan\beta$ is large.  In either case,
the final state reached through an intermediate $H^\pm$ will cause the
original $t\bar{t}$ event to fail the usual cuts for the lepton + jets
channel.  A reduced rate in this channel can therefore signal the
presence of a light charged scalar.  As shown in Figure
\ref{fig:chgddh}, D\O\ has set a limit\cite{chgdhiggsd0} on
$M_{H^\pm}$ as a function of $\tan\beta$ and $\sigma_{tt}$.  In Run II
the limits should span a wider range of $\tan\beta$ and reach nearly
to the kinematic limit.

\subsection{SUSY Models}\label{subsec:susy}

The heavy top quark plays a role in several interesting issues related to
Higgs and sfermion masses in supersymmetric models.

\smallskip 
\noindent{\tt Scalar mass-squared}
\smallskip

SUSY models need to explain why the scalar Higgs boson acquires a negative
mass-squared (breaking the electroweak symmetry) while the scalar fermions
do not (preserving color and electromagnetism).  In a number of SUSY
models, such as the MSSM with GUT unification or models with dynamical SUSY
breaking, the answer involves the heavy top quark.\cite{susyrev}  In these
theories, the masses of the Higgs bosons and sfermions are related at a
high energy scale $M_\chi$:
\begin{equation}
M_{h,H}^2 (M_\chi) = m_0^2 + \mu^2\ \ \ \ \ \ \ \ \ \ \ \ M_{\tilde{f}}^2
(M_\chi) = m_0^2
\end{equation}
where the squared masses are all positive so that the vacuum preserves the
color and electroweak symmetries.  To find the masses of the scalar
particles at lower energy scales, one studies the renormalization group
running of the masses.\cite{ref39}  The large mass of the top quark makes
significant corrections to the running masses.  Comparing the evolution
equations\cite{ref40} for the Higgs, the scalar partner of $t_R$ and the
scalar partner of $Q_L \equiv (t,b)_L$,
\begin{equation}
{d\over{d \ln(q)}}
\pmatrix{{M_h^2} \cr {\tilde{M}_{t_R}^2}\cr
{\tilde{M}_{Q^3_L}}^2\cr} = - {8\alpha_s\over{ 3\pi}} M_3^2
\pmatrix{0 \cr 1 \cr 1 \cr} +
{{\lambda_t^2}\over{8\pi^2}}\, 
({\tilde{M}_{Q^3_L}}^2 +
{\tilde{M}_{t_R}}^2+ M_h^2 + A_{o,t}^2) \pmatrix{3 \cr 2\cr 1\cr}
\end{equation}
it is clear that the influence of the top quark Yukawa coupling is greatest
for the Higgs.  At scale $q$, the approximate solution for $M_h$ is
\begin{equation}
M_h^2 (q) = M_h^2 (M_X) - {3\over{8\pi^2}} 
{\lambda_t^2} { \left( {\tilde{M}_{Q^3_L}}^2 +
{\tilde{M}_{t_R}}^2 + M_h^2 + A_{o,t}^2\right)} ln \left({M_X \over q }
\right) 
\end{equation}
and $\lambda_t$ is seen to be reducing $M_h^2$.  For $m_t \sim 175$ GeV,
this effect drives the Higgs mass, and only the Higgs mass, negative --
just as desired.\cite{ref41}

\smallskip
\noindent{\tt Light Higgs mass}
\smallskip

The low-energy spectrum of the MSSM includes a pair of neutral scalars
$h^0$ (by convention, the lighter one) and $H^0$.  At tree level, $M_h <
M_Z \vert\cos(2\beta)\vert$ where $\tan\beta$ is the ratio of the vev's of
the two Higgs doublets.\cite{higgshunt}  Searches for light Higgs bosons
then appear to put low values of $\tan\beta$ in jeopardy.  In fact
experiment has now pushed the lower bound on $M_h$ well above the $Z$ mass:
$M_h \gae 107.7$ GeV. \cite{lepewwg}

Enter the top quark.  Radiative corrections to $M_h$ involving virtual top
quarks introduce a dependence on the top mass.  For large $m_t$, this can
raise the upper bound on $M_h$ significantly.\cite{ref28}  When $\tan\beta > 1$,
\begin{equation}
M_h^2 < M_Z^2 \cos^2(2\beta) + {3 G_f
    \over{\sqrt{2}\pi^2}}\ {m_t^4}\ ln
\left({\tilde m^2 \over m_t^2}\right)
\end{equation}
and the $m_t^4$ term raises the upper bound well above $M_Z$.  Including
higher-order corrections, the most general limit\cite{ref28} appears to be
$M_h < 130$ GeV, well above the current bounds but in reach of upcoming
experiment.

\smallskip
\noindent{\tt Light top squarks}
\smallskip

\begin{figure}[bt]
\begin{center}
\scalebox{.35}{\includegraphics*[25mm,95mm][190mm,240mm]{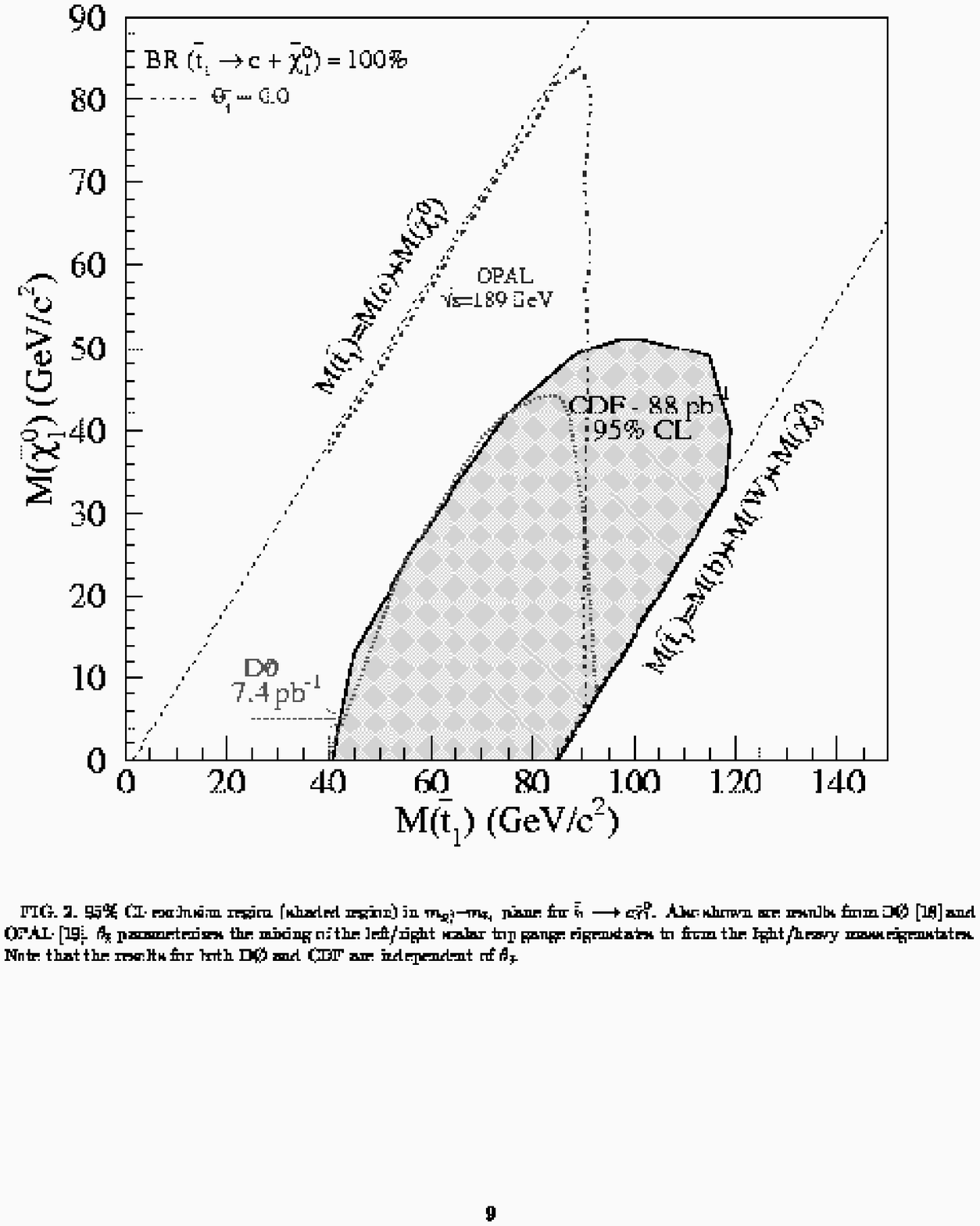}}  
\hspace{.1cm}
\scalebox{.35}{\includegraphics*[35mm,95mm][190mm,240mm]{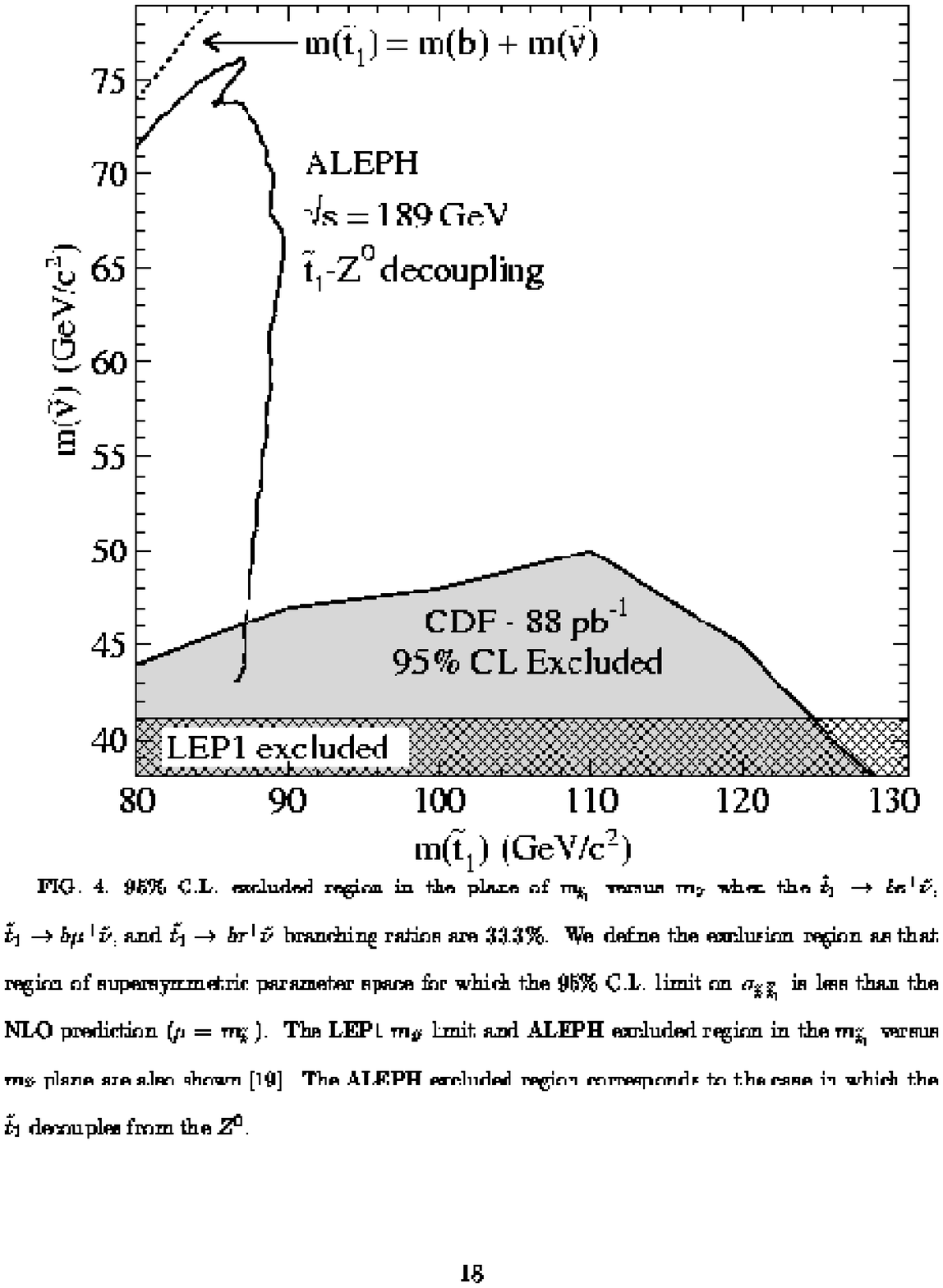}}
\end{center} 
\vspace{.5cm}
\caption[bb]{ Limits\protect\cite{run1stop} on Light Stop (left) via
$\tilde{t}_1 \to c\,\tilde{\chi}^0_1$.  (right) via $\tilde{t}_1 \to b\,
\tilde{\chi}^+_1 \to b\, \ell \tilde{\nu}$ or direct $\tilde{t}_1 \to b\,
\ell \tilde{\nu} $ assuming equal branching to all lepton flavors. }
\label{fig:stopz}
\end{figure}

\begin{figure}[tb]
\begin{center}
\scalebox{.35}{\includegraphics{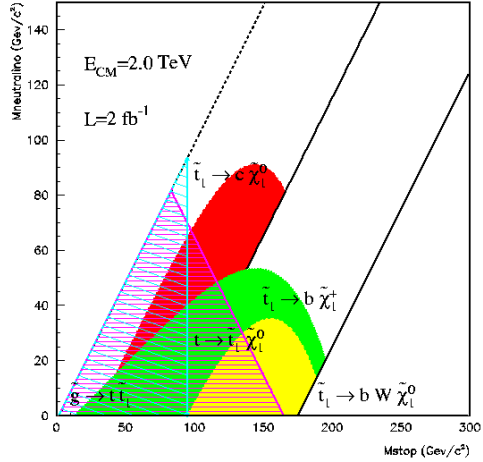}}
\end{center}
\caption[ee]{Anticipated Run II Stop limits from various decay channels.\protect\cite{run2stop}}
\label{fig:stopII}
\end{figure}

Since SUSY models include a bosonic partner for each SM fermion, there
is a pair of complex scalar top squarks affiliated with the top quark
(one for $t_L$, one for $t_R$).  A look at the mass-squared matrix for
the stops\cite{susyrev} (in the $\tilde{t}_L, \tilde{t}_R$ basis)
\begin{equation}\null\hspace{-1cm}\tilde{m}_t ^2 = 
\pmatrix{\tilde{M}^2_Q + m_t^2 &\ & 
{m_t}(A_t + \mu\cot\beta)\cr 
+ M_Z^2(\frac12 - \frac23 \sin^2\theta_W)\cos2\beta &\ &\cr
\ &\ &\ \cr  {m_t}(A_t +
    \mu\cot\beta)& &\tilde{M}^2_U + m_t^2 \cr
&\ & + \frac23 M_Z^2 \sin^2\theta_W
    \cos2\beta\cr} 
\end{equation}
reveals that the off-diagonal entries are proportional to $m_t$.  Hence, a
large $m_t$ can drive one of the top squark mass eigenstates to be
relatively light.  Experiment still allows a light stop,\cite{run1stop} as
may be seen in Figure \ref{fig:stopz}; Run II will be sensitive to higher
stop masses in several decay channels\cite{run2stop} (Figure
\ref{fig:stopII}).

Perhaps some of the Run I ``top'' sample included top
squarks.\cite{lightstop}   If the top squark is not much heavier than the
top quark, it is possible that $\tilde{t}\tilde{t}$ production occurred in
Run I, with the top squarks subsequently decaying to top plus neutralino or
gluino (depending on the masses of the gauginos).  If the top is a bit
heavier than the stop, some top quarks produced in $t\bar{t}$ pairs in Run
I may have decayed to top squarks via $t \to \tilde{t} \tilde{N}$ with the
top squarks' subsequent decay being either semi-leptonic $\tilde{t} \to b
\ell \tilde\nu$ or flavor-changing $\tilde t \to c \tilde{N}, c \tilde{g}$.
With either ordering of mass, it is possible that gluino pair production
occurred, followed by $\tilde{g} \to t \tilde{t}$.  These ideas can be
tested using the rate, decay channels, and kinematics of top quark
events.\cite{symptom}  For example, stop or gluino production could
increase the apparent $t\bar t$ production rate above that of the SM.  Or
final states including like-sign dileptons could result from gluino decays.

\subsection{Dynamical Electroweak Symmetry Breaking}\label{subsec:dewsb}

Extended technicolor (ETC) is an explicit realization of dynamical
electroweak symmetry breaking and fermion mass
generation.\cite{dynamrev} One starts with a strong gauge group
(technicolor) felt only by a set of new massless fermions (technifermions)
and extends the technicolor gauge group to a larger (ETC) group under which
ordinary fermions are also charged.  At a scale $M$, ETC breaks to its
technicolor subgroup and the gauge bosons coupling ordinary fermions to
technifermions acquire a mass of order $M$.  At a scale $\Lambda_{TC} < M$,
a technifermion condensate breaks the electroweak symmetry as described
earlier.  The quarks and leptons acquire mass because massive ETC gauge
bosons couple them to the condensate.  The top quark's mass, e.g., arises
when the condensing technifermions transform the scattering diagram in
Figure \ref{fig:etccond} (left) into the top self-energy diagram shown at
right.  Its size is
\begin{equation}
m_t \approx (g_{ETC}^2/M^2)\langle T\bar T\rangle \approx 
(g_{ETC}^2/M^2)(4\pi v^3)\ \ \ .
\label{eq:mttopp}
\end{equation}
Thus $M$ must satisfy $M/g_{ETC} \approx 1.4$ TeV in order to
produce $m_t = 175$ GeV.

{\begin{figure}[t]
\begin{center}
\scalebox{.2}{\includegraphics{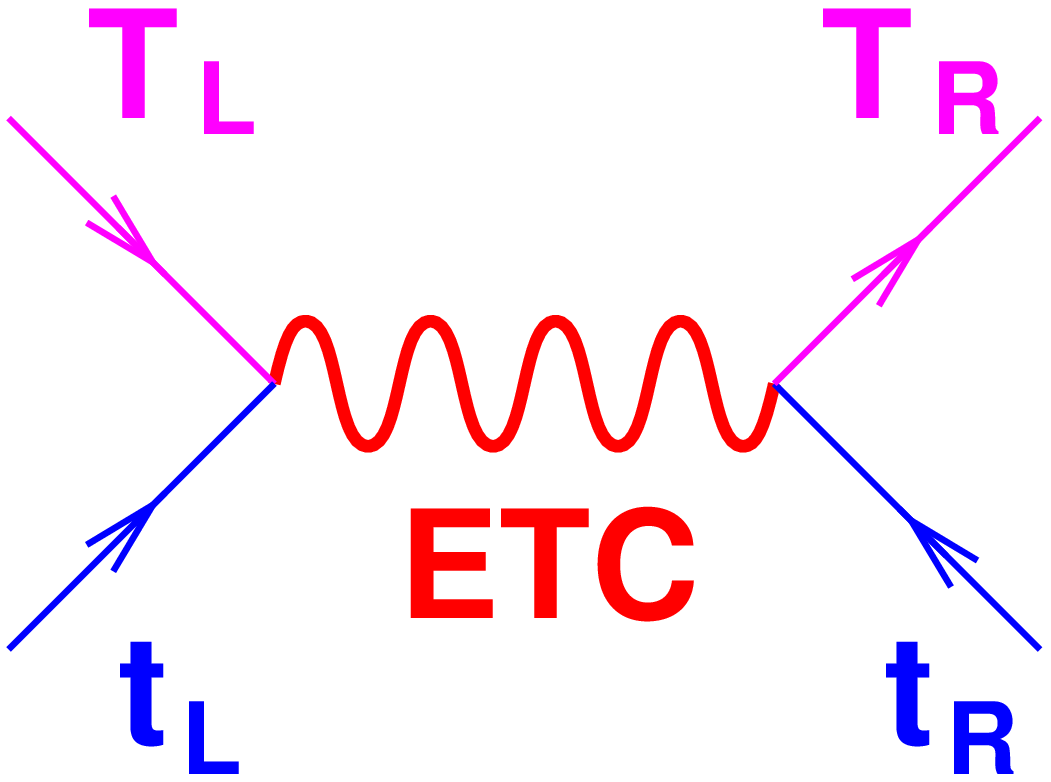}}
\hspace{2cm} \scalebox{.2}{\includegraphics{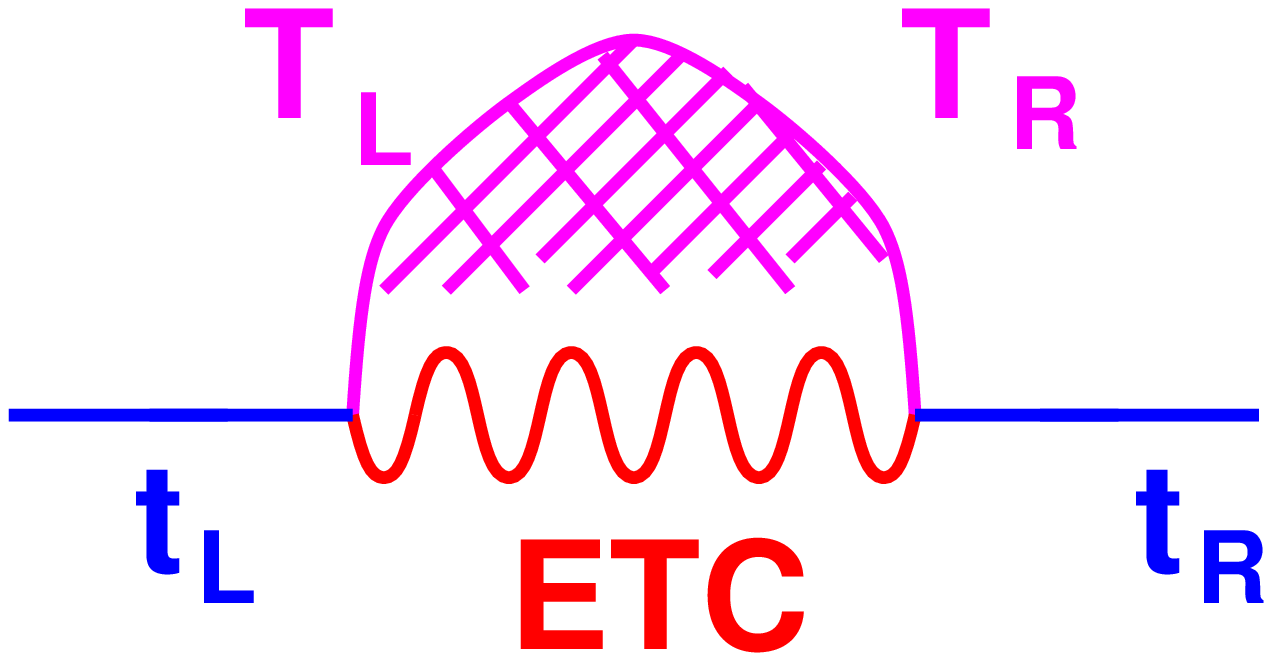}}
\end{center}
\caption{(left) Top-technifermion scattering mediated by a heavy ETC gauge
boson. (right) Technifermion condensation creates the top quark mass.}
\label{fig:etccond}
\end{figure}}

While this mechanism works well in principle, it is difficult to construct
a complete model that can accommodate a large value for $m_t$ while
remaining consistent with precision electroweak data.  Two key challenges
have led model-building in new and promising directions.  First, the
dynamics responsible for the large value of $m_t$ must couple to $b_L$
because $t$ and $b$ are weak partners.  How, then, can one obtain a
predicted value of $R_b$ that agrees with experiment?  Attempts to answer
this question have led to models in which the weak interactions of the top
quark\cite{ncetc,topflavor} (and, perhaps, all third generation fermions)
are non-standard.  Second, despite the large mass splitting $m_t \gg m_b$,
the value of the rho parameter is very near unity.  How can dynamical
models accommodate large weak isospin violation in the $t-b$ sector without
producing a large shift in $M_W$?  This issue has sparked theories in which
the strong (color) interactions of the top quark\cite{tc2} (and possibly
other quarks\cite{futc2}) are modified from the predictions of
QCD.

In the remainder of this talk, we explore the theoretical and
experimental implications of having non-standard gauge interactions for the
top quark.

\subsection{New Top Weak Interactions}\label{subsec:newweak}

In classic ETC models, the large value of $m_t$ comes from ETC dynamics at
a relatively low scale $M$ of order a few TeV.  At that scale, the weak
symmetry is still intact so that $t_L$ and $b_L$ function as weak partners.
Moreover, experiment tells us that $\vert V_{tb} \vert \approx 1$.  As a
result, the ETC dynamics responsible for generating $m_t$ must couple with
equal strength to $t_L$ and $b_L$.  While many properties of the top quark
are only loosely constrained by experiment, the $b$ quark has been far more
closely studied.  In particular, the LEP measurements of the $Zb\bar{b}$
coupling are precise enough to be sensitive to the quantum corrections
arising from physics beyond the SM.  As we now discuss, radiative
corrections to the $Zb\bar{b}$ vertex from low-scale ETC dynamics can be so
large that new weak interactions for the top quark are required to make the
models consistent with experiment.\cite{etcrb,ncetc}

\begin{figure}[tb]
\begin{center}
\scalebox{.2}{\includegraphics{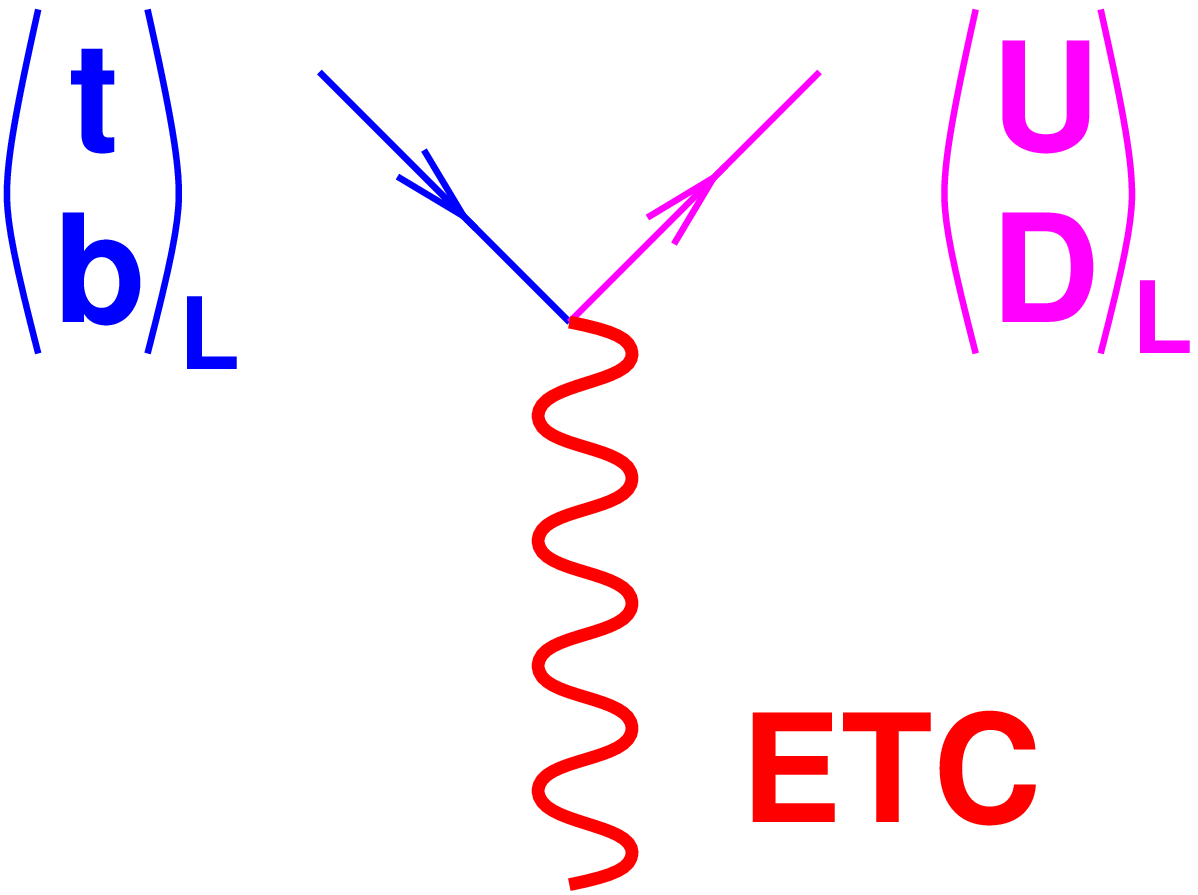}}
\hspace{2cm} \scalebox{.2}{\includegraphics{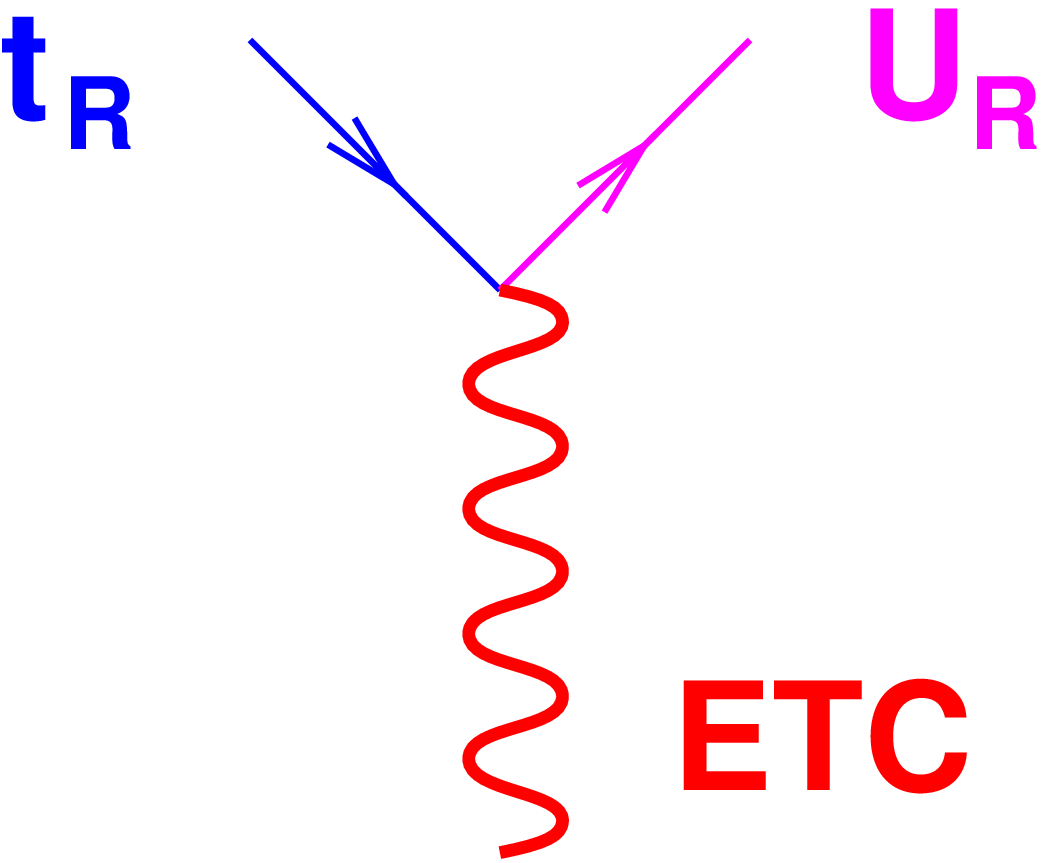}}
\end{center}
\caption{Fermion currents coupling to the weak-singlet ETC boson that
generates $m_t$.}
\label{fig:etcfer}
\end{figure}

To begin, consider the usual ETC models in
which the extended technicolor and weak gauge groups commute, so that the
ETC gauge bosons carry no weak charge.  In these models, the ETC gauge
boson whose exchange gives rise to $m_t$ couples to the fermion
currents\cite{etcrb} 
\begin{equation}
\xi \left(\bar\psi^i_L\ \gamma^\mu\ T^{ik}_L\right)\  +\ 
\xi^{-1} \left(\bar t_R\ \gamma^\mu\ U^k_R\right)
\end{equation}
where $\xi$ is a Clebsh of order 1 (see Figure \ref{fig:etcfer}).
Then the top quark mass arises from technifermion condensation and ETC
boson exchange as in Figure \ref{fig:etccond}, with the relevant
technifermions being $U_L$ and $U_R$.

\begin{figure}[tb]
\begin{center}
\scalebox{.2}{\includegraphics{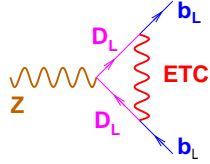}}
\end{center}
\caption[gr]{Direct correction to the $Zb\bar{b}$ vertex from the ETC gauge
boson responsible for $m_t$ in a commuting model.}
\label{fig:etc-vv}
\end{figure}

Exchange of the same\cite{etcrb} ETC boson causes a
direct (vertex) correction to the $Z\to b\bar{b}$ decay as shown in Figure
\ref{fig:etc-vv}; note that it is $D_L$ technifermions with 
$I_3 = -\frac{1}{2}$ which enter the loop.
This effect reduces the magnitude of the $Zb\bar{b}$ coupling by
\begin{equation}
\delta g_L = { e \over {4 \sin{\theta} \cos{\theta}}} 
\left({g^2 v^2 \over {M^2}}\right) 
\label{eq:delgg}
\end{equation}
Given the relationship between $M$ and
$m_t$ from Eq. \ref{eq:mttopp}, we find
\begin{equation}
{{g^2 v^2 \over {M^2}} \approx {m_t \over {4 {\pi} v}}}
\end{equation}
so that the top quark mass sets the size of the coupling shift.  

\begin{figure}[tb]
\begin{center}
\scalebox{.3}{\includegraphics{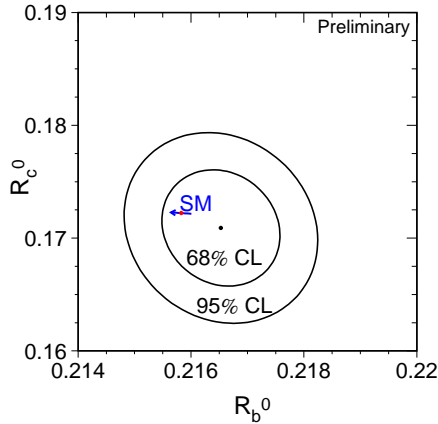}}
\end{center}
\vspace{-.5cm}
\caption[gr]{Data\cite{lepewwg} on $R_b$ and $R_c$ showing experimental
best fit (dot) and SM prediction (arrow).  A 5\% negative shift in $R_b$ is
clearly excluded. }
\label{fig:leprbdata}
\end{figure}

How to observe the shift in the couplings?  The vertex correction will
certainly produce a correction $\delta\Gamma_b$ to the $Z$ decay width
$\Gamma(Z \to b\bar{b})$.  But since $\Gamma_b$ also receives oblique
radiative corrections, $\Gamma_b^{corr.} = (1 + \Delta\rho) (\Gamma_b +
\delta\Gamma_b)$, a measurement of $\Gamma_b$ is not the best way to track
$\delta g_b$.  The ratio of $\Gamma_b$ to the hadronic decay width of the
$Z$
\begin{equation}
R_b \equiv \Gamma(Z \to b\bar{b}) / \Gamma(Z \to {\rm hadrons})
\end{equation}
is also proportional to $\delta g_L$ and has the additional advantage that
oblique and QCD radiative corrections each cancel in the ratio (up to
factors suppressed by small quark masses).  One finds\cite{etcrb}
\begin{equation}
{{\delta R_b \over R_b} \approx - 5.1\%\cdot \xi^2\cdot  
\left(\frac{m_t}{175{\rm GeV}}\right) }
\label{eq:delrbb}
\end{equation}
Such a large shift in $R_b$ is excluded\cite{lepewwg} by the data (see Figure
\ref{fig:leprbdata}).  Then the ETC models whose dynamics produces this
shift are likewise excluded.

This suggests one should consider an alternative class of ETC
models\cite{ncetc} in which the weak group $SU(2)_W$ is embedded in
$G_{ETC}$, so that the weak bosons carry weak charge.  Embedding the weak
interactions of all quarks in a low-scale ETC group would produce masses of
order $m_t$ for all up-type quarks.  Instead, one can extend $SU(2)$ to a
direct product group $SU(2)_h \times SU(2)_\ell$ such that the third
generation fermions transform under $SU(2)_h$ and the others under
$SU(2)_\ell$.  Only $SU(2)_h$ is embedded in the low-scale ETC group; the
masses of the light fermions will come from physics at higher scales.
Breaking the two weak groups to their diagonal subgroup ensures approximate
Cabibbo universality at low energies.  The electroweak and technicolor
gauge structure of these non-commuting models is sketched below\cite{ncetc}:
\begin{eqnarray}
G_{ETC}  &\times& SU(2)_{light} \times U(1) \nonumber\\
&\downarrow& f \nonumber \\
G_{TC} &\times& SU(2)_{heavy}  \times SU(2)_{light}  \times
U(1)_Y \nonumber\\
&\downarrow& u \nonumber\\
G_{TC}  &\times& SU(2)_{weak} \times U(1)_Y\nonumber\\
&\downarrow& v\nonumber\\
G_{TC} & \times&  U(1)_{EM}
\end{eqnarray}

\begin{figure}[tb]
\begin{center}
\scalebox{.2}{\includegraphics{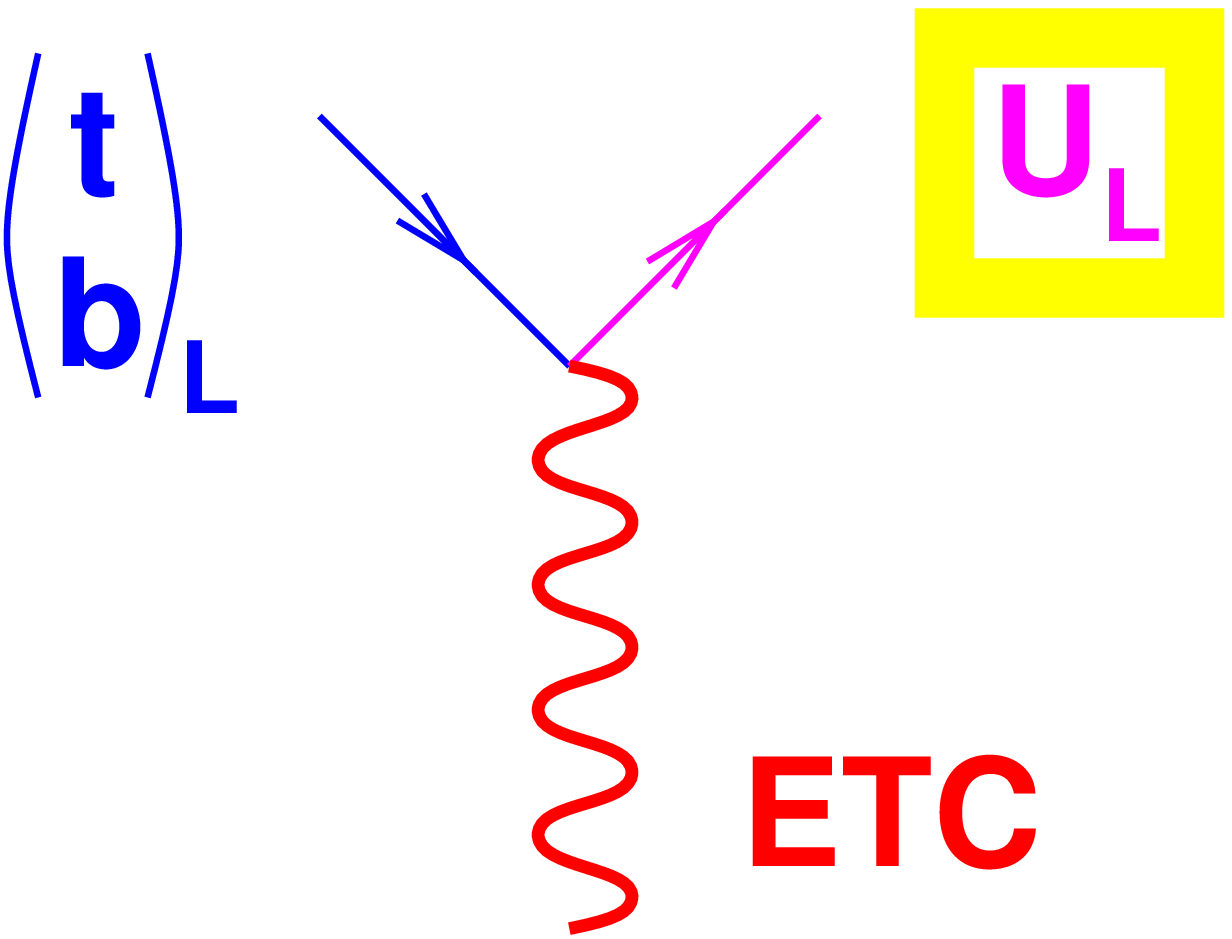}}
\hspace{2cm} \scalebox{.2}{\includegraphics{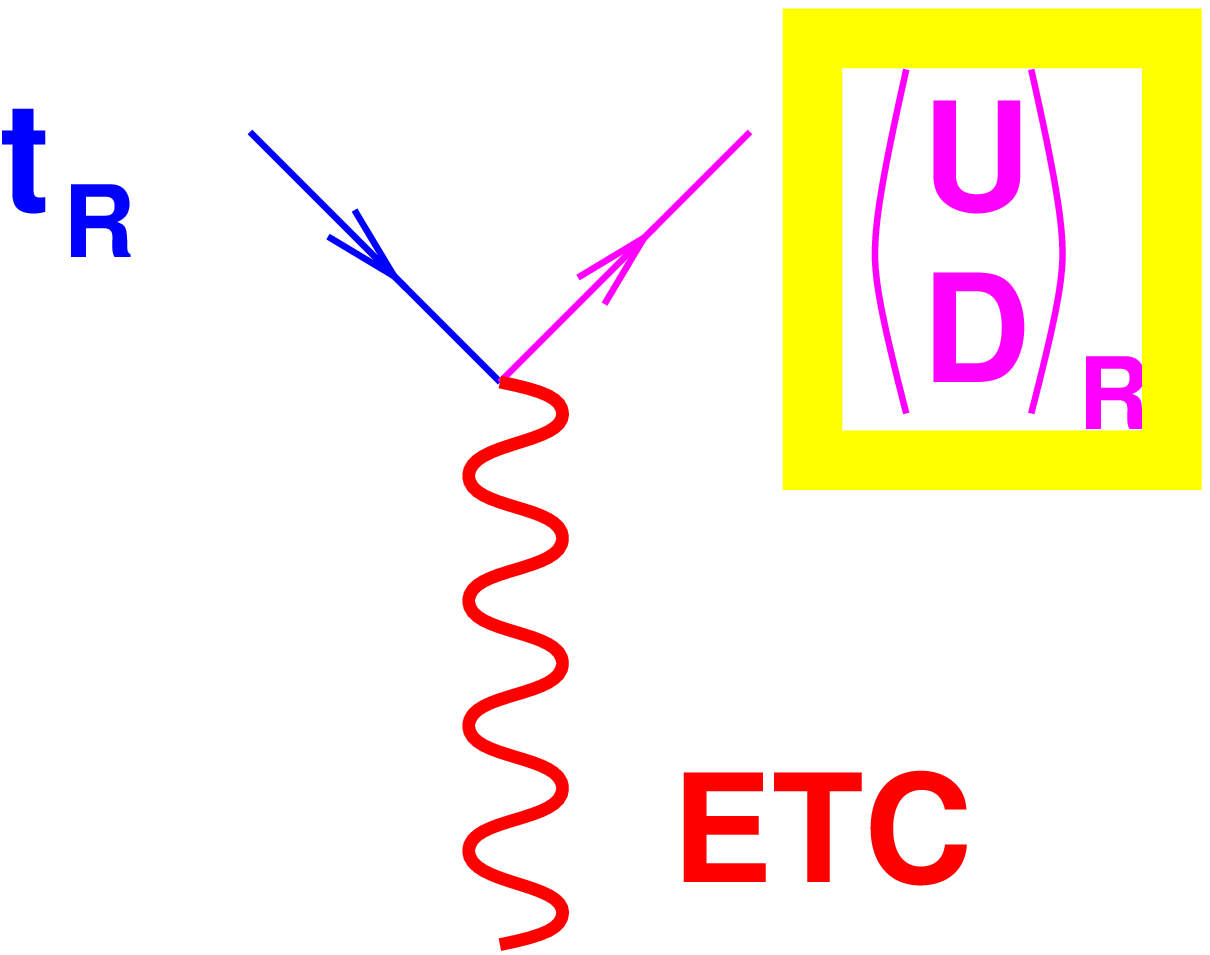}}
\end{center}
\caption{Fermion currents coupling to the weak-doublet ETC boson that
generates $m_t$ in non-commuting ETC models.}
\label{fig:ncetc-vv}
\end{figure}

Due to the extended gauge structure, there are now two non-standard
contributions to $R_b$, one from the dynamics that generates $m_t$ and the
other from the mixing of the two $Z$ bosons from the two weak groups.  The
ETC boson responsible for $m_t$ now couples weak-double fermions to
weak-singlet technifermions (and vice versa) as in Figure
\ref{fig:ncetc-vv}.  The radiative correction to the $Zb\bar{b}$ vertex is
as in Figure \ref{fig:etc-vv} except that the technifermions involved are
now $U_L$ with $T_3 = +\frac{1}{2}$.  As a result, the shift in $\delta
g_L$ and $R_b$ have the same size the results in Eqs. \ref{eq:delgg} and
\ref{eq:delrbb} -- but the opposite sign.\cite{ncetc}  Were these the only
contributions to $R_b$, this class of models would be excluded.

Consider, however, what happens when the $SU(2)_h \times SU(2)_\ell \times
U(1)_Y$ bosons mix to form mass eigenstates.\cite{ncetc}  The result is heavy
states $W^H, Z^H$ that couple mainly to the third generation, light
states $W^L, Z^L$ resembling the standard $W$ and $Z$, and a massless
photon $A^\mu = \sin\theta [\sin\phi \,W_{3\ell}^\mu +
\cos\phi\,W_{3h}^\mu] +\cos\theta X^\mu $ coupling to $Q = T_{3h} +
T_{3\ell} + Y $.  Here, $\phi$ describes the mixing between the two
weak groups and $\theta$ is the usual weak angle.  In terms of a rotated
basis (with $s\equiv\sin\phi,\ c\equiv\cos\phi$)
\begin{eqnarray}
W^{\pm}_1 &=& s\,W^{\pm}_\ell+c\,W^{\pm}_h \ \ \ \ \ \ \  W^{\pm}_2 =
c\,W^{\pm}_\ell-s\,W^{\pm}_h\nonumber \\
D^\mu &=&\partial^\mu + ig\left( T_\ell^\pm + T_h^\pm \right)
W^{\pm\,\mu}_1 +ig\left( {c \over s}T_\ell^\pm - {s \over c}T_h^\pm \right)
W^{\pm\,\mu}_2 \nonumber\\
Z_1 &=& \cos\theta (s\,W_{3\ell}+c\,W_{3h}) - \sin\theta X \ \ \ \ \ \ \ 
Z_2 = c\,W_{3\ell}-s\,W_{3h} \nonumber \\
D^\mu &=&\partial^\mu + i{g\over \cos\theta}\left( T_{3\ell} + T_{3h} - \sin^2\theta Q \right)Z^{\mu}_1
\end{eqnarray}
where $W_1, Z_1$ have SM couplings and all non-standard couplings
accrue to $W_2, Z_2$,
the mass eigenstates are (with $x \equiv u^2/v^2$)
\begin{eqnarray}
W^L &\approx&  W_1-{c^3 s \over x}\,W_2~,\ \ \ \ W^H \approx W_2 + {c^3
s \over {x}}\,W_1 \nonumber \\
Z^L &\approx&  Z_1-{c^3 s \over {x\cos\theta}}\,Z_2~,\ \ \ \ \ \ Z^H \approx
Z_2 + {c^3 s \over {x\cos\theta}}\,Z_1
\end{eqnarray}
and the heavy boson masses are degenerate: $M_{W^H} \approx M_{Z^H} \approx
M_W \sqrt{x} / s c$.   The $Z^L$ coupling to
quarks thus differs from the SM value by $\delta g_L = (c^4/x) T_{3\ell} -
(c^2s^2/x) T_{3h}$ which reduces $R_b$ by\cite{ncetc}
\begin{equation}
\frac{\delta R_b}{R_b} \approx - 5.1\% \ [\sin^2\phi \frac{f^2}{u^2}]
\end{equation}
where the term in square brackets is ${\cal O}(1)$.  

As the ETC and $ZZ'$ mixing contributions to $R_b$ are of the same
magnitude, but opposite size, $R_b$ can be consistent with experiment in
non-commuting ETC models.  The key element that permits a large $m_t$ and a
small value of $R_b$ to co-exist is the presence of non-standard weak
interactions for the top quark.\cite{ncetc}  This is something experiment
can test, and has since been incorporated into models such as
topflavor\cite{topflavor} and top seesaw.\cite{topseesawext}

\begin{figure}[tb]
\begin{center}
\scalebox{.5}{\includegraphics{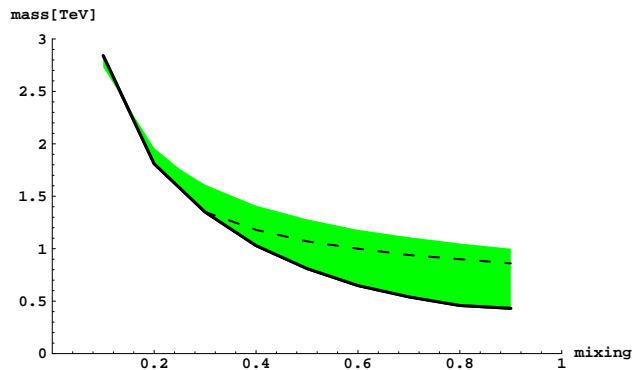}}
\end{center}
\caption{FNAL Run II single top production can explore the shaded region of
the $M_{W'}$ vs. $\sin^2\phi$ plane.\protect\cite{testvtb}  The area below
the solid curve is excluded by precision electroweak
data.\protect\cite{ncetclim}  In the shaded region $R_\sigma$ increases by
$\geq$ 16\%; below the dashed curve, by $\geq$ 24\%.}
\label{fig:singtp}
\end{figure}

There are several ways to test whether the high-energy weak interactions
have the form $SU(2)_h \times SU(2)_\ell$.  One possibility is to search
for the extra weak bosons.  The bosons' predicted effects on precision
electroweak data gives rise to the exclusion curve\cite{ncetclim} in Figure
\ref{fig:singtp}. Low-energy exchange of $Z^H$ and $W^H$ bosons would cause
apparent four-fermion contact interactions; LEP limits on $eebb$ and
$ee\tau\tau$ contact terms imply\cite{zprimetautau} $M_{Z^H} \gae 400$ GeV.
Direct production of $Z^H$ and $W^H$ at Fermilab is also feasible; a Run II
search for $Z^H \to \tau\tau \to e\mu X$ will be
sensitive\cite{zprimetautau} to $Z^H$ masses up to 750 GeV.  Another
possibility is to measure the top quark's weak interactions in single top
production.  Run II should measure the ratio of single top and single
lepton cross-sections $R_\sigma \equiv \sigma_{tb}/\sigma_{\ell\nu}$ to
$\pm 8\%$ in the $W^*$ process.\cite{boos}  A number of systematic
uncertainties, such as those from parton distribution functions, cancel in
the ratio.  In the SM, $R_\sigma$ is proportional to the square of the
$Wtb$ coupling. Non-commuting ETC models affect the ratio in two ways:
mixing of the $W_h$ and $W_\ell$ alters the $W^L$ coupling to fermions, and
both $W^L$ and $W^H$ exchange contributes to the
cross-sections\footnote{The ETC dynamics which generates $m_t$ has no
effect on the $Wtb$ vertex because the relevant ETC boson does not couple
to $b_R$.}. Computing the shift in $R_\sigma$ from these effects reveals
(Figure \ref{fig:singtp}) that Run II will be sensitive\cite{testvtb} to
$W^H$ bosons up to masses of about 1.5 TeV.

\subsection{New Top Strong Interactions}\label{subsec:newstrong}

\begin{figure}[bt]
\begin{center}
\scalebox{.2}{\includegraphics{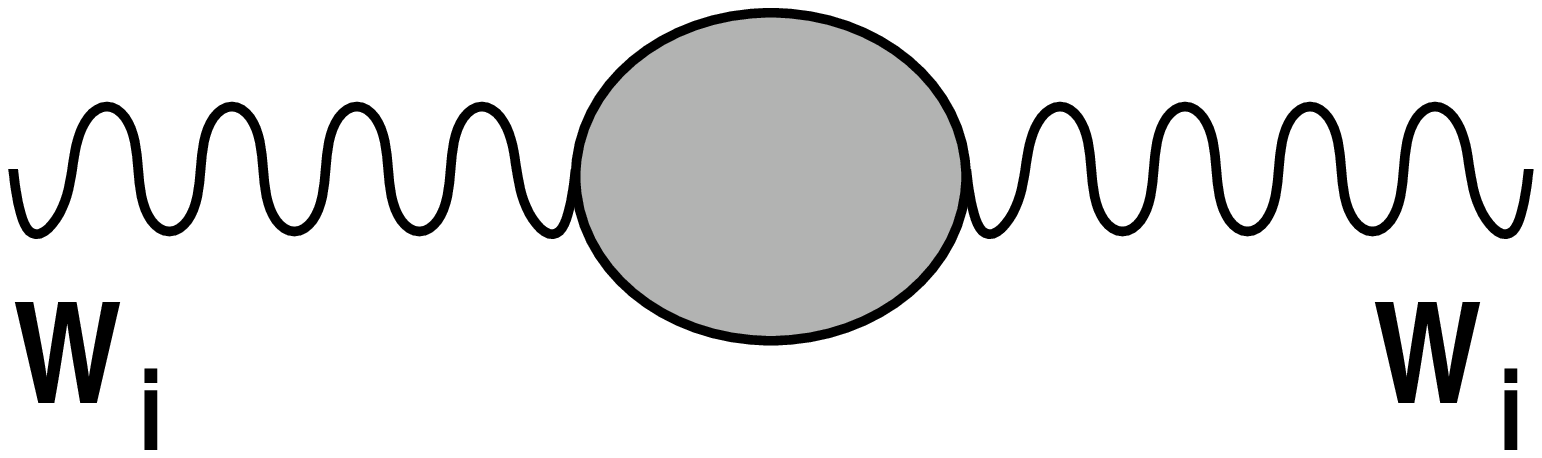}}\hspace{1cm}\raise15pt\hbox{=$\
  \ i
  \Pi_{ii} g^{\mu\nu} + ...$}
\end{center}
\vspace{-.2cm}
\caption{Electroweak boson propagator used in calculation of $\Delta\rho$.}
\label{fig:rhoprop}
\end{figure}

At tree-level in the SM, $\rho \equiv M_W^2 / M_Z^2 \cos^2\theta_W \equiv
1$ due to a ``custodial'' global $SU(2)$ symmetry relating members of a
weak isodoublet.  Because the two fermions in each isodoublet have
different masses and hypercharges, however, oblique\cite{lsp} radiative
corrections to the $W$ and $Z$ propagators alter the value of $\rho$.  The
one-loop correction from the (t,b) doublet is particularly large because
$m_t \gg m_b$. The shift in $\rho$ is computed from the propagators in
Figure \ref{fig:rhoprop} as\cite{lsp}
\begin{equation}
\Delta\rho(0) \equiv \rho(0) - 1 = \frac{e^2}{\sin^2\theta_W
  \cos^2\theta_W 
  M_Z^2} [\Pi_{11}(0) - \Pi_{33}(0)]
\end{equation}
Experiment\cite{pdg} finds $\vert\Delta\rho\vert \leq 0.4\% $, a stringent
constraint on isospin-violating new physics.  For example, a heavy lepton
doublet (N,E) with standard weak couplings and mass $\gg M_Z$ would
add\cite{lsp} 
\begin{equation}
\Delta\rho_{N,E} \approx
\frac{\alpha_{EM}}{16\pi\sin^2\theta_W\cos^2\theta_W M_Z^2}
[m_N^2 + m_E^2 - \frac{2m_N^2
  m_E^2}{m_N^2-m_E^2}log(\frac{m_N^2}{m_E^2})]
\end{equation}
and a new quark doublet, three times as much; the data forces the
new fermions to be nearly degenerate.

\begin{figure}[tb]
\begin{center}
\scalebox{.4}{\includegraphics{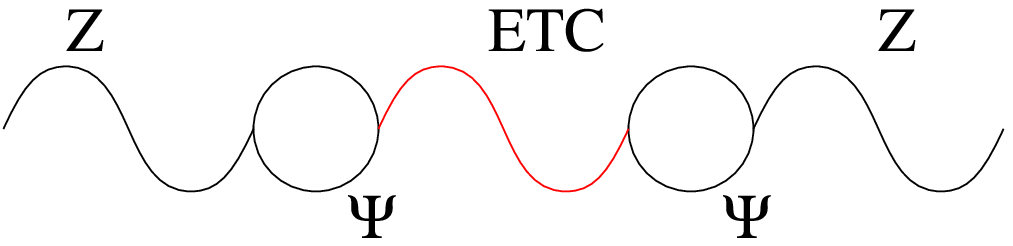}}
\hspace{1cm} \scalebox{.4}{\includegraphics{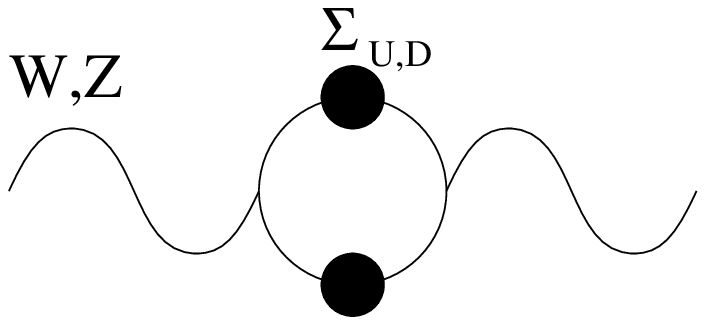}}
\end{center}
\vspace{-.2cm}
\caption{ ETC contributions to $\Delta\rho$: (left) direct, from gauge boson
mixing (right) indirect, from technifermion mass splitting.}
\label{fig:zetcmix}
\end{figure}

\begin{figure}[tb]
\begin{center}
\scalebox{.45}{\includegraphics{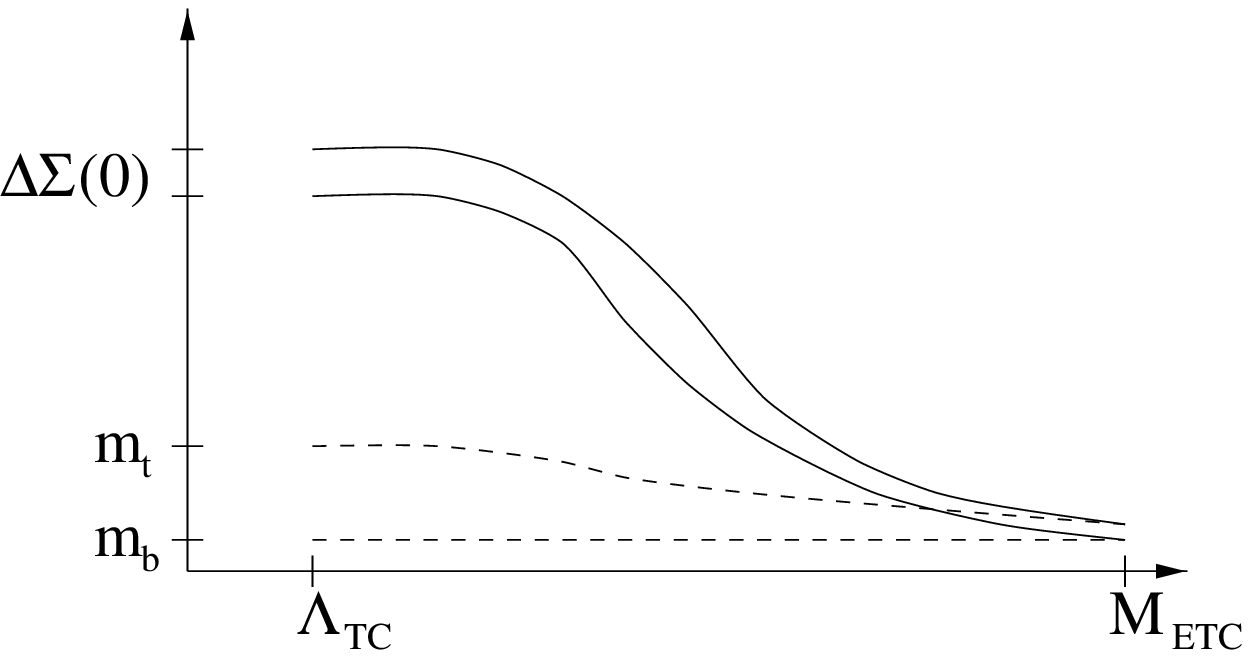}} 
\scalebox{.45}{\includegraphics{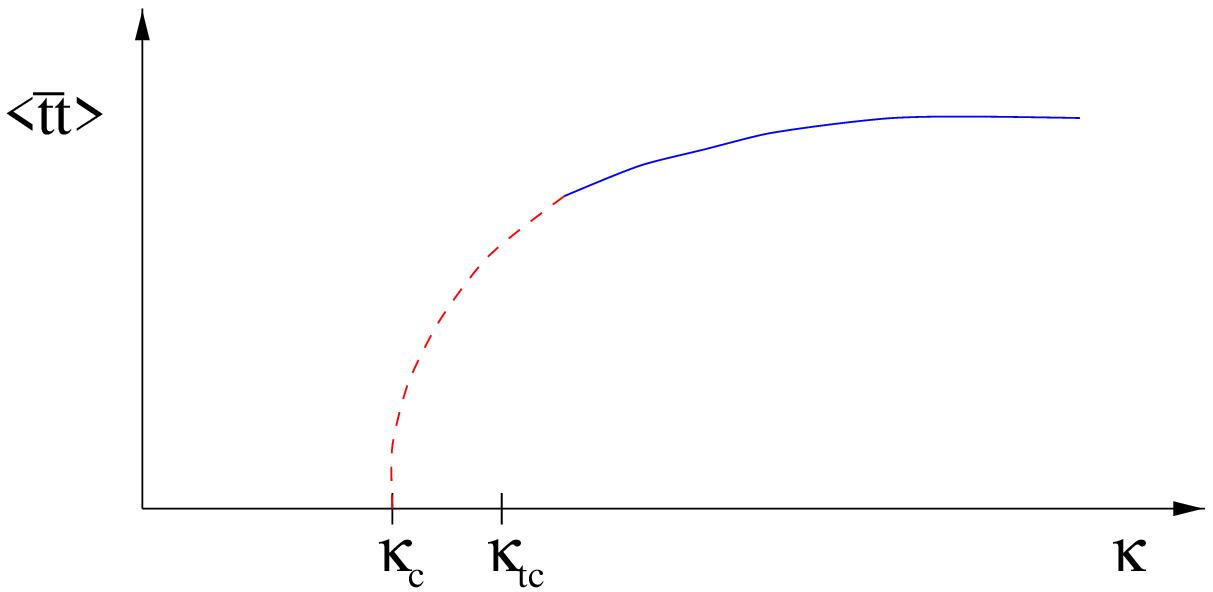}}
\end{center}
\caption{(left) ETC and new top dynamics generate masses for
technifermions, $t$ and $b$. (right) Second-order phase transition forms a
top condensate for $\kappa > \kappa_c$.}
\label{fig:dynmasss}
\end{figure}

Dynamical theories of mass generation like ETC must break weak isospin in
order to produce the large top-bottom mass splitting.  However, the new
dynamics may also cause large contributions to $\delta\rho$.  Direct
mixing between and ETC gauge boson and the Z (Figure \ref{fig:zetcmix})
induces the dangerous effect\cite{rho12}
\begin{equation}
\Delta\rho \approx  12\% \cdot 
\left({\sqrt{N_D} F_{TC} \over 250 {\rm\ GeV}}\right)^2
\cdot \left({1 {\rm\ TeV} \over M_{ETC}/g_{ETC}}\right)^2
\end{equation}
in models with $N_D$ technifermion doublets and technipion decay
constant $F_{TC}$.  To avoid this, one could make the ETC boson heavy;
however the required $M_{ETC}/g_{ETC} >\ 5.5{\rm TeV} (\sqrt{N_D}
F_{TC} / 250\ {\rm GeV})$ is too large to produce $m_t = 175$ GeV.
Instead, one must obtain $N_D F^2_{TC} \ll (250{\rm GeV})^2$ by
separating the ETC sectors responsible for electroweak symmetry
breaking and the top mass.  A second contribution comes
indirectly\cite{wtcprl} through the technifermion mass splitting:
$\Delta\rho \sim (\Sigma_U(0) - \Sigma_D(0))^2/M_Z^2$, as in Figure
\ref{fig:zetcmix}.  Again, a cure\cite{tc2,strongetc} is to arrange
for the $t$ and $b$ to get only part of their mass from technicolor.
As sketched in Figure \ref{fig:dynmasss}, suppose $M_{ETC}$ is large
and ETC makes only a small contribution to the fermion and
technifermion masses.  At a scale between $M_{ETC}$ and $\Lambda_{TC}$
new strong dynamics felt only by (t,b) turns on and generates $m_t \gg
m_b$.  The technifermion mass splitting is small, $\Delta\Sigma(0)
\approx m_t(M_{ETC} - m_b(M_{ETC}) \ll m_t$, and no large
contributions to $\Delta\rho$ ensue.

The realization that new strongly-coupled dynamics for the (t,b) doublet
could be so useful has had a dramatic effect on model-building.  Models in
which some (topcolor\cite{tc2}) or even all (top mode,\cite{topmode} top
seesaw\cite{topseesaw}) of electroweak symmetry breaking is due to a top
condensate have proliferated.  One physical realization of a new
interaction for the top is a spontaneously broken extended gauge group
called topcolor\cite{tc2}: $SU(3)_h \times SU(3)_\ell \to SU(3)_{QCD}$.
The (t,b) doublet transforms under $SU(3)_h$ and the light quarks, under
$SU(3)_\ell$.  Below the symmetry-breaking scale $M$, the spectrum includes
massive topgluons which mediate vectorial color-octet interactions among
top quarks: $ -(4\pi\kappa/M^2) (\bar{t}\gamma_\mu \frac{\lambda^a}{2}
t)^2$.  If the coupling $\kappa$ lies above a critical value ($\kappa_c =
3\pi/8$ in the NJL\cite{NJL} approximation), a top condensate forms (Figure
\ref{fig:dynmasss}).  For a second-order phase transition, $\langle \bar{t}
t \rangle / M^3 \propto (\kappa - \kappa_c) / \kappa_c$, so the top quark
mass generated by this dynamics can lie well below the symmetry breaking
scale; so long as $M$ is not too large, the scale separation need not imply
an unacceptable degree of fine tuning.

A more complete model incorporating these ideas is topcolor-assisted
technicolor\cite{tc2} (TC2).  The symmetry-breaking structure is:
\begin{eqnarray} 
G_{TC} &\times& {SU(3)_h \times SU(3)_\ell} \times SU(2)_W
  \times  {U(1)_h \times U(1)_\ell} \nonumber \\
&\downarrow&\ \ \ M \gae 1\ {\rm TeV} \nonumber \\
G_{TC} &\times& {SU(3)_{QCD}}  \times SU(2)_{W} \times
  { U(1)_Y} \nonumber \\
&\downarrow&\ \ \ \Lambda_{TC}\sim 1\ {\rm  TeV}\nonumber \\
G_{TC} &\times& SU(3)_{QCD} \times {U(1)_{EM}}
\end{eqnarray}
Below the scale $M$, the heavy topgluons and Z' mediate
new effective interactions\cite{tc2,tc2phase} for the (t,b) doublet
\begin{equation} 
-{4\pi \kappa_3\over{M^2}}\left[\overline{\psi}\gamma_\mu  
{{\lambda^a}\over{2}} \psi \right]^2 - {4\pi
\kappa_1\over{M^2}}
\left[{1\over3}\overline{\psi_L}\gamma_\mu  \psi_L
+{4\over3}\overline{t_R}\gamma_\mu  t_R -{2\over3}\overline{b_R}\gamma_\mu
b_R \right]^2 
\label{eq:tc2outline}
\end{equation}
where the $\lambda^a$ are color matrices and $g_{3h} \gg g_{3\ell}$,
$g_{1h} \gg g_{1\ell}$.  The $\kappa_3$ terms are uniformly attractive;
were they alone, they would generate large $m_t$ {\bf and} $m_b$.  The
$\kappa_1$ terms, in contrast, include a repulsive component for $b$.  As a
result, the combined effective interactions\cite{tc2,tc2phase}
\begin{equation}
\kappa^t = \kappa_3 +{1\over3}\kappa_1 >
\kappa_c  > \kappa_3 -{1\over 6}\kappa_1 =\kappa^b
\end{equation}
can be super-critical for top, causing $\langle\bar{t}t\rangle \neq 0$ and
a large $m_t$, and sub-critical for bottom, leaving $\langle\bar{b}b\rangle
= 0$. 

The benefits of including new strong dynamics for the top quark are clear
in TC2 models.\cite{tc2phase}  Because technicolor is responsible for most
of electroweak symmetry breaking, $\Delta\rho \approx 0$.  Direct
contributions to $\Delta\rho$ are avoided because the top condensate
provides only $f\sim 60$ GeV; indirect contributions are not an issue if
the technifermion hypercharges preserve weak isospin.  The top condensate
yields a large top mass.  ETC dynamics at $M_{ETC} \gg 1$ TeV generate the
light $m_f$ without large FCNC and contribute only $\sim 1$ GeV to the
heavy quark masses so there is no large shift in $R_b$.

\subsection{Phenomenology of Strong Top Dynamics}\label{subsec:pheno}

Models with new strong top dynamics fall into three general classes with
distinctive spectra and phenomenology: topcolor,\cite{tc2,tc2phase}
flavor-universal extended color,\cite{futc2} and top
seesaw.\cite{topseesaw} These theories include a variety of new states that
can weigh less than a few TeV.  A generic feature is colored gauge bosons
with generation-specific (topgluon) or flavor-universal (coloron) couplings
to quarks.  The strongly-bound quarks may also form composite scalar
states.  Many models include color-singlet (Z') bosons with
generation-dependent couplings.  Some theories generate masses with the
help of exotic fermions (usually, but not always weak-singlets).  In this
section of the talk, we review experimental searches for these new states.

\begin{figure}[tb]
\begin{center}
\scalebox{.55}{\includegraphics{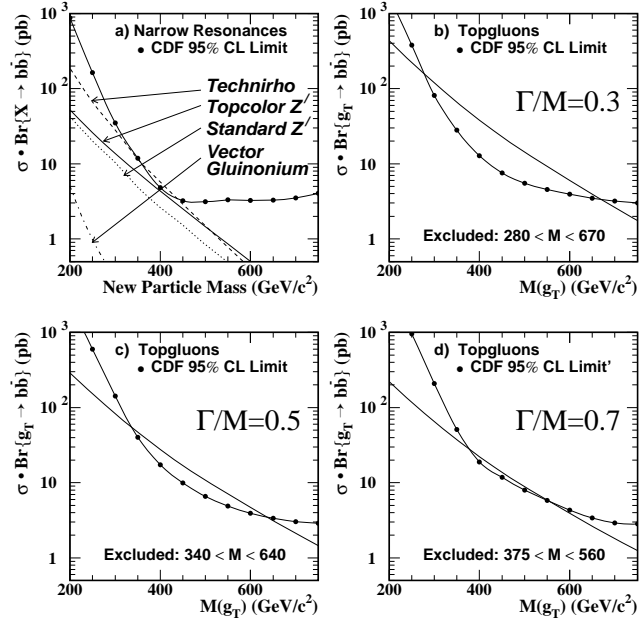}}
\end{center}
\caption{Results of CDF searches\protect\cite{topglucdf} for topgluons and $Z'$
decaying to $b\bar{b}$.}
\label{fig:cdfbb}
\end{figure}

\smallskip
\noindent{\tt Topcolor Models}
\smallskip

\begin{figure}[tb]
\begin{center}
\scalebox{.4}{\includegraphics{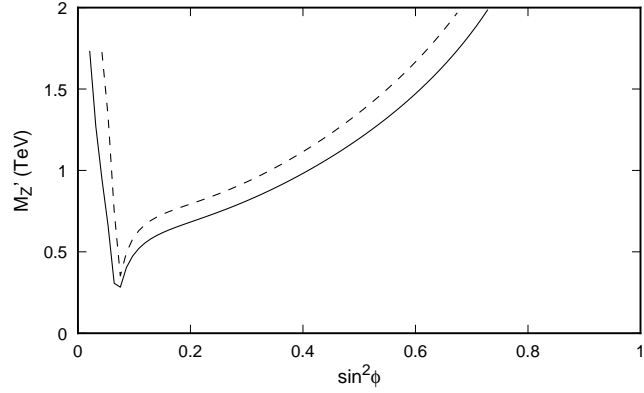}}
\end{center}
\caption{Lower bounds on the mass of topcolor $Z'$ from precision
electroweak data.\protect\cite{tc2zp}} 
\label{fig:chte}
\end{figure}

The gauge structure of topcolor\cite{tc2,tc2phase} models, as outlined in
section \ref{subsec:newstrong}, generally includes extended color and
hypercharge sectors (as in Eq. \ref{eq:tc2outline}) and a standard weak
gauge group.  The third-generation fermions transform under the more
strongly-coupled $SU(3)_h \times U(1)_h$ group, so that after the extended
symmetry breaks to the SM gauge group the heavy topgluons and $Z'$ couple
preferentially to the third generation.  The light fermions transform under
$SU(3)_\ell \times U(1)_\ell$.  CDF's search\cite{topglucdf} for topgluons
decaying to $b\bar{b}$ has put constraints on the topgluon mass for three
different assumed widths (Figure \ref{fig:cdfbb}); the topgluon's strong
coupling to quarks ensures that it will be a rather broad resonance.  Run
II and the LHC should be sensitive to topgluons in $b\bar{b}$ or $t\bar{t}$
final states.  The $Z'$, being more weakly coupled is narrow; CDF's limit
on $\sigma\cdot B$ for narrow states\cite{topglucdf} decaying to $b\bar{b}$
just misses being able to constrain this state (Figure \ref{fig:cdfbb}).  A
more recent CDF search\cite{leptophobe} for a leptophobic topcolor $Z'$
decaying to top pairs excludes bosons weighing less than 480 (780) GeV
assuming $\Gamma/M = 0.012\ (0.04)$.  Precision electroweak data
constrains\cite{tc2zp} topcolor $Z'$ bosons as shown in Figure
\ref{fig:chte}; light masses are still allowed if the $Z'$ couples almost
exclusively to the third generation.  As mentioned earlier, FNAL Run II
will be sensitive\cite{zprimetautau} to topcolor $Z'$ bosons as heavy as
750 GeV in the process $Z' \to \tau\tau \to e\mu X$.  Ultimately, an NLC
would be capable of finding a 3-6 TeV $Z'$ decaying to taus.\cite{futc2}

\begin{figure}[tb]
\begin{center}
\null\hspace{-1.5cm}
\scalebox{.35}{\includegraphics{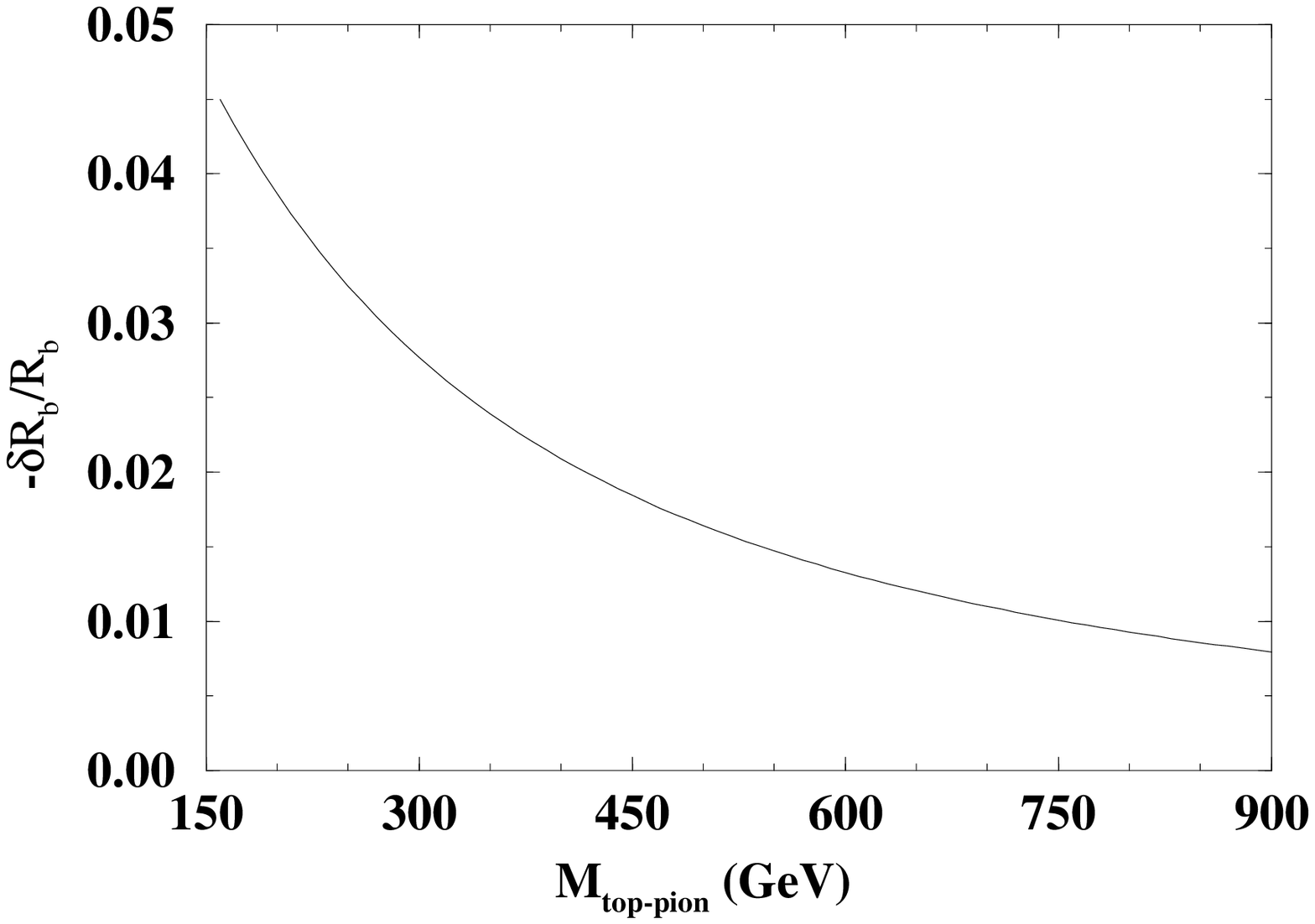}}\hspace{-1cm} 
\rotatebox{90}{\scalebox{.33}{\includegraphics{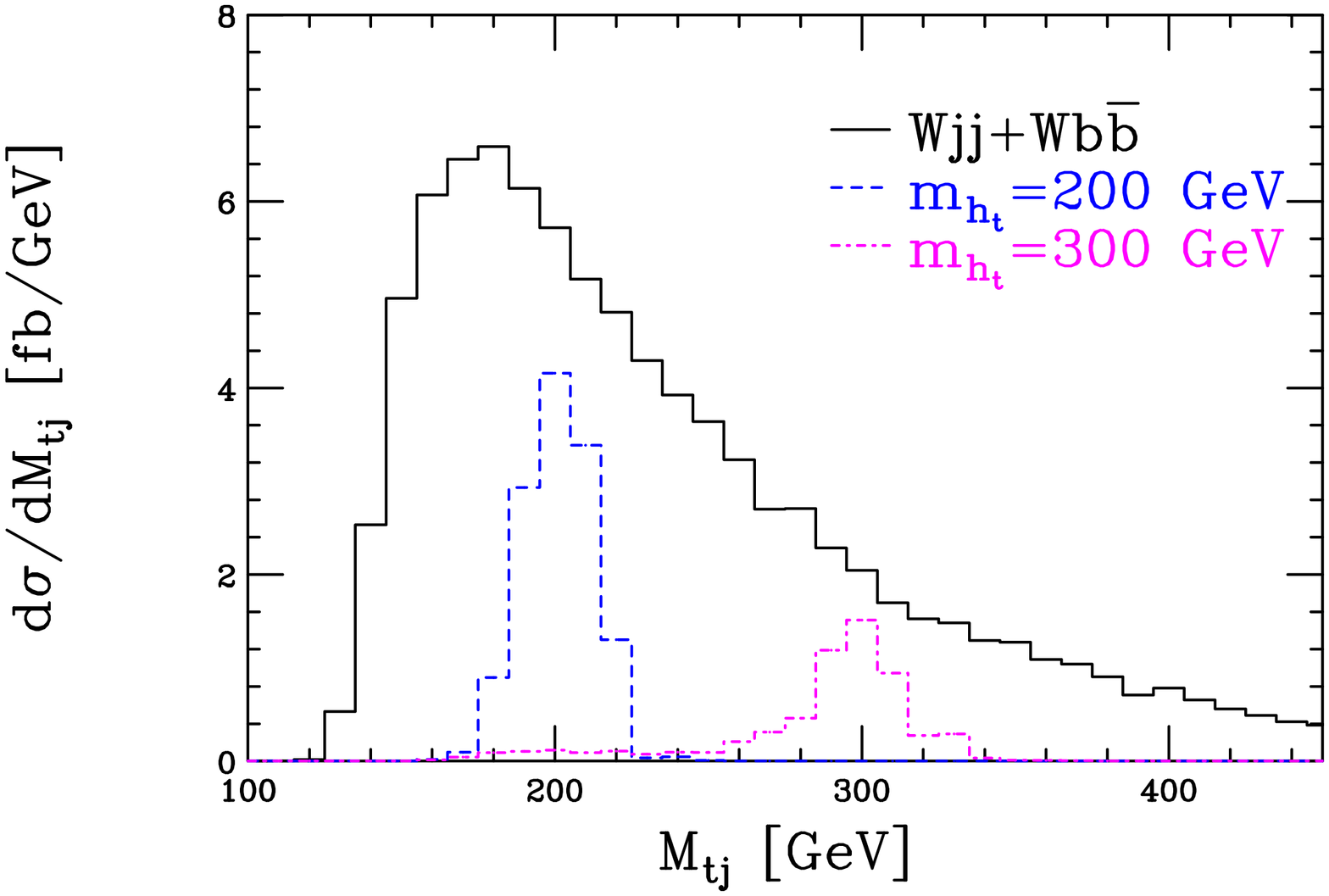}}}
\end{center}
\caption{(left) Fractional reduction in $R_b$ as a function of top-pion
mass.\protect\cite{burdmanrb} (right) Simulated signal and background for
charged top-pions in the single top sample at the
Tevatron.\protect\cite{singtpion}}
\label{fig:toppi}
\end{figure}

The strong topcolor dynamics binds top and bottom quarks into a set of
top-pions\cite{tc2,tc2phase} $t\bar{t}, t\bar{b}, b\bar{t}$ and $b\bar{b}$.
It has been observed\cite{burdmanrb} that top-pion exchange in loops would
noticeably decrease $R_b$ (Figure \ref{fig:toppi}) and this implies that
the top-pions must be quite heavy unless other physics cancels this
effect\cite{modrb}.  Several searches for top-pion and top-higgs ($\sigma$)
states have been proposed.  A singly-produced neutral top-higgs can be
detected\cite{fctpion} through its flavor-changing decays to $tc$ at Run
II.  Charged top-pions, on the other hand, would be visible\cite{singtpion}
in single top production, as in Figure \ref{fig:toppi}, up to masses of 350
GeV at Run II and 1 TeV at LHC.

\smallskip
\noindent{\tt Flavor-Universal Coloron Models}
\smallskip

\begin{figure}[tb]
\begin{center}
\scalebox{.4}{\includegraphics{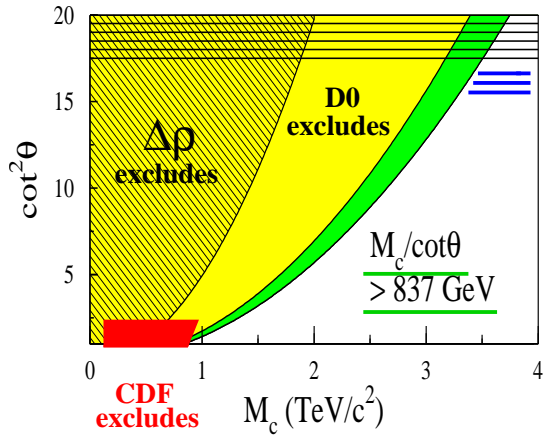}}
\end{center}
\caption{Limits\protect\cite{collim} on the mass and mixing angle of
flavor-universal colorons.}
\label{fig:coloron}
\end{figure}

The gauge structure of these models\cite{futc2} is
identical with that of the topcolor\cite{tc2} models; they differ only in
fermion charge assignments.  The fermion hypercharges are as in topcolor
models; hence, the $Z'$ phenomenology is also the same.  But as the model's
name suggests, all quarks transform under the more strongly-coupled
$SU(3)_h$ group; none transform under $SU(3)_\ell$.  As a result, the heavy
coloron bosons in the low-energy spectrum couple with equal strength to all
quarks.  Several experimental limits\cite{collim} have been placed on these
color-octet states, as shown in Figure \ref{fig:coloron}.  CDF has excluded
narrow colorons with masses below about 900 GeV by searching for resonances
decaying to dijets.  The bounds on $\Delta\rho$ exclude light colorons
which could be exchanged across quark loops in weak boson propagators.
Heavier colorons tend to be broad ($\Gamma \propto \kappa_3 M_c$) and
therefore produce a distortion of the dijet angular distribution or excess
events at high invariant mass, rather than a bump in the dijet spectrum.  A
D\O\ study of the dijet angular distribution eliminated the light-shaded
region of Figure \ref{fig:coloron} and a study\cite{collim} of the D\O\
invariant mass distribution eliminated the darker-shaded slice, giving the
limit $M_c / \cot\theta > 837$ GeV (where $\theta$ is the mixing angle
between the two $SU(3)$ groups.  This implies $M_c \gae 3.4$ TeV in
dynamical models of mass generation where the coloron coupling is strong.

\begin{figure}[tb]
\begin{center}
\scalebox{.5}{\includegraphics{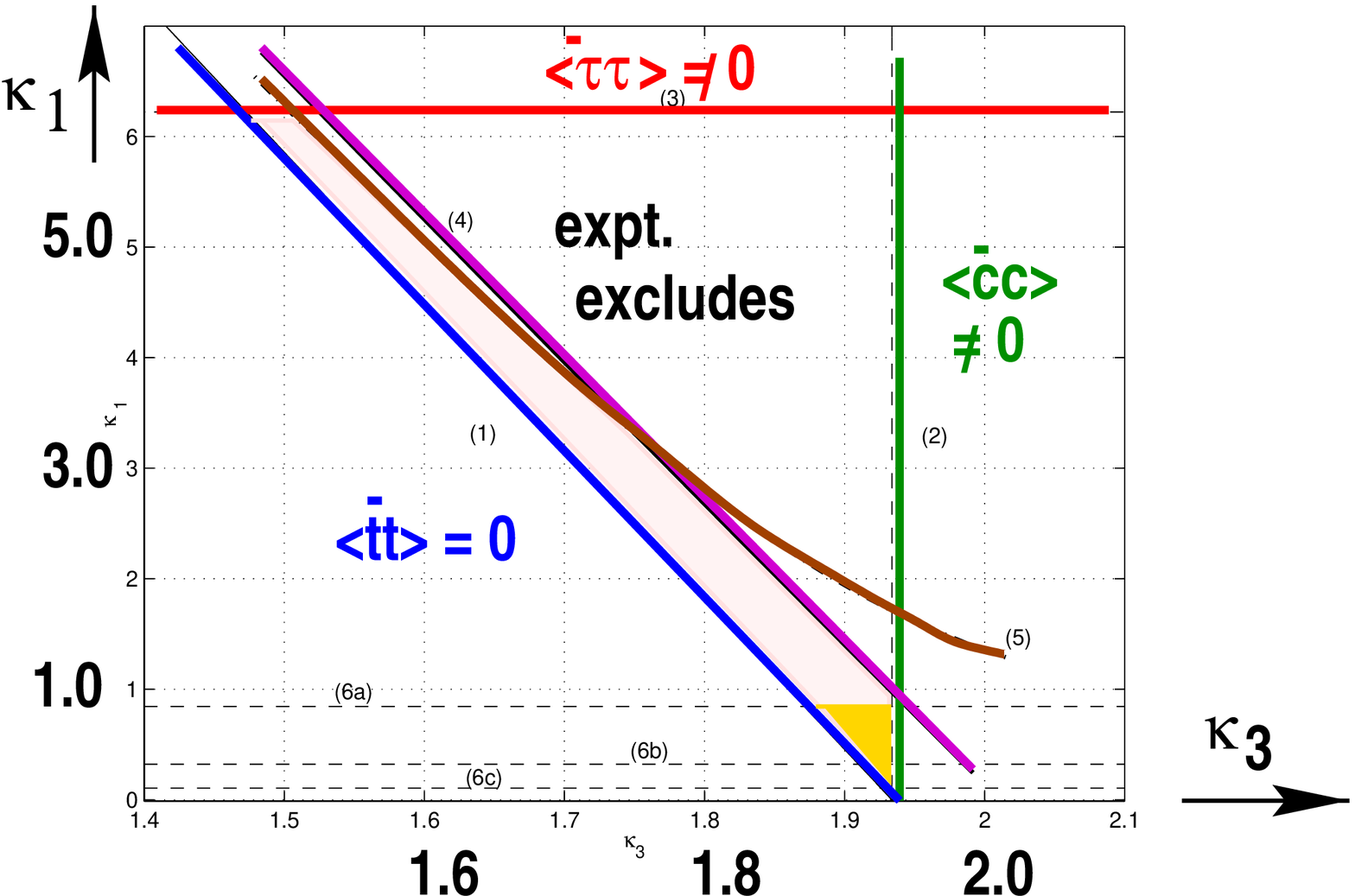}}
\end{center}
\caption{Limits on the coupling strengths $\kappa_3$ and $\kappa_1$ in
flavor-universal coloron models.\protect\cite{futc2}} 
\label{fig:phaseplane}
\end{figure}

\begin{figure}[tb]
\begin{center}
\scalebox{.35}{\includegraphics{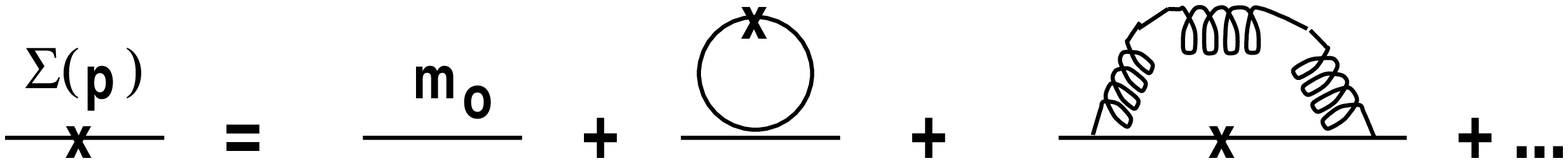}}
\end{center}
\caption{NJL gap equation for dynamical generation of fermion mass.} 
\label{fig:njlgap}
\end{figure}

In a TC2-like model incorporating flavor-universal colorons,\cite{futc2} the gauge couplings $\kappa_3 \equiv \alpha_s \cot^2\theta_3$
and $\kappa_1 \equiv \alpha_Y \cot^2\theta_1$ must satisfy several
constraints which are summarized in Figure \ref{fig:phaseplane}.  Requiring
solutions to the gauged NJL gap equations for dynamical fermion masses
(Figure \ref{fig:njlgap}) such that only the top quark condenses leads to
the inequalities\cite{futc2}
\begin{eqnarray}
&\kappa_{3}& + {2\over 27}\kappa_{1} \geq {{2 \pi}\over 3} -
{4\over3}\alpha_s - {4\over 9}\alpha_Y \ \ \ \ \ \ \langle t\bar{t}\rangle
\neq 0\nonumber \\
&\kappa_{3}& + {2\over 27}{\alpha_{Y}^{2}\over \kappa_{1}} < {{2 \pi}\over
3} - {4\over3}\alpha_s - {4\over 9}\alpha_Y\ \ \ \ \ \ \langle
c\bar{c}\rangle = 0 \nonumber \\
&\kappa_{1}&\ < 2 \pi - 6 \alpha_Y \ \ \ \ \ \ \langle \tau\tau \rangle = 0
\end{eqnarray}
which form the outer triangle in Figure \ref{fig:phaseplane}.  Mixing
between the Z and $Z'$ alters the $Z\tau\tau$ coupling by\cite{tc2zp,futc2}
\begin{equation}
\delta g_{\tau_{L}}={1\over 2}\delta 
g_{\tau_{R}}=\sin^2\theta_{W}{M_{Z}^2 
\over M_{Z'}^2} \bigg{[}1-{f_{t}^2 \over v^2}({\kappa_{1}\over 
\alpha_{Y}}+1)\bigg{]}
\end{equation}
where the top-pion decay constant is $f_{t}^2={3 \over {8\pi^2}} m_{t}^2
\ln\bigg{(}{\Lambda^2 \over m_{t}^2}\bigg{)} $.  Keeping $Z\to \tau\tau$
consistent with experiment yields the upper bound labeled (5). Both ZZ'
mixing and coloron exchange contribute\cite{tc2zp,futc2} to $\Delta\rho$
\begin{eqnarray}
\Delta \rho_{\ast}^{(C)} &\approx& {{16 {\pi}^2 \alpha_{Y}} \over {3
\sin^2\theta_{W}}} \bigg{(}{f_{t}^2 \over {M_{C} M_{Z}}}\bigg{)}^2
\kappa_{3} \nonumber \\
\Delta \rho_{\ast}^{(Z')} &\approx& {\alpha_{Y}\sin^2\theta_{W} \over
\kappa_{1}} {M_{Z}^2 \over M_{Z'}^2}\bigg{[}1-{f_{t}^2 \over
v^2}({\kappa_{1}
\over \alpha_{Y}}+1)\bigg{]}^2
\end{eqnarray}
yielding upper bound (4).  Finally, requiring that the Landau pole of
the strongly-coupled $U(1)_h$ group lie sufficiently far above the
symmetry-breaking scale $M$ yields the curves labeled (6a,b,c)
according to whether the separation of scales is by a factor of 10,
$10^2$, or $10^5$.  The combined limits\cite{futc2} indicate that the
coloron coupling is not far below critical ($\kappa_3 \sim 1.9$) while
$\kappa_1 \lae 1$.  Similar constraints exist \cite{tc2phase} for the
original TC2 models.

\smallskip
\noindent{\tt Top Seesaw Models}
\smallskip

\begin{table}[tb]
\caption{Third generation quark charge assignments in top seesaw
models.\protect\cite{topseesaw}}
\begin{center}
\begin{tabular}{|c|c|c|c|}\hline
 &  $ SU(3)_h$ & $ SU(3)_\ell$ &  $ SU(2)$ \\ \hline 
$ (t,\, b)_L$   &  3 &  1 &  2 \\
$ t_R,\ b_R$  &  1 &  3 &  1 \\
$ \chi_L$ &  1 &  3 &  1 \\
$ \chi_R$ &  3 &  1 &  1 \\
\hline
\end{tabular}
\end{center}
\label{tab:quarkchg}
\end{table}

Top seesaw models\cite{topseesaw} include an extended $SU(3)_h \times
SU(3)_\ell$ color group which spontaneously breaks to $SU(3)_{QCD}$ while
the electroweak gauge sector is standard.  In addition to the ordinary
quarks, there exist weak-singlet quarks $\chi$ which mix with the top
quark; some variants\cite{georgigrant,topseesawext} include weak-singlet
partners for the b, or for all quarks, or weak-doublet partners for some
quarks.  The color and weak quantum numbers of the third-generation quarks
are shown in Table \ref{tab:quarkchg}.  When the $SU(3)_h$ coupling becomes
strong, the dynamical mass of the top quark is created through a
combination of $t_L \chi_R$ condensation and seesaw mixing:
\begin{equation}
\scalebox{.2}{\includegraphics{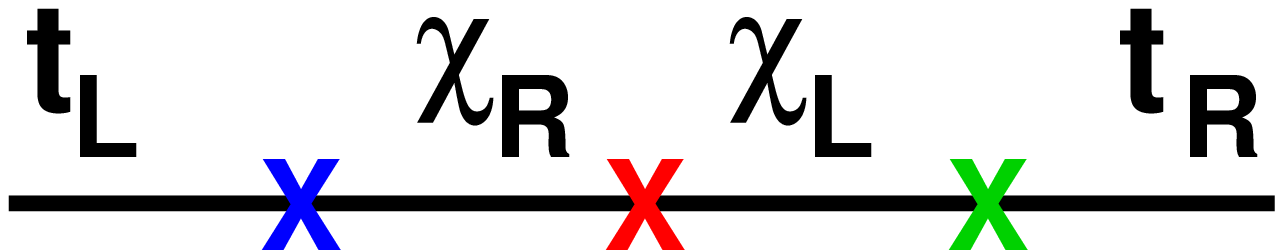}} \hspace{1cm}
\left( \begin{array}{cc} \bar{t}_L & \bar{\chi}_L \end{array} \right)
\left( \begin{array}{cc} 0 & {m_{t \chi}} \\ 
{\mu_{\chi t}} & {\mu_{\chi \chi}} \end{array} \right) 
\left( \begin{array}{c} t_R \\ \chi_R \end{array} \right)
\end{equation}
Composite scalars $\bar{t}_L \chi_R$ are also created by the strong
dynamics.

\begin{figure}[tb]
\begin{center}
\scalebox{.4}{\includegraphics{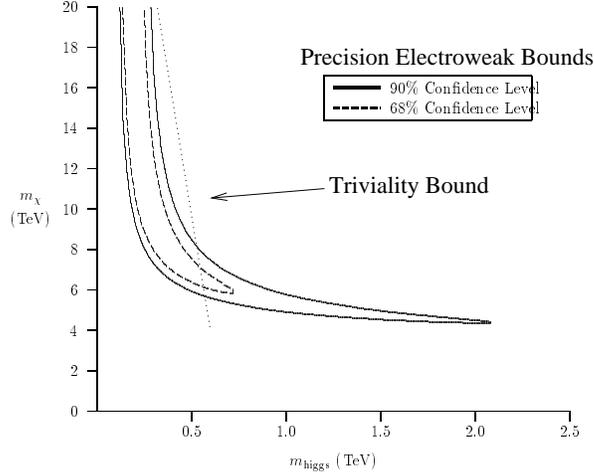}}
\end{center}
\caption{Electroweak\protect\cite{georgigrant} and
triviality\protect\cite{topseetriv} bounds on the masses of the exotic
quarks and composite scalars in a top seesaw model.  The allowed region is
within the banana-shaped region and to the left of the diagonal line.}
\label{fig:banana}
\end{figure}

The phenomenology of the weak singlet quarks has received some
attention in the literature.  Experimental limits on weak isospin
violation ($\Delta\rho$) provide a key constraint on models in which
top has a weak-singlet partner and bottom does not.  Even including a
weak-singlet partner for the b quark cannot altogether alleviate this,
as data on $R_b$ limits the mixing between b and its partner.  A
combination of precision electroweak bounds and triviality
considerations limits the $\chi$ quarks and the composite scalar to
the mass range shown Figure \ref{fig:banana}.  The exotic quarks are
required\cite{georgigrant} to have masses in excess of about 5 TeV.
Note that the upper bound on the scalar mass from electroweak
constraints at lower values of $M_\chi$ is looser than in the SM
because\cite{topseetriv} the model constrains extra contributions to
$\Delta\rho$.  

Direct searches for weak-singlet quarks are limited to
lower mass ranges; while they cannot probe the partner of the top,
they are potentially sensitive to weak-singlet partners of the lighter
quarks.  For example, a heavy mostly-weak-singlet quark $q^H$ could
contribute\cite{wsingferm} to the FNAL top dilepton sample via
\begin{equation}
 p \bar{p} \to q^H \bar{q}^H \to q^L W \bar{q}^L W 
\to q^L \bar{q}^L \ell \nu_\ell \ell' \nu_{\ell'}
\end{equation}
Comparing the number of dilepton events to the SM prediction yields a lower
bound on $M_{q^H}$.  The limits will be weaker than that for a sequential
4th generation quark because the mostly-singlet $q^H$ do not always decay
via the charged-current weak interactions.  The $d^H$ branching fraction to
$d^H \to W u^H$ is only about 60\% due to competition from the
flavor-conserving neutral current process $d^H \to Z d^L$.  In the case of
$b^H$, the cross-generation charged-current decay is also Cabibbo
suppressed and the channel $b^H \to Z b^L$ dominates. As a result, Run 1
data places the limit\cite{wsingferm} $M_{s^H, d^H} \gae 140$ GeV, but
cannot directly constrain $M_{b^H}$.  In models where all three generations
of quarks have weak-singlet partners, self-consistency
requires\cite{wsingferm} $M_{b^H}\gae 160$ GeV.

\subsection{Summary}\label{subsec:summary-bsm}

The quest for understanding electroweak symmetry breaking and fermion
masses points to physics beyond the SM.  In many theories, the top
quark is predicted to have unusual properties accessible to
experiments at the Fermilab Tevatron's Run II, the LHC or an NLC.  New
physics associated with the top quark might include new gauge
interactions or decay channels, exotic fermions mixing with top, a
light supersymmetric partner, strongly-bound top-quark states, or
something not yet even imagined.  Studying the top quark clearly
has tremendous potential to produce results that will be surprising
and enlightening.

\section*{Acknowledgments}
The author thanks R.S. Chivukula and S. Willenbrock for very useful
comments on the manuscript.  She also acknowledges the support of the NSF
POWRE and RAIS Bunting Fellowship programs.  {\it This work was supported
in part by the National Science Foundation under grant PHY-0074274 and by
the Department of Energy under grant DE-FG02-91ER40676.}

\end{document}